\documentclass[aps,pra,superscriptaddress,showpacs,showkeys]{revtex4}
\usepackage{epsfig}
\usepackage{graphicx}
\usepackage{bm}
\usepackage{amssymb}
\newcommand{\zh}{\bm}
\newcommand{\real}{\mathop{\rm Re}\nolimits}

\newcommand{\Ordnung}{\mathop{\rm O}\nolimits}

\newcommand{\oes}{{a}}

\newcommand{\oeg}{{a_0}}
\newcommand{\oegg}{{a_{\star}}}
\newcommand{\meg}{{A_0}}

\newcommand{\oee}{{\epsilon}}
\newcommand{\mee}{{E}}

\newcommand{\dee}{{\varepsilon}}

\newcommand{\omegares}{\omega^{\txt{res}}}
\newcommand{\Sigmareg}{{\hat\Sigma}}
\newcommand{\addi}{{\lambda_a}}
\newcommand{\addih}{{{\lambda'}_a}}
\newcommand{\jj}{{\mbox{$j$--$j$} }}
\newcommand{\eg}{{\it e.g.}}
\newcommand{\ie}{{\it i.e.}}
\newcommand{\etc}{{\it etc}}
\newcommand{\etal}{{\it et al.}\ }

\newcommand{\exc}{{a}}
\newcommand{\Ini}{{I}}
\newcommand{\Fin}{{F}}

\newcommand{\zhr}{{\zh r}}

\newcommand{\zhn}{{\zh n}}

\newcommand{\zhk}{{\zh k}}

\newcommand{\zhnu}{{\zh\nu}}

\newcommand{\Lg}{{N}}
\newcommand{\Vg}{{T}}
\newcommand{\setg}{{g}}
\newcommand{\bpsi}{{\bar{\psi}}}

\newcommand{\Br}[1]{(\ref{#1})}
\newcommand{\Eq}[1]{Eq.\ (\ref{#1})}

\newcommand{\Fig}[1]{Fig.\ \ref{#1}}
\newcommand{\txt}[1]{{\rm #1}}

\newcommand{\confaso}{{{}^1 S_0}}
\newcommand{\confapa}{{{}^1 P_1}}
\newcommand{\confcpa}{{{}^3 P_1}}
\newcommand{\confcpb}{{{}^3 P_2}}
\newcommand{\confcsa}{{{}^3 S_1}}
\newcommand{\xmatrix}[4]{
        \left[
        \begin{array}{cc}
        #1  & #2  \\
        #3  & #4  \\
        \end{array}
        \right]
        }
\begin{document}

\title{QED calculation of transition probabilities in
       two-electron ions}

\author{Oleg Yu.\ Andreev}
\affiliation{V.\ A.\ Fock Institute of Physics, Faculty of Physics,
             St.\ Petersburg State University, Ulyanovskaya 1,
             198504,
             Petrodvorets, St.\ Petersburg, Russia}
\affiliation{Institut f\"ur Theoretische Physik,
             Technische Universit\"at Dresden,
             Mommsenstra{\ss}e 13, D-01062, Dresden, Germany}
\author{Leonti N.\ Labzowsky}
\affiliation{V.\ A.\ Fock Institute of Physics, Faculty of Physics,
             St.\ Petersburg State University, Ulyanovskaya 1,
             198504,
             Petrodvorets, St.\ Petersburg, Russia}
\affiliation{Petersburg Nuclear Physics Institute,
             188300,
             Gatchina, St.\ Petersburg, Russia}
\author{G\"unter Plunien}
\affiliation{Institut f\"ur Theoretische Physik,
             Technische Universit\"at Dresden,
             Mommsenstra{\ss}e 13, D-01062, Dresden, Germany}

\date{\today}

\begin{abstract}
An accurate QED calculation of transition probabilities for the low-lying two-electron configurations of multicharged ions is presented.
The calculation is performed for the nondegenerate states $(1s2s)\,\confcsa$, $(1s2p_{3/2})\,\confcpb$ ($M 1$ and $M 2$ transitions, respectively) and for the quasidegenerate states $(1s2p)\,\confapa$, $(1s2p)\,\confcpa$ ($E 1$ transitions) decaying to the ground state $(1s1s)\,\confaso$.
Two-electron ions with nuclear charge numbers $Z=10-92$ are considered.
The line profile approach is employed for the description of the process in multicharged ions within the framework of QED.
\end{abstract}


\maketitle

\section{Introduction}
\label{introduction}
Highly charged ions (HCI), in particular, two-electron HCI considered in the present work are under intensive experimental and theoretical investigation during the last decades.
In HCI the electrons are propagating in the field of the nucleus,
which exceeds in strength all other external electric fields accessible in laboratories.
This allows for tests of QED in strong electric fields.
The most precise experimental data have reached a relative level of accuracy of about $1\,\%$ in one-electron ions (the measurement of the binding energy shift, \ie, the difference between the electron binding energy and the Dirac point-nucleus value for this energy, ground state, $Z=92$
\cite{stohlker07p012008}),
about $0.4\,\%$ in two-electron ions (the measurement of binding energy of one electron in the two-electron ion, ground state, $Z=92$
\cite{gumberibze04})
and about $0.03\,\%$ in three-electron ions (the measurements of the $2p_{1/2}-2s_{1/2}$ energy difference for the first excited and ground states, $Z=92$)
\cite{schweppe91}.
In the theoretical studies of the few-electron HCI such a level of accuracy requires the inclusion of the second-order (two-loop) radiative corrections as well as the screening of the first order (one-loop) radiative corrections.
The nuclear size, nuclear recoil and even the nuclear polarization corrections appear to be of importance as well.
What concerns the interelectron interaction corrections, the first-, second- and partly third-order corrections should be accounted for in high-$Z$ HCI.
For intermediate $Z$ values and, especially for the quasidegenerate energy levels (see below), the inclusion of the interelectron interaction corrections to all-orders, at least within a simplified treatment, becomes necessary.
The existing experimental data for the transition probabilities are less accurate ($3\,\%$ for $Z=54$
\cite{marrus89}),
but also require from theory to take into account the interelectron interaction and the lowest order radiative corrections.
Moreover, HCI can be used for the investigation of fundamental problems beyond QED:
First, for testing the Standard Model via the observation of Parity Nonconservation (PNC) effects in HCI.
Various suggestions on this subject were made in
\cite{zolotorev97}
for one-electron HCI, in
\cite{gorshkov74,schaeter89,dunford96,labzowsky01p054105,nefiodov02p52,gribakin05}
for two-electron HCI and in
\cite{maul97,maul96}
for four- and five-electron HCI, respectively.
Regarding two-electron HCI such proposals are based exclusively on the crossings of the energy levels with opposite parity at some $Z$ values.
For example, according to the recent calculation
\cite{artemyev05}
the energy level splitting $\Delta \mee=\mee(2{}^3P_0)-\mee(2{}^1S_0)$ for He-like Gd ($Z=64$), amounts to $\Delta\mee=0.04\pm 0.74\,\mbox{eV}$.
Since the PNC effect is proportional to $\eta=\real\{\Delta\mee-i\Gamma/2\}^{-1}=\Delta\mee/(\Delta\mee^2 + \Gamma^2/4)$, where $\Gamma=0.016\,\mbox{eV}$ is the $2{}^3P_0$ level width, the relative PNC effect can be unprecedently large ($\eta=0.05$ for $\Delta\mee=\Gamma/2$) or exactly zero for ($\Delta\mee=0$).
To eliminate the uncertainty in calculations of the energy splitting $\Delta E$ the full account for the two-loop radiative corrections and the more accurate treatment of the interelectron interaction become indispensable.
The evaluation of the PNC effects also demands a most precise knowledge of the transition probabilities (level width).
Second, it was proposed to use HCI for the search of the variation of fundamental constants
\cite{andreev05,schiller07}.
Again the precise knowledge of the level crossings and transition probabilities is needed for this purpose
\cite{andreev05,schiller07}.
In many cases the two-electron ions are preferable for performing the corresponding experiments
(see, \eg,
\cite{labzowsky01p054105,andreev05}).
Therefore, it is necessary to develop adequate and accurate methods, which allow for predictions of energy levels, transition probabilities and other characteristics of two-electron HCI with utmost precision.

\par
The most investigated properties of HCI are the energy levels of the electron configurations.
Since more than 40 years in the numerous theoretical works a large variety of different methods based on the Relativistic Many-Body Theory (RMBT) and QED were suggested and employed in practical calculations.
A short survey of these methods in a historical retrospective was presented recently in
\cite{andreev08pr}.
A common property of all these methods is that they are exact to all-orders in the parameter $\alpha Z$ ($\alpha$ is the fine-structure constant and $Z$ is the nuclear charge number).
In many-electron systems the expansion in $\alpha Z$ implies an expansion with respect to the relativistic parameter $\bar{v}/c$ ($\bar{v}$ is the mean velocity of an atomic electron, $c$ is the speed of light), so that the $\alpha Z$-expansion methods can be only applied to  nonrelativistic systems (low-$Z$ atoms and ions).
The Relativistic Dirac-Hartree-Fock (RDHF) method
\cite{desclaux02b}
and its natural extensions like Multi-Configurational RDHF (MC-RDHF)
\cite{grant02b}
or a coupled-cluster method based on the RDHF approximation (CC-RDHF)
\cite{kaldor03b}
were widely used in calculations performed for particular ion species.
Within these methods the one-electron part of the many-body Hamiltonian is treated exactly as well as the Coulomb part of the interelectron interaction Hamiltonian.
Only the Breit part of the interaction Hamiltonian is treated approximately.
The validity of this approach was analyzed thoroughly in
\cite{brown51,grant88,mittleman72}.
For few-electron ions with high-$Z$ values the application of the perturbation theory with respect to the interelectron interaction becomes possible since the interelectron interaction is of the order of $1/Z$ compared to the binding energy
\cite{labzowsky93b}.
This feature of the many-electron atomic systems is exploited in the most powerful non-QED methods for the evaluation of the properties of HCI, in the Relativistic Many Body Perturbation Theory approach (RMBPT)
\cite{johnson92p2197}.
By means of this method the most extensive calculations of the energy levels in two-electron HCI within a wide range of nuclear charge 
numbers $Z$ were performed
\cite{plante94p3519}.
Still, compared to the exact QED theory this method suffers from the lack of the negative energy contributions (though these contributions can be introduced with the special nontrivial efforts
\cite{derevianko98,chen01,savukov05}),
from the approximate treatment of the Breit interaction (without retardation) and from the neglect of the cross-photons interactions (which represent a special QED effect) in higher orders of perturbation theory.
Moreover, the inclusion of radiative QED corrections within the RMBPT approach is possible only with the use of the $\alpha Z$-expansion expressions
\cite{eides01}
which, strictly speaking, are not valid for HCI.
During the last few decades several rigorous QED approaches for the evaluation of the various properties of the HCI were formulated.
Unlike the non-QED treatments, the application of QED allows for the consequent improvement of the accuracy of calculations.
The first QED methods, based on the adiabatic S-matrix approach and the energy shift formula by Gell-Mann and Low
\cite{gellmann51}
(this formula was later adjusted by Sucher to the QED applications
\cite{sucher57})
was introduced in
\cite{labzowsky70}
and later applied to the various QED calculations by many authors.
However, in higher orders of perturbation theory the adiabatic S-matrix approach becomes rather cumbersome due to the necessity of explicit evaluation of the adiabatic limit (when adiabatic parameter tends to zero).
More advantageous for these purposes appeared to be the Two-Time Green's Function (TTGF) method first formulated in
\cite{shabaev90p43,shabaev91p5665}
(see also the recent review
\cite{shabaev02}).
With this approach a large number of calculations concerning the higher-order (two-loop) radiative corrections to the energy levels
\cite{yerokhin01p062507}
as well as the first-order radiative corrections to the hyperfine splittings in HCI
\cite{shabaev97p252}
and to the bound-electron g-factors in HCI
\cite{yerokhin02p143001}
was performed.
An original approach with the covariant generalization of the evolution operator was recently developed in
\cite{lindgren01,lindgren04}.
A special QED approach for the evaluation of the different characteristics of the HCI originates from the QED theory of the spectral line profile first developed by F.\ Low
\cite{low52}.
The application of the Line Profile Approach (LPA) to the evaluation of the energy levels shifts was first formulated in
\cite{labzowsky93karasiev},
simple examples were presented in
\cite{labzowsky98aqc}.
The LPA possesses all the advantages of the other methods and allows for the evaluation of any higher-order corrections.
A most general formulation of the LPA was given in
\cite{andreev01,andreev03,andreev04}
with application to the energy level calculations in HCI (see also the review
\cite{andreev08pr}).
In the present paper we apply the LPA to the high precision calculations of the transition probabilities in HCI.

\par
Because of inherent difficulties, the transition probabilities are less investigated than the energies.
This can be explained by the presence of an extra photon line (emitted photon) in the corresponding Feynman graphs and by the poorer convergence of the QED perturbation theory.
A calculation of the transition probabilities with respect to the relativistic corrections has been performed by Drake
\cite{drake71,drake79}
within the unified method.
The work
\cite{drake71}
presents the first relativistic calculation of transition probabilities for the $(1s2s)\,\confcsa\to(1s1s)\,\confaso$ transition.
The work
\cite{drake79}
presents the calculation of transition probabilities for the $(1s2p)\,\confcpa,\,\confapa\to(1s1s)\,\confaso$ transitions.
A comprehensive review on the transition probabilities for two-electron ions has been presented by Johnson \etal
\cite{johnson95}
about one decade ago, where the transitions probabilities for low-lying two-electron configurations have been calculated for ions within the entire range of nuclear-charge numbers $Z$.
In
\cite{johnson95}
the RMBPT approach was employed.
The contribution of the negative-energy states was discussed in
\cite{derevianko98,chen01,savukov05}.
The first complete QED evaluation of transition probabilities in HCI with the account for the interelectron interaction and radiative corrections has been presented in
\cite{sapirstein04p022113,indelicato04p062506}.
The calculation was performed for nondegenerate states for transitions with emission of electric
\cite{sapirstein04p022113}
and magnetic
\cite{indelicato04p062506}
photons, respectively.

\par
In this work we present a calculation of the transition probabilities for two-electron ions with nuclear charge $Z=10-100$.
The calculation is performed rigorously within the framework of QED.
We also present a special technique developed to master the slow convergence of the QED perturbation theory in the case of the ions with intermediate $Z$ values.
The calculation is performed for the nondegenerate states $(1s2s)\,\confcsa$, $(1s2p_{3/2})\,\confcpb$ ($M 1$ and $M 2$ transitions, respectively) and for the quasidegenerate levels $(1s2p)\,\confapa$, $(1s2p)\,\confcpa$ ($E 1$ transitions), decaying to the ground state $(1s1s)\,\confaso$.
In the present work we apply the line profile approach (LPA) for the derivation of all necessary formulas and develop it for the evaluation of transition probabilities for quasidegenerate levels in the framework of QED.
In this paper we focus on the interelectron interaction corrections and leave the inclusion of the radiative corrections to subsequent studies.

\par
Our paper is organized as follows.
In
Section~\ref{lpa}
we present the general formulation of the LPA discussing its foundations and justification.
A novel development of the LPA for the description of transition probabilities is presented in
Section~\ref{tpformulas}.
In
Subsection~\ref{tpformulasx}
we consider transition probabilities for one-electron ions.
The next subsections are devoted to two-electron ions.
In
Subsection~\ref{tpformulas0}
the generic expressions for the transition probabilities in zeroth order (\ie, neglecting the interelectron interaction) are presented.
The corresponding first-order expressions, where the one-photon exchange between the electrons is included, are given in
Subsection~\ref{tpformulas1}.
The formulas employed for the evaluation of amplitudes and transition probabilities are presented in
Section~\ref{tpevaluation}.
The formulas for the nondegenerate case
(Subsection~\ref{tpevaluation1})
and for the degenerate case
(Subsection~\ref{tpevaluation2})
are described separately.
In
Subsection~\ref{tpevaluation3}
we derive the formulas for transition probabilities as they are applied in numerical calculations.
Section~\ref{methods}
is devoted to the description of the computational methods.
The discussion and analysis of the results, their comparison with the another available data and conclusions are found in the final
Section~\ref{results}.

%
%
\section{Line profile approach}
\label{lpa}
\label{section02}
The LPA is the version of QED perturbation theory (PT) which starts from the description of the atomic electrons as a set of noninteracting particles moving in the field of the nucleus $V^\txt{nuc}$ (Furry picture) and described by the solutions of the Dirac equation
\begin{eqnarray}
(\gamma_{\mu}\hat{p}^{\mu} - \gamma_0 V^\txt{nuc} - m)\psi
&=&\label{eq080729n01}
0
\,.
\end{eqnarray}
Here, $\hat{p}^{\mu}$ are the components of the momentum 4-vector, $p^0=\dee$ is the bound electron energy, $\gamma_{\mu}$ are the conventional Dirac matrices.
Throughout this paper we use the relativistic units where $\hslash=c=1$ and the fine-structure constant $\alpha=e^2/(\hslash c)$.
The charge of the electron is $e=-|e|$.
In this paper we designate the eigenvalues of
\Eq{eq080729n01}
as $\dee$ while the physical one-electron energies as $\oee=\dee+\Delta \oee$.
The idea of the LPA is to evaluate the corrections to energy ($\Delta \oee$) as the shift of position of resonance in some scattering process due to the interaction with the quantized electromagnetic field.
This shift up to the very high orders of QED PT does not depend on the particular resonance process and when this dependence appears the concept of the energy level for the excited states cannot be strictly defined anymore
\cite{andreev08pr}.
For the practical implementation of the LPA the process of the elastic photon scattering on atomic electron was employed.
This procedure in the lowest QED PT order is depicted in
\Fig{figure161}.

\par
According to the standard Feynman rules (see, \eg,
\cite{labzowsky93b}),
the S-matrix element for the graph depicted in
\Fig{figure161}
reads
\begin{eqnarray}
S^{(2)}
&=&\label{eq080729n02}
(-ie)^2
\int d^4 x_u d^4 x_d\,
{\bar\psi}_\oeg(x_u)
\gamma^{\mu_u}A^{(k',\lambda')*}_{\mu_u}(x_u)
S(x_u,x_d)
\gamma^{\mu_d}A^{(k,\lambda)}_{\mu_d}(x_d)
\psi_\oeg(x_d)
\,,
\end{eqnarray}
where $x^{\mu}=(t,\zhr)$ denotes a spacetime point, $\psi_\oeg(x)=\psi_\oeg(\zhr)e^{-i\dee t}$ is the one-electron wave function, ${\bar\psi}=\psi^{+}\gamma^0$ is the Dirac conjugated wave function and $A^{(k,\lambda)}_{\mu}(x)=A^{(k,\lambda)}_{\mu}(\zhr)e^{-i\omega t}$ is the 4-vector of the electromagnetic field potential (photon wave function), $k^{\mu}=(\omega,\zhk)$, $\lambda$ are the wave vector and polarization.
The frequency of the absorbed and emitted photons are $\omega=|\zhk|$ and $\omega'=|\zhk'|$, respectively.
We employ the standard covariant notations for 4-vectors together with the sign convention for the metric tensor $(g_{\mu\nu})=\text{diag}(1,-1,-1,-1)$.
Einstein's sum convention is implied.
The 4-dimensional volume is $d^4x=dtd^3\zhr$.

\par
We employ the notations $x_u$, $x_d$ for the ``up'' and ``down'' vertex coordinates in
\Fig{figure161}.
These nonstandard notations will be convenient for the more complicated graphs considered below.
The bound-electron propagator is represented in terms of an eigenmode decomposition with respect to one-electron eigenstates of
\Eq{eq080729n01}:
\begin{eqnarray}
S(x_u,x_d)
&=&\label{eq080729n03}
\frac{i}{2\pi}
\int d\omega_n e^{-i\omega(t_u-t_d)}
\sum\limits_n
\frac{\psi_n(\zhr_u){\bar\psi}_n(\zhr_d)}{\omega_n - \dee_n(1-i0)}
\,.
\end{eqnarray}
The sum over $n$ runs over the entire Dirac spectrum.
Note, that the subscript at the integration variable ($\omega_n$) is not the subject of summation over $n$.
These indices are introduced again for the convenience in handling more complicated graphs.

\par
Inserting the expressions for the propagator and wave functions in
\Eq{eq080729n02}
and integrating over time and frequency variables we arrive at
\begin{eqnarray}
S^{(2)}
&=&\label{eq080729n04}
(-2\pi i)\delta(\omega'-\omega) e^2
\sum\limits_n
\frac{A^{(k',\lambda')*}_{\oeg n}A^{(k,\lambda)}_{n\oeg}}
{\omega+\dee_\oeg-\dee_n}
\,.
\end{eqnarray}
Here we employed the shorthand notation
\begin{eqnarray}
A^{(k,\lambda)}_{ab}
&=&\label{eq080729n05}
\int d^3\zhr\, {\bar\psi}_a(\zhr) \gamma^{\mu}A^{(k,\lambda)}_{\mu}(\zhr)\psi_b(\zhr)
\,.
\end{eqnarray}
The amplitude ($U$) of the process of elastic photon scattering is related to the S-matrix element via
\begin{eqnarray}
S
&=&\label{eq080729n06}
(-2\pi i)\delta(\omega'-\omega) U
\,,
\end{eqnarray}
where $\omega$ and $\omega'$ represent the energies of the initial and final states, respectively.

\par
The resonance scattering means that the photon frequency $\omega$ is close to the energy difference between two atomic levels $\omega=\dee_a-\dee_\oeg+\Ordnung(\alpha)$, where $\alpha$ is the fine-structure constant.
Accordingly, we have to retain only one term in the sum over $n$ in
\Eq{eq080729n04}.
Then the resonance amplitude looks like
\begin{eqnarray}
U^{(2)}
&=&\label{eq080729n07}
e^2
\frac{A^{(k',\lambda')*}_{\oeg a}A^{(k,\lambda)}_{a\oeg}}
{\omega+\dee_\oeg-\dee_a}
\,.
\end{eqnarray}
This amplitude has a singularity at $\omega=-\dee_\oeg+\dee_a $.
To avoid this singularity and to obtain the Lorentz profile for the photon absorption and photon emission processes we have to consider the radiative insertions into the internal electron line in
\Fig{figure161}.
This insertion is depicted in
\Fig{figure164}.

\par
The corresponding matrix element can be written down as
\begin{eqnarray}
S^{(4)}
&=&\nonumber
(-ie)^4
\int d^4 x_u d^4 x_1 d^4 x_2 d^4 x_d\,
{\bar\psi}_\oeg(x_u)
\gamma^{\mu_u}A^{(k',\lambda')*}_{\mu_u}(x_u)
S(x_u,x_1)
\\\label{eq080729n08}
&&
\times
\gamma^{\mu_1}
S(x_1,x_2)
\gamma^{\mu_2}
S(x_2,x_d)
\gamma^{\mu_d}A^{(k,\lambda)}_{\mu_d}(x_d)
D_{\mu_1\mu_2}(x_1,x_2)
\psi_\oeg(x_d)
\,,
\end{eqnarray}
where $D_{\mu_1\mu_2}(x_1,x_2)$ denotes the photon propagator which in the Feynman gauge reads
\begin{eqnarray}
D_{\mu_1\mu_2}(x_1,x_2)
&=&\label{eq080729n09}
\frac{i}{2\pi}
\int d\Omega\, I_{\mu_1\mu_2}(|\Omega|,r_{12}) e^{-i\Omega(t_1-t_2)}
\,,
\\
I_{\mu_1\mu_2}(\Omega,r_{12})
&=&
g_{\mu_1\mu_2}
\frac{1}{r_{12}} e^{i\Omega r_{12}}
\,,
\end{eqnarray}
with $r_{12}=|\zhr_1-\zhr_2|$ and the metric tensor $g_{\mu_1\mu_2}$.

\par
Integration over the time and frequency variables and employment of
\Eq{eq080729n06}
leads to the following expression for the amplitude in the resonance approximation
\begin{eqnarray}
U^{(4)}
&=&\label{eq080729n10}
U^{(2)}
\frac{\Sigmareg_{aa}(\omega+\dee_\oeg)}
{\omega+\dee_{\oeg}-\dee_a}
\,.
\end{eqnarray}
Here we introduced the energy-dependent matrix element of the electron self-energy
\begin{eqnarray}
\Sigmareg_{ud}(\xi)
&=&\label{eq080729n11a}
e^2
\sum\limits_{n}
\frac{i}{2\pi}
\int d\Omega\,
\frac{I_{unnd}(|\Omega|)}{\xi-\Omega-\dee_n(1-i0)}
\end{eqnarray}
together with the shorthand notation
\begin{eqnarray}
I_{u_1u_2d_1d_2}(\Omega)
&=&\label{eq080729n11b}
\sum\limits_{\mu_1\mu_2}
\int d\zhr_1 d\zhr_2\,
{\bar\psi}_{u_1}(\zhr_1)
{\bar\psi}_{u_2}(\zhr_2)
\gamma^{\mu_1}\gamma^{\mu_2}
I_{\mu_1\mu_2}(\Omega,r_{12})
\psi_{d_1}(\zhr_1)
\psi_{d_2}(\zhr_2)
\,.
\end{eqnarray}
It is assumed that the ultraviolet divergent matrix element
\Br{eq080729n11a}
are renormalized in a standard way for the tightly bound electrons in atoms (see, \eg,
\cite{labzowsky93b}).

\par
Repeating these insertions within the resonance approximation leads to the geometric progression.
The resummation of this progression yields
\begin{eqnarray}
U
&=&\label{eq080729n12}
e^2
\frac{A^{(k',\lambda')*}_{\oeg\exc} A^{(k,\lambda)}_{\exc\oeg}}
{\omega+\dee_\oeg-V_\exc(\omega)}
\,,
\end{eqnarray}
where
\begin{eqnarray}
V_\exc(\omega)
&=&\label{eq080729n13}
V^{(0)}_\exc + \Delta V_\exc(\omega)
\,,
\\
V^{(0)}_\exc
&=&\label{eq080729n14}
\dee_\exc
\,,
\\
\Delta V_\exc(\omega)
&=&\label{eq080729n15}
\Sigmareg_{\exc\exc}(\omega+\dee_\oeg)
\,.
\end{eqnarray}
Within the resonance approximation we can replace $\omega+\dee_\oeg$ by $\dee_\exc$.
Then $\Delta V_\exc=\Sigmareg_{\exc\exc}(\dee_\exc)$ represents the energy shift due to the lowest order electron self-energy correction.
It is convenient to decompose explicitly the electron self-energy matrix element into real and imaginary parts
\begin{eqnarray}
\Sigmareg_{aa}(\dee_\exc)
&=&\label{eq080729n16}
L_\exc -\frac{i}{2}\Gamma_\exc
\,,
\end{eqnarray}
where $L_\exc$ is the electron self-energy correction to the energy of the excited level $\exc$ and $\Gamma_\exc$ is the width of this level.
Thus, the pole in
\Eq{eq080729n12}
is shifted into the complex plane and the singularity on the real axis is avoided.
The real parts of the resonance frequency in zeroth- and first-order of PT are given by
\begin{eqnarray}
\omega^{\txt{res},(0)}
&=&\label{eq080729n17}
-\dee_\oeg+\dee_\exc
\,,
\\
\omega^{\txt{res},(0+1)}
&=&\label{eq080729n18}
-\dee_\oeg+\dee_\exc+L_\exc
\,.
\end{eqnarray}

\par
Taking the amplitude
\Eq{eq080729n12}
by square modulus, integrating over photon directions and summing over photon polarizations we arrive at the probability (cross-section) for the process of the photon scattering on an one-electron atom or ion.
In the resonance approximation this probability factorizes into the product of absorption and emission probabilities with the same Lorentz-profile factor
\cite{andreev08pr}
\begin{eqnarray}
{\mathcal L}(\omegares)
&=&\label{eq080729n19}
\frac{\Gamma_{\exc\oeg}}{(\omega-\omegares)^2+\frac{1}{4}\Gamma^2_\exc}
\,,
\end{eqnarray}
where $\Gamma_{\exc\oeg}$ is the partial width associated with the transition $\exc\to\oeg$.

\par
One of the most important problems within the LPA is the treatment of the ground state.
An insertion of the electron self-energy correction into the outer electron lines in the standard Feynman graphs describing the elastic photon scattering on an atomic electron leads unavoidably to singularities.
Since these outer lines correspond to the initial and final electron state, the problem of improving the energy of the ground state arises.
Moreover the singularities leave the whole theory incomplete so far that the S-matrix for the bound electrons actually does not exist.
One way to circumvent these difficulties was suggested decades ago by Sucher and Barbieri
\cite{barbieri78}
whose idea was to evaluate the corrections to the transition probabilities via the imaginary part of the two-loop diagonal self-energy corrections for the ground state of an atom.
Within the LPA we propose another solution of the problem
\cite{labzowsky02p22,andreev08pr}.
To introduce the radiative corrections to the ground state we consider the more complicated two-photon process of the excitation of the resonant level $a$.
This process starts from the one-photon absorption by an artificial ``lower than ground'' state $\oegg$ (see
\Fig{figure85}).
This state plays the role of a regulator, which can be again removed at the end of evaluations.
The insertions of the electron self-energy correction into the lower (or into the upper) electron propagator lead finally to the Lorentz line profile of the form
\cite{labzowsky02p22,andreev08pr}:
\begin{eqnarray}
{\mathcal L}(\omegares)
&=&\label{eq080729n20}
\frac{\Gamma_{\exc\oeg}+\Gamma_{\oeg\oegg}}
{(\omega-\omegares)^2+\frac{1}{4}(\Gamma_\exc+\Gamma_\oeg)^2}
\end{eqnarray}
in the emission process $\exc\to\oeg+\gamma$.
In the most simple case of one channel decay the partial widths $\Gamma_{\exc\oeg}$, $\Gamma_{\oeg\oegg}$ can be substituted by $\Gamma_{\exc}$, $\Gamma_{\oeg}$, respectively.
In
\Eq{eq080729n20}
the integration over the second photon frequency $\omega_0$ is performed and $\Gamma_\oeg$ represents the width of the level $\oeg$.
The presence of the width $\Gamma_\oeg$ is the only remnant of the introduced artificial ``lower than ground'' state $\oegg$.
In
\Eq{eq080729n20}
$\omegares=-\oee_\oeg+\oee_\exc$ includes both corrections $L_\exc$ to the energy $\dee_\exc$ and $L_\oeg$ to the energy $\dee_\oeg$.
If both states $\exc$ and $\oeg$ are excited states, the formula
\Br{eq080729n20}
represents the known expression for the Lorentz line shape in case of the decaying final state.
In case of the ground $\oeg$ state, we may consider $\Gamma_\oeg$ as the regulating parameter.
Setting $\Gamma_\oeg=0$ at the end of the calculation, we obtain from
\Eq{eq080729n20}
the emission line profile for the transition from the excited state $\exc$ to the ground state $\oeg$ with the radiative corrections for the ground state included in the definition of $\omegares$.

\par
Now being able to evaluate any desired property of an atom including the ground state energy corrections, we can disregard any insertions in the outer electron lines with their singularities and define the bound electron S-matrix in this way.
Heuristically, this approach seems to be convincing, however, it looks unsatisfactory from the formal point of view.
Therefore, we present a direct formal proof of the existence of the S-matrix for the bound electrons, based on the adiabatic approach.
This proof was given earlier for the $S(\infty,0)$ matrix
\cite{labzowsky83}.
However, $S(\infty,0)$ matrix in QED provides difficulties with renormalization.
Here we give the proof for $S(\infty,-\infty)$ matrix for the first time.
Our starting point is the Sucher adiabatic $S_{\addi}(\infty,-\infty)$ matrix
\cite{sucher57}
instead of Gell-Mann and Low adiabatic $S_{\addi}(\infty,0)$ matrix
\cite{gellmann51},
employed in
\cite{labzowsky83}.

\par
The standard description of an arbitrary process in the free-electron QED starts from the time evolution of an initial state to a final state governed by the evolution operator within the interaction representation based on the relation
\begin{eqnarray}
|\Phi(\infty)\rangle
&=&\nonumber
T \exp\left\{-i\int^{\infty}_{-\infty}dt\, {\hat H}^\txt{int}(t)\right\}|\Phi(-\infty)\rangle
\\
&=&\label{eq080729n21}
{\hat S}(\infty,-\infty)|\Phi(-\infty)\rangle
\,.
\end{eqnarray}
Here, ${\hat H}^\txt{int}(t)$ is the interaction Hamiltonian in the interaction representation, $|\Phi(\pm\infty)\rangle$ are the state vectors at asymptotic times $t=\pm\infty$ and ${\hat S}(\infty,-\infty)$ is the evolution operator,
usually called S-matrix.
The interaction between the particles involved is assumed to be absent at $t=\pm\infty$ and the transition probabilities due to the particle interactions at the finite time moments can be expressed in terms of matrix elements of S-matrix.

\par
Contrary to this, the interaction between the bound electrons is permanently present.
Within the adiabatic formalism of Gell-Mann and Low
\cite{gellmann51}
the interaction Hamiltonian ${\hat H}^\txt{int}(t)$ is replaced by the operator
\begin{eqnarray}
{\hat H}^\txt{int}(t)
&=&\label{eq080729n22}
e^{-\addi |t|}
{\hat H}^\txt{int}(t)
\,,
\end{eqnarray}
where $\addi$ is the adiabatic parameter.
Then, at the time moment $t=\pm\infty$ the interaction is switched off and at $t=0$ is fully switched on.
Using the interaction operator
\Eq{eq080729n22}
one can perform the QED calculations in the usual manner and then set $\addi=0$ at the end, thus restoring the full interaction for the entire time intervals.
This allows for the extension of the established techniques for calculating free-electron S-matrix elements to bound electrons in atoms.

\par
Gell-Mann and Low
\cite{gellmann51}
derived a formula which yields the energy shift of bound electron states due to the interaction
\Eq{eq080729n22}
in terms of the evolution operator ${\hat S}_{\addi}(0,-\infty)$.
Sucher
\cite{sucher57}
derived a symmetrized version of the energy shift formula, containing the matrix elements of the evolution operator ${\hat S}_{\addi}(\infty,-\infty)$.
On the basis of the Sucher's formula an adiabatic S-matrix approach for the evaluation of the energy corrections in bound electron QED was later developed
\cite{labzowsky70}.

\par
Here we apply the adiabatic approach for another purpose.
We will show that the singularities, arising after the electron self-energy insertions both in the initial and final outer electron lines in the S-matrix element corresponding to the Feynman graph
\Fig{figure161},
can be converted to the phase factor in the following way:
\begin{eqnarray}
S_{\addi}
&=&\label{eq080729n23}
(-2\pi i)\delta(\omega'-\omega)
U
\exp\left(\frac{\Sigmareg_{\oeg\oeg}(\dee_\oeg)}{i\addi}\right)
\,.
\end{eqnarray}
This is an asymptotic equation ($\addi\to+0$).
The $\addi$-dependence is located in the imaginary exponent.
Here the amplitude $U$ differs from
Eqs.\ \Br{eq080729n06},\Br{eq080729n07}
by the replacement of $\dee_\oeg$ to $\oee_\oeg=\dee_\oeg+\Sigmareg_{\oeg\oeg}(\dee_\oeg)$, where $\Sigmareg_{\oeg\oeg}(\dee_\oeg)$ is the diagonal matrix element of the lowest order electron self-energy operator.
As for the ground state $\oeg$ this matrix element is real, the phase factor does not contribute to the absolute value of amplitude defined by
\Eq{eq080729n06},
and, accordingly, to the line profile.
The proof is given in Appendix.
This proof can be repeated for any QED correction of any order.
This result justifies the employment of the energy $\oee_\oeg$ (with the QED corrections included) instead of $\dee_\oeg$ in
\Eq{eq080729n12}.
The remaining finite contributions from all the insertions in the outer lines present the QED corrections to the outer wave functions.
Moreover, this proof can be repeated also for the few-electron atoms.
In this case the ground state will be corrected not only for the QED corrections, but for the interelectron interaction as well.
Thus, the problem of the existence of the bound electron S-matrix for the scattering process described by the Feynman graph
\Fig{figure161}
is solved in any order of QED perturbation theory.

\par
Now we go over to the most general formulation of the LPA.
We will use the matrix formulation, that allows for the extension of the LPA to the case of quasidegenerate states.
This formulation is also most suitable for the application of the LPA to the evaluation of transition probabilities.
Within this formulation the generalization of
\Eq{eq080729n12}
looks like
\begin{eqnarray}
U
&=&\label{ln01}
T^{+} \frac{1}{D(\omega)}T
\,,
\end{eqnarray}
where the matrix $T$ describes the absorption of the photon by the electron in the ground state $\oeg$ with the excitation to the resonance (intermediate) state $\exc$.
The matrix $T^{+}$ describes the emission of the photon with the transition $\exc\to\oeg$.
The diagonal matrix (energy denominator) $D(\omega)$ is defined as
\begin{eqnarray}
D(\omega)
&=&\label{ln02}
\omega+\oee_\oeg-V^{(0)}
\,,
\end{eqnarray}
where $\omega$ is the photon frequency.
The resonance condition reads
\begin{eqnarray}
\omegares
&=&\label{ln03}
-\oee_\oeg+\dee_\exc
\,,
\end{eqnarray}
where $\oee_\oeg$ is the energy of the ground state, $\dee_\exc$ is the Dirac energy of the state $\exc$.
The energy $\oee_\oeg$ is not necessarily equal to the Dirac energy, it may include already the radiative corrections (see discussion above).
In
\Eq{ln02}
the diagonal matrix $D(\omega)$ involves
\begin{eqnarray}
V^{(0)}
&=&\label{ln04}
\dee_\exc
\,.
\end{eqnarray}
We employ here the matrix formulation of the single photon scattering amplitude
\Br{ln01}
on the one-electron ion in view of the further generalization of our approach to the quasidegenerate states in two-electron ions.
This amplitude is described by the Feynman graph
\Fig{figure82}.
In this diagram we describe the photon interaction with the ground state by the boxes and deliberately omit the outer electron lines describing the ground state wave functions.
The justification of this approach was given above.

\par
The next step is the insertion of the electron self energy corrections in the internal electron line within the resonance approximation (see
\Fig{figure83}).
The corresponding amplitude reads
\cite{andreev04}
\begin{eqnarray}
U
&=&\label{ln05}
T^{+}
\frac{1}{D(\omega)}{\Sigmareg}(\omega+\oee_\oeg)
\frac{1}{D(\omega)}T
\,,
\end{eqnarray}
where $\Sigmareg$ is the diagonal matrix corresponding to the regularized electron self-energy operator.
In the case under consideration this matrix reduces to the diagonal matrix element of the electron self-energy operator for the state $\exc$.

\par
Continuing recursively this process, \ie, inserting two, three, \etc, self-energies in the internal electron line in
\Fig{figure82}
and summing the geometric progression, we obtain finally
\begin{eqnarray}
U
&=&\label{ln06}
T^{+} \frac{1}{D(\omega)-\Delta V(\omega)}T
\,,
\end{eqnarray}
where $\Delta V(\omega)=\Sigmareg(\omega+\oee_\oeg)$.
Evaluation of the corresponding matrix element of the one-loop self-energy insertion at $\omega=\omegares$ leads back to
\Eq{eq080729n16}.

\par
\Eq{ln06}
illustrates the main idea of the LPA: the radiative corrections to the energy arise as the shifts of the resonances frequency due to the various insertions in the internal electron line in
\Fig{figure82}
in the resonance approximation.
Graphically
\Eq{ln06}
can be represented by the Feynman graph
\Fig{figure84}.
Instead of the correction $\Sigmareg$ in the box in graph
\Fig{figure84}
any irreducible correction can be inserted; the corresponding energy shift will arise as the resonance frequency shift in
\Eq{ln06}.

\par
In
\cite{andreev01,andreev03,andreev04}
the LPA was generalized to the few-electron ions, in particular for quasidegenerate states.
The applications to two- and three-electron ions were presented.
The general features of the LPA application to the evaluation of the transition probabilities were formulated in
\cite{andreev08pr}
and exemplified with some numerical studies.

\par
In the next section of this paper we provide the detailed derivations of transition probabilities in the few-electron ions within the framework of the LPA.

%
%
\section{Transition probabilities}
\label{tpformulas}
We are going to evaluate the transition probability for the process
\begin{eqnarray}
&&\label{tpIF}
\Ini \stackrel{\omega_{0}}{\longrightarrow} \Fin
\,,
\end{eqnarray}
where
$\Ini$ is the initial two-electron state decaying to the final state
$\Fin$ with emission of the photon $\omega_0$.
Within the framework of the LPA the state of an ion is associated with
a position of the resonance.
Therefore, we will consider a more general process which
incorporates the transition
\Br{tpIF}:
\begin{eqnarray}
&&\label{tpAIFA}
\meg
\stackrel{\omega}{\longrightarrow} \Ini
\stackrel{\omega_{0}}{\longrightarrow} \Fin
\stackrel{\omega'}{\longrightarrow}
\meg
\,,
\end{eqnarray}
\ie, a transition from the state $\meg$ (let $\meg$ be the ground state) to the state $\Ini$ with absorption of a 
photon $\omega$.
Then, the state $\Ini$ decays to the state $\Fin$ with emission of the 
photon $\omega_{0}$ and, finally, the state $\Fin$ decays back to
the state $\meg$ with emission of a photon $\omega'$.
The initial state ($\Ini$) is associated with the resonance near
$\omega=-\mee_\meg+\mee^{(0)}_{\Ini}$, where $\mee^{(0)}_{\Ini}$ is
the zero-order energy of the state $\Ini$ (sum of the Dirac energies).
The final state ($\Fin$) is defined by the resonance near
$\omega'=-\mee_\meg+\mee^{(0)}_{\Fin}$.
The energy of the ground state $\meg$ is given by $\mee_\meg$.

\par
It will be shown below that in the resonance approximation the amplitude of the scattering process
(\ref{tpAIFA})
can be written as
\begin{eqnarray}
U
&=&\label{tpu0}
T^{+} \,
\frac{1}{D(\omega')-\Delta V(\omega')} \,\, \Xi(\omega_{0}) \,\,
\frac{1}{D(\omega)-\Delta V(\omega)} \, T
\,.
\end{eqnarray}
The matrix $T$ describes the absorption of the photon $\omega$ by the ground state $\meg$, the matrix $T^{+}$ describes the emission of the photon $\omega'$ with the transition to the ground state $\meg$.
The matrix $D(\omega)$ is defined by
\Eq{ln02},
where $V^{(0)}$ is now the sum of the Dirac energies for the electrons which belong to the state $\Ini$.
The matrix $D$ is diagonal in the basis of the two-electron functions in the \jj coupling scheme.
The matrix of the interaction operator $\Delta V(\omega)$ was investigated in
\cite{andreev04}.
Here we will construct it in the first order of the perturbation theory.

\par
The right denominator corresponds to the resonance associated with
the state $\Ini$ and the left one defines the resonance for the state
$\Fin$.
The function $\Xi(\omega_{0})$ is a complicated vertex
which describes the emission of photon $\omega_0$ by the ion in the
state $\Ini$ decaying to the state $\Fin$.
The matrix element of the vertex $\Xi(\omega_{0})$ calculated on
the eigenvectors $\Phi_{\Ini}$, $\Phi_{\Fin}$ of the matrices $D(\omega)-\Delta V(\omega)$ and $D(\omega')-\Delta V(\omega')$ corresponding to the states $\Ini$ and $\Fin$, respectively,
represents the amplitude of the decay process
(\ref{tpIF})
\begin{eqnarray}
U_{\Ini\to \Fin}
&=&\label{tpuif}
\left(\Xi(\omega_{0})\right)_{\Phi_{\Fin}\Phi_{\Ini}}
\,.
\end{eqnarray}
The eigenvectors $\Phi_{\Ini}$, $\Phi_{\Fin}$ and the vertex $\Xi(\omega_0)$ can be constructed order by order employing perturbation theory.
This procedure is formulated consistently in following sections.

\par
Below in this section we will derive general formulas for the transition probabilities in two-electron ions in zeroth and first order of the QED perturbation theory considering the interelectron interaction as perturbation.

\subsection{One-electron ion}
\label{tpformulasx}
\label{subsubsection03040200}
In oder to introduce our notations we start from the one-photon transition in a one-electron ion within zeroth-order QED perturbation theory.
This process with transition from the initial state $\Ini$ into the final state $\Fin$ is described by
\Eq{tpIF}.

In zeroth-order QED perturbation theory the corresponding S-matrix element is given by Feynman graph depicted in
\Fig{figure77}
and reads
\begin{eqnarray}
S
&=&\label{tp080526n01}
 \int d^4 x\,
{\bar{\psi}_{\Fin}}(\zhr)
e^{it\dee_{\Fin}}
(-ie)\gamma^{\mu}A^{(k_0,\lambda_0) *}_{\mu}(\zhr)
e^{i\omega_0 t}
e^{-it\dee_{\Ini}}
\psi_{\Ini}(\zhr)
\,.
\end{eqnarray}
The Dirac functions $\psi_{\Ini}(\zhr)$, ${\bar{\psi}_{\Fin}}(\zhr)$ and the Dirac energies $\dee_{\Ini}$, $\dee_{\Fin}$ characterize the initial and final one-electron states.
The emitted photon is described by the momentum 4-vector $k_0$ and the polarization $\lambda_0$.

\par
In the coordinate representation the photon wave function
\begin{eqnarray}
A^{\mu(k,\lambda)}(\zhr)e^{-i\omega t}
&=&\label{photonwavefunction}
\sqrt{\frac{2\pi}{\omega}}
\epsilon^{\mu(\lambda)}
e^{-i(\omega t-\zhk\zhr)}
\,,\,\,
\mu=1,2,3
\end{eqnarray}
describes a photon with the momentum $k$ and polarization $\lambda$ ($\epsilon^{(\lambda)}$ is the polarization 4-vector).
Here, the photon wave function is understood within the ``transverse'' gauge, which in the work
\cite{johnson95}
is referred as
the ``velocity'' gauge,
\begin{eqnarray}
A^{0(k,\lambda)}(\zhr)e^{-i\omega t}
&=&\label{photonwavefunction0}
0
\,.
\end{eqnarray}

\par
Performing the integration over the time variable yields
\begin{eqnarray}
S
&=&\label{tp080526n02}
(-2\pi i)\delta(\dee_{\Fin}+\omega_0-\dee_\Ini) \,e
\int d^3 \zhr\,
{\bar{\psi}_{\Fin}}(\zhr)
\gamma^{\mu}A^{(k_0,\lambda_0) *}_{\mu}(\zhr)
\psi_{\Ini}(\zhr)
\,.
\end{eqnarray}
The expression for the amplitude ($U$) of the process is defined by
\Eq{eq080729n06}.
Then, the amplitude corresponding to
\Eq{tp080526n02}
reads
\begin{eqnarray}
U
&=&\label{tp080526n04}
\,e
\int d^3 \zhr\,
{\bar{\psi}_{\Fin}}(\zhr)
\gamma^{\mu}A^{(k_0,\lambda_0) *}_{\mu}(\zhr)
\psi_{\Ini}(\zhr)
\,.
\end{eqnarray}

\par
Within the framework of the line profile approach we consider a process described by
\Eq{tpAIFA}.
This process is depicted in
\Fig{figure512}.
The S-matrix element corresponding to
\Fig{figure512}
is written as
\begin{eqnarray}
S
&=&\nonumber
\int d^4 x_{u} d^4 x_{c}
d^4 x_{d} d \omega_{u} d \omega_{d} \,
{\bar{\psi}_\oeg}(\zhr_{u})
e^{it_{u}(\dee_\oeg)}
(-ie)\gamma^{\mu_{u}}A^{(k',\lambda') *}_{\mu_{u}}(\zhr_{u})
e^{i\omega' t_{u}}
\\
&&\nonumber
\times
\frac{i}{2\pi} \sum\limits_{u}
\frac{\psi_{u}(\zhr_{u}) {\bar{\psi}}_{u}(\zhr_{c})}
     {\omega_{u}-\dee_{u}(1-i0)}
e^{-i\omega_{u}(t_{u}-t_{c})}
(-ie)\gamma^{\mu_{c}}A^{(k_0,\lambda_0) *}_{\mu_{c}}(\zhr_{c})
e^{i\omega_0 t_{c}}
\\
&&\nonumber
\times
\frac{i}{2\pi} \sum\limits_{d} \frac{\psi_{d}(\zhr_{c})
{\bar{\psi}}_{d}(\zhr_{d})}
     {\omega_{d}-\dee_{d}(1-i0)}
e^{-i\omega_{d}(t_{c}-t_{d})}
\\
&&\label{tp080526n05}
\times
(-ie)\gamma^{\mu_{d}}A^{(k,\lambda)}_{\mu_{d}}(\zhr_{d})
e^{-i\omega t_{d}}
e^{-it_{d}(\dee_\oeg)}
\psi_\oeg(\zhr_{d})
\,.
\end{eqnarray}
We employ notations ``$u$'', ``$c$'', ``$d$'' for the upper, central and lower vertices of Feynman graphs.
Note, that subscripts at the integration variables $\omega_{u}$, $\omega_{d}$ refer to the corresponding vertices.
After integration over the time variables and over the frequencies ($\omega_{u}$, $\omega_{d}$) we get the expression defining the amplitude ($U$) of the scattering process
\begin{eqnarray}
S
&=&\nonumber
(-2\pi i) \delta(\omega'+\omega_0-\omega)
e^3
\int d^3 \zhr_{u} d^3 \zhr_{c}
d^3 \zhr_{d} \,
{\bar{\psi}_\oeg}(\zhr_{u})
\gamma^{\mu_{u}}A^{(k',\lambda') *}_{\mu_{u}}(\zhr_{u})
\sum\limits_{u}
\frac{\psi_{u}(\zhr_{u}) {\bar{\psi}}_{u}(\zhr_{c})}
     {\dee_\oeg+\omega'-\dee_{u}}
\\
&&\label{tp080526n06}
\times
\gamma^{\mu_{c}}A^{(k_0,\lambda_0) *}_{\mu_{c}}(\zhr_{c})
\sum\limits_{d} \frac{\psi_{d}(\zhr_{c})
{\bar{\psi}}_{d}(\zhr_{d})}
     {\dee_\oeg+\omega-\dee_{d}}
\gamma^{\mu_{d}}A^{(k,\lambda)}_{\mu_{d}}(\zhr_{d})
\psi_\oeg(\zhr_{d})
\\
&&\label{tp080526n062}
=
(-2\pi i) \delta(\omega'+\omega_0-\omega)
\,U
\,.
\end{eqnarray}

\par
Being interested in the transition between $\Ini$ and $\Fin$ states we consider the frequencies of the absorbed and emitted photons satisfying the conditions
\begin{eqnarray}
\omega
&=&\label{tp080527n01}
-\dee_\oeg+\dee_{\Ini}
+\Ordnung(\alpha)
\\
\omega'
&=&\label{tp080527n02}
-\dee_\oeg+\dee_{\Fin}
+\Ordnung(\alpha)
\,.
\end{eqnarray}
Let us also assume that the states $\Ini$, $\Fin$ are well isolated.
Hence, we can rewrite the amplitude in
\Eq{tp080526n062}
as
\begin{eqnarray}
U
&=&\nonumber
e^3
\int d^3 \zhr_{u} d^3 \zhr_{c}
d^3 \zhr_{d} \,
{\bar{\psi}_\oeg}(\zhr_{u})
\gamma^{\mu_{u}}A^{(k',\lambda') *}_{\mu_{u}}(\zhr_{u})
\frac{\psi_{\Fin}(\zhr_{u}) {\bar{\psi}}_{\Fin}(\zhr_{c})}
     {\dee_\oeg+\omega'-\dee_{\Fin}}
\\
&&\label{tp080527n03}
\times
\gamma^{\mu_{c}}A^{(k_0,\lambda_0) *}_{\mu_{c}}(\zhr_{c})
\frac{\psi_{\Ini}(\zhr_{c})
{\bar{\psi}}_{\Ini}(\zhr_{d})}
     {\dee_\oeg+\omega-\dee_{\Ini}}
\gamma^{\mu_{d}}A^{(k,\lambda)}_{\mu_{d}}(\zhr_{d})
\psi_\oeg(\zhr_{d})
+R
\,,
\end{eqnarray}
where $R$ denotes the terms regular at $\omega$, $\omega'$ given by
Eqs.\ \Br{tp080527n01}, \Br{tp080527n02}.
The first term is singular and it defines the resonances corresponding to the initial state $\Ini$ and to the final state $\Fin$.
In the resonance approximation we retain only the terms singular at the positions of resonances what means that we neglect the terms denoted by $R$.
The corresponding corrections are called nonresonant corrections.
These corrections give rise to an asymmetry of the line profile and define the level of accuracy at which the concept of energy levels itself becomes inadequate for the analysis of experimental data.
They are investigated in
\cite{labzowsky94,labzowsky97p271}
for highly charged ions and in
\cite{labzowsky01p143003,jentschura02,labzowsky02p1187,labzowsky02p054502,labzowsky02p024503,labzowsky03p15,labzowsky04p3271}
for the hydrogen atom.

\par
Aiming for the application of the LPA to two-electron ions we introduce the following notations.
The vertex functions $\Phi_{\oeg}$, $\Phi^{+}_{\oeg}$ representing absorption of the photon by the electron in the state $\oeg$ and  emission of the photon with subsequent decay of an atom into the state $\oeg$, respectively, are
\begin{eqnarray}
\Phi_{\oeg}(\zhr)
&=&\label{tp080527n10}
e
\gamma^{\mu}A^{(k,\lambda)}_{\mu}(\zhr)
\psi_\oeg(\zhr)
\,
\\
\Phi^{+}_{\oeg}(\zhr)
&=&\label{tp080527n11}
e
{\bar{\psi}_\oeg}(\zhr)
\gamma^{\mu}A^{(k',\lambda') *}_{\mu}(\zhr)
.
\end{eqnarray}
Then, the expression for the amplitude takes the form
\begin{eqnarray}
U
&=&\nonumber
e
\int d^3 \zhr_{u} d^3 \zhr_{c}
d^3 \zhr_{d} \,
\Phi^{+}_{\oeg}(\zhr_u)
\frac{\psi_{\Fin}(\zhr_{u}) {\bar{\psi}}_{\Fin}(\zhr_{c})}
     {\dee_\oeg+\omega'-\dee_{\Fin}}
\\
&&\label{tp080527n12}
\times
\gamma^{\mu_{c}}A^{(k_0,\lambda_0) *}_{\mu_{c}}(\zhr_{c})
\frac{\psi_{\Ini}(\zhr_{c})
{\bar{\psi}}_{\Ini}(\zhr_{d})}
     {\dee_\oeg+\omega-\dee_{\Ini}}
\Phi_{\oeg}(\zhr_d)
+R
\,.
\end{eqnarray}

\par
Introducing notations
\begin{eqnarray}
T_{n\oeg}
&=&\label{tp080527n04}
-
\int d^3 \zhr\,
{\bar{\psi}}_{n}(\zhr)
\Phi_{\oeg}(\zhr)
\,,
\\
T^{+}_{\oeg n}
&=&\label{tp080527n05}
-
\int d^3 \zhr\,
\Phi^{+}_{\oeg}(\zhr)
\psi_{n}(\zhr)
\end{eqnarray}
we can reexpress
\Eq{tp080527n03}
as
\begin{eqnarray}
U
&=&\nonumber
T^{+}_{\oeg \Fin}
\frac{1}{\dee_\oeg+\omega'-\dee_{\Fin}}
\\
&&\label{tp080527n06}
\times
e
\int  d^3 \zhr\,
{\bar{\psi}}_{\Fin}(\zhr)
\gamma^{\mu}A^{(k_0,\lambda_0) *}_{\mu}(\zhr)
\psi_{\Ini}(\zhr)
\frac{1}{\dee_\oeg+\omega-\dee_{\Ini}}
T_{\Ini\oeg}
+R
\,.
\end{eqnarray}
Introducing the matrices
\begin{eqnarray}
\Xi_{n_1 n_2}(\omega_0)
&=&\label{tp080527n07}
e \int  d^3 \zhr\,
{\bar{\psi}}_{n_1}(\zhr)
\gamma^{\mu}A^{(k_0,\lambda_0) *}_{\mu}(\zhr)
\psi_{n_2}(\zhr)
\,,
\\
D_{n_1 n_2}(\omega)
&=&\label{tp080527n08}
(\omega+\dee_\oeg-\dee_{n_1})\delta_{n_1, n_2}
\,,
\end{eqnarray}
where $\delta_{n_1, n_2}$ means the Kronecker symbol, we can write the expression for the amplitude in matrix form
\begin{eqnarray}
U
&=&\label{tp080527n09x2}
T^{+}_{\oeg}
D^{-1}(\omega')
\,\Xi(\omega_0)\,
D^{-1}(\omega)
T_{\oeg}
\\
&=&\label{tp080527n09}
T^{+}_{\oeg}
\frac{1}{D(\omega')}
\,\Xi(\omega_0)\,
\frac{1}{D(\omega)}
T_{\oeg}
\,.
\end{eqnarray}
Expression
\Br{tp080527n09}
coinside with
\Eq{tpu0}
in zeroth order: $\Delta V=0+\Ordnung(\alpha)$.
Taking into account the radiative corrections, the matrix $\Delta V$ will contain radiative insertions such as the self-energy and vacuum-polarization operators.
As it was mentioned above in the present studies we will neglect the influence of the radiative corrections.

\par
According to
\Eq{tpuif}
the amplitude of the process described by
\Eq{tpIF}
reads in the zeroth order of the perturbation theory
\begin{eqnarray}
U
&=&\label{tp080527n13}
\left( \Xi \right)_{\Fin\Ini}
\,.
\end{eqnarray}
As the matrix $D(\omega)$ is diagonal on the Dirac functions and $\Delta V=0$, the Dirac functions corresponding to the initial state $\Ini$ and to the final state $\Fin$ are the eigenvectors for the matrices $D(\omega)-\Delta V(\omega)$ and $D(\omega')-\Delta V(\omega')$, respectively.

\par
As a consequence of the application of the resonance approximation amplitude
\Eq{tp080527n13}
does not depend on the particular choice for the functions $\Phi_\oeg$, $\Phi^{+}_\oeg$, what means that the amplitude
\Br{tp080527n13}
does not depend on how the initial state $\Ini$ was excited and how the final state $\Fin$ decayed.
Accordinly, the state $\oeg$ can be an arbitrary state.
In the further derivations we will chose states $\Phi_\oeg$, $\Phi^{+}_\oeg$ calculated within lowest orders of the perturbation theory.

\subsection{Two-electron ions: zeroth-order perturbation theory}
\label{tpformulas0}
\label{subsubsection03040201}
In the zeroth order of the QED perturbation theory the S-matrix element for the process
\Eq{tpIF}
in two-electron ions is given by the Feynman graphs
\Fig{figure80}.
Within the framework of the line profile approach we consider process given by
\Eq{tpAIFA}.
We assume the state $\meg$ as being the ground state.

\par
In the approximation of noninteracting electrons the S-matrix element corresponding to the scattering process
\Br{tpAIFA}
is given by the Feynman graphs in
\Fig{figure81}.
The graphs (a) and (b)
yield the same contribution, so we will consider twice the graph
\Fig{figure81} (a).

\par
In zeroth order the wave function of the ground state can be taken as Slater determinant
\begin{eqnarray}
\Psi^{(0)}_\meg(x_1,x_2)
&=&\label{detx1}\label{defpsi2el}
\frac{1}{\sqrt{2}}
\det\{\psi_{a_0}(x_1)\psi_{b_0}(x_2)\}
=
\Psi^{(0)}_\meg(\zhr_1,\zhr_2)
e^{-i \dee_{1s} (t_1+t_2)}
\,.
\end{eqnarray}
Here, $\psi_{a_0}(x)=\psi_{1s+}(x)=\psi_{1s+}(\zhr)e^{-i\dee_{1s} t}$, $\psi_{b_0}(x)=\psi_{1s-}(x)$
refer to Dirac one-electron functions with different projections of the total one-electron angular momentum (identical with electron spin in case of $1s$-state).
In zeroth order the ground state energy ($\mee_\meg$) is $\mee^{(0)}_\meg=2\dee_{1s}$.
The S-matrix element represented by the diagrams in
\Fig{figure81}
is the same as the S-matrix element represented by
\Fig{figure512}
and, accordingly, it is given by
\Eq{tp080526n05}.

\par
With the purpose of employment of the matrix $\Xi(\omega_0)$ introduced in
\Eq{tpu0}
we rewrite the S-matrix element corresponding to the Feynman graph
\Fig{figure81} (a)
in the form
\begin{eqnarray}
S
&=&\nonumber
(-i)^2 \int d^4 x_{u_1} d^4 x_{u_2} d^4 x_{c_1}
d^4 x_{d_1} d^4 x_{d_2} d \omega_{u_1} d \omega_{d_1} d \omega_{n}
\\
&&\nonumber
\times
{\bar{\Phi}^{(0)}_\meg}(\zhr_{u_1},\zhr_{u_2})
e^{it_{u_1}(\mee^{(0)}_\meg+\omega')} \delta(t_{u_1}-t_{u_2})
\\
&&\nonumber
\times
\frac{i}{2\pi} \sum\limits_{u_1}
\frac{\psi_{u_1}(\zhr_{u_1}) {\bar{\psi}}_{u_1}(\zhr_{c_1})}
     {\omega_{u_1}-\dee_{u_1}(1-i0)}
e^{-i\omega_{u_1}(t_{u_1}-t_{c_1})}
\\
&&\nonumber
\times
(-ie)\gamma^{\mu_{c_1}}A^{(k_0,\lambda_0) *}_{\mu_{c_1}}(\zhr_{c_1})
e^{i\omega_0 t_{c_1}}
\frac{i}{2\pi} \sum\limits_{d_1} \frac{\psi_{d_1}(\zhr_{c_1})
{\bar{\psi}}_{d_1}(\zhr_{d_1})}
     {\omega_{d_1}-\dee_{d_1}(1-i0)}
e^{-i\omega_{d_1}(t_{c_1}-t_{d_1})}
\\
&&\nonumber
\times
\frac{i}{2\pi} \sum\limits_{n}
\frac{\psi_{n}(\zhr_{u_2}) {\bar{\psi}}_{n}(\zhr_{d_2})}
     {\omega_{n}-\dee_{n}(1-i0)}
e^{-i\omega_n (t_{u_2}-t_{d_2})}
\\
&&\label{tpS0x1x2el}
\times e^{-it_{d_1}(\mee^{(0)}_\meg+\omega)}
\delta(t_{d_1}-t_{d_2}) \Phi^{(0)}_\meg(\zhr_{d_1},\zhr_{d_2})
\,,
\end{eqnarray}
where we have introduced the two-electron vertex functions
\begin{eqnarray}
\Phi^{(0)}_\meg(\zhr_1,\zhr_2)
&=&\label{eq080527n22}
e\gamma^{\mu}A^{(k,\lambda)}_{\mu}(\zhr_1)
\Psi^{(0)}_\meg(\zhr_1,\zhr_2)
\,,
\\
{\bar\Phi}^{(0)}_\meg(\zhr_1,\zhr_2)
&=&\label{eq080527n23}
{\bar\Psi}^{(0)}_\meg(\zhr_1,\zhr_2)
e\gamma^{\mu}A^{\ast(k',\lambda')}_{\mu}(\zhr_1)
\end{eqnarray}
and function conjugated to the function
\Br{defpsi2el}
\begin{eqnarray}
\bar{\Psi}^{(0)}_\meg(x_1,x_2)
&=&\label{detx2}
\frac{1}{\sqrt{2}}
\det\{\bar{\psi}_{a_0}(x_1)\bar{\psi}_{b_0}(x_2)\}
=
\bar\Psi^{(0)}_\meg(\zhr_1,\zhr_2)
e^{i\dee_{1s} (t_1+t_2)}
\,.
\end{eqnarray}
As in the case of the one-electron ion the functions $\Phi^{(0)}_\meg$, ${\bar\Phi}^{(0)}_\meg$ describe the properties of the scattering process (compare
Eqs.\ (\ref{tp080527n10}, \ref{tp080527n11})).

\par
The S-matrix element, corresponding to the Feynman graph
\Fig{figure81} (a)
contains two electron propagators.
In
\Eq{tpS0x1x2el}
we formally introduced the third propagator $S(x_{u_2},x_{d_2})$.
We have to show that the expression
\Br{tpS0x1x2el}
leads to
\Eq{tp080526n06}.
With the aid of the Dirac-Sokhotsky formulas we can write
\begin{eqnarray}
[\omega_{n}-\dee_{n}(1-i0)]^{-1}
&=&\label{eqd061204n5}
\frac{2\pi}{i} \delta(\omega_{n}-\dee_{n})
-
[-\omega_{n}+\dee_{n}+i0\dee_{n})]^{-1}
\,.
\end{eqnarray}
Employing this identity and integrating over variables $t_{u_2}, t_{d_2}$ and $\omega_n$, because of orthogonality of the Dirac functions, reduces the sum over $n$ to terms $n=1s$ only.
The first term in the right-hand side of
\Eq{eqd061204n5}
yields equation
\Br{tp080526n06},
while the second term vanishes after integration over $\omega_{u_1}$ (because $\dee_{n}>0$ and the both poles lie in the same complex half plane).

\par
Going beyond the approximation of non-interacting electrons, the functions
$\Phi_\meg$, ${\bar{\Phi}_\meg}$ will be more complicated functions than Slater determinants
Eqs.\ \Br{eq080527n22}, \Br{eq080527n23}.
However, when calculating transition probability we can ignore the details involved in the preparation of the initial state ($\Ini$) and the further decay of the final state ($\Fin$).
Accordingly, in the resonance approximation, there is no need to specify the functions $\Phi_\meg$, ${\bar{\Phi}_\meg}$.

\par
In order to utilize efficiently the Feynman graphs technique within the framework of the line profile approach in the case of the two-electron ions we introduced two new elements: lower and upper boxes with letters $\meg$ inside.
These boxes describe the two-electron state $\meg$ absorbing or emitting a photon.
These lower and upper boxes correspond to the following expressions in the S-matrix elements (see
\Eq{tpS0x1x2el})
\begin{eqnarray}
&&\nonumber
e^{-it_{d_1}(\mee_\meg+\omega)}
\delta(t_{d_1}-t_{d_2}) \Phi_\meg(\zhr_{d_1},\zhr_{d_2})
\,,
\\
&&\nonumber
{\bar{\Phi}_\meg}(\zhr_{u_1},\zhr_{u_2})
e^{it_{u_1}(\mee_\meg+\omega')} \delta(t_{u_1}-t_{u_2})
\,,
\end{eqnarray}
respectively.
Accordingly, in zeroth-order of the perturbation theory the S-matrix element for the scattering process
\Eq{tpAIFA}
is represented by the graphs in
\Fig{figure29}
and is given by
\Eq{tpS0x1x2el}.

\par
Integration over the time variables in
\Br{tpS0x1x2el}
yields the following expression
\begin{eqnarray}
S
&=&\nonumber
(-i)^3 \left(\frac{i}{2\pi}\right)^3
(2\pi)^3
\sum\limits_{u_1 d_1 n}
\int d \omega_{u_1} d \omega_{d_1} d \omega_{n}
\Delta_{\ref{tpdefDelta}}
\\
&&\nonumber
\times
T^{+}_{\meg u_1 n}
e(A^{(k_0,\lambda_0) *})_{u_1 d_1} T_{d_1 n \meg }
\\
&&\label{tp2elx050503x1}
\times
[\omega_{u_1}-\dee_{u_1}(1-i0)]^{-1}
[\omega_{d_1}-\dee_{d_1}(1-i0)]^{-1}
[\omega_{n}-\dee_{n}(1-i0)]^{-1}
\,,
\end{eqnarray}
which involves the following shorthand notations
\begin{eqnarray}
\Delta_{\ref{tpdefDelta}}
&=&\label{tpdefDelta}
\delta(\mee_\meg+\omega'-\omega_{u_1}-\omega_{n})
\delta(\omega_{u_1}-\omega_{d_1}+\omega_0)
\delta(\omega_{d_1}+\omega_{n}-\mee_\meg-\omega)
\,,
\end{eqnarray}
the complicated vertex
\begin{eqnarray}
T_{n_1n_2\meg}
&=&
- \int d^{3}\zhr_1 d^{3}\zhr_2
\,\, \bar{\psi}_{n_1}(\zhr_1) \bar{\psi}_{n_2}(\zhr_2)
\Phi_\meg(\zhr_1,\zhr_2)
\end{eqnarray}
and the one-electron matrix element
\begin{eqnarray}
A^{(k,\lambda)}_{n_1 n_2}
&=&\label{defmea}
\int d^{3}\zhr\,
{\bar \psi}_{n_1}(\zhr)
\gamma^{\mu} A^{(k,\lambda)}_{\mu}(\zhr)
\psi_{n_2}(\zhr)
\,.
\end{eqnarray}

\par
We are interested in the transition probability at the frequencies corresponding to the positions of resonances
\begin{eqnarray}
\omega
&=&
-\mee_\meg+\mee^{(0)}_{\Ini}+\Ordnung (\alpha)
\,,
\\
\omega'
&=&
-\mee_\meg+\mee^{(0)}_{\Fin}+\Ordnung (\alpha)
\,.
\end{eqnarray}
To zeroth order $\mee^{(0)}_{\Ini}=\dee_{\Ini_1}+\dee_{\Ini_2}$ and
$\mee^{(0)}_{\Fin}=\dee_{\Fin_1}+\dee_{\Fin_2}$ determine the
positions of the resonances (sum of the Dirac energies)
corresponding to the initial and final states, respectively.
Within the framework of the resonance approximation one can retain only
the terms which are singular at the positions of resonances.
The Dirac energies $\dee_{\Ini_1}$, $\dee_{\Ini_2}$, $\dee_{\Fin_1}$, $\dee_{\Fin_2}$ correspond to the positive-energy electron states, accordingly, within the resonance approximation we can omit all terms
$\dee_{u_1}<0$, $\dee_{d_1}<0$, $\dee_{n}<0$,
what fixes the signs of the imaginary part of the poles in
\Eq{tp2elx050503x1}.

\par
Applying
\Eq{eqd061204n5}
we can write the following equality
\begin{eqnarray}
&&\nonumber
\Delta_{\ref{tpdefDelta}}
[\omega_{u_1}-\dee_{u_1}(1-i0)]^{-1}
[\omega_{d_1}-\dee_{d_1}(1-i0)]^{-1}
[\omega_{n}-\dee_{n}(1-i0)]^{-1}
\\
&&\nonumber
=
\Delta_{\ref{tpdefDelta}}
[\mee_\meg+\omega'-\dee_{u_1}-\dee_{n}]^{-1}
[\mee_\meg+\omega-\dee_{d_1}-\dee_{n}]^{-1}
\frac{2\pi}{i}
\delta(\mee_\meg+\omega-\omega_{d_1}-\dee_{n})
\\
&&\label{eq061108n01}
+
\Delta_{\ref{tpdefDelta}} R
\,.
\end{eqnarray}
The abbreviation $\Delta_{\ref{tpdefDelta}} R$ for the product between $\Delta_{\ref{tpdefDelta}}$ as given by
\Eq{tpdefDelta}
and the quantity $R$ referring exclusively to the terms which are regular at the positions of the resonances is employed.
These terms are regular, because the imaginary parts of the poles
enter with equal signs.

\par
We employ
\Eq{eq061108n01}
for the evaluation of
\Eq{tp2elx050503x1}.
Moreover, applying the resonance approximation in
\Eq{tp2elx050503x1}
all the regular terms $R$ in
\Br{eq061108n01}
can be omitted and the sum over ${u_1},{d_1},{n}$
reduces to the terms only, which satisfy the conditions
\begin{eqnarray}
\dee_{d_1}+\dee_n
&=&\label{eqd061204n1}
\mee^{(0)}_{\Ini}=\dee_{\Ini_1}+\dee_{\Ini_2}
\,,
\\
\dee_{u_1}+\dee_n
&=&\label{eqd061204n2}
\mee^{(0)}_{\Fin}=\dee_{\Fin_1}+\dee_{\Fin_2}
\,.
\end{eqnarray}
Accordingly, for the contribution of the Feynman graph
\Fig{figure29} (a)
we can write
\begin{eqnarray}
S^\txt{l}
&=&\nonumber
(-2\pi i)
\delta(\omega-\omega_0 -\omega')
\\
&&\label{tp2elx050503x2}
\times
T^{+}_{\meg u_1 n}
[\mee_\meg+\omega'-\dee_{u_1}-\dee_{n}]^{-1}
e A^{(k_0,\lambda_0) *}_{u_1 d_1}
[\mee_\meg+\omega -\dee_{d_1}-\dee_{n}]^{-1} T_{d_1 n \meg }
\,.
\end{eqnarray}
Here, we suppose that $u_1,d_1,n$ match with the conditions
Eqs.\ \Br{eqd061204n1}, \Br{eqd061204n2},
so that the index $d_1$ runs over $\Ini_1,\Ini_2$,
index $u_1$ runs over $\Fin_1$, $\Fin_2$ and index $n$ runs over
$\Ini_1,\Ini_2,\Fin_1,\Fin_2$.
This yields a nonvanishing contribution only if $\Ini_2=\Fin_2$
(single excitation).

\par
In the same way the expression for S-matrix element
corresponding to graph
\Fig{figure29}(b)
can be derived
\begin{eqnarray}
S^\txt{r}
&=&\nonumber
(-2\pi i)
\delta(\omega-\omega_0 -\omega')
\\
&&\label{tp2elx050503x3}
\times
T^{+}_{\meg n u_2 }
[\mee_\meg+\omega'-\dee_{n}-\dee_{u_2}]^{-1}
e A^{(k_0,\lambda_0) *}_{u_2 d_2}
[\mee_\meg+\omega-\dee_{n}  -\dee_{d_2}]^{-1} T_{n d_2\meg }
\,,
\end{eqnarray}
where the states $u_2,d_2,n$ now satisfy the conditions
\begin{eqnarray}
\dee_n + \dee_{d_2}
&=&\label{eqd061204n3}
\mee^{(0)}_I
\,,
\\
\dee_n + \dee_{u_2}
&=&\label{eqd061204n4}
\mee^{(0)}_F
\,.
\end{eqnarray}
One can verify that the results for the graphs
\Fig{figure29}(a)
and
\Fig{figure29}(b)
are equal, \ie,
\begin{eqnarray}
S^\txt{l}
&=&\label{x050525x01}
S^\txt{r}
\,.
\end{eqnarray}

\par
Our goal is to present expressions
Eqs.\ \Br{tp2elx050503x2}, \Br{tp2elx050503x3}
in the form of
\Eq{tpu0}.
In doing so, we consider the graph depicted on
\Fig{figure60}.
The block $\Xi$ represents a complicated vertex
describing the emission of the photon $\omega_0$.
This vertex can be written in the form
\begin{eqnarray}
\Xi(x_{c_1},x_{c_2},x_{s_1},x_{s_2})
&=&\label{eqd061215n01}
\Xi(\zhr_{c_1},\zhr_{c_2},\zhr_{s_1},\zhr_{s_2})
e^{it_{c_1}\omega_0}
\delta(t_{c_2}-t_{c_1}) \delta(t_{s_1}-t_{c_1})
\delta(t_{s_2}-t_{c_1})
\,.
\end{eqnarray}
The function $\Xi$ is a generic but yet unknown function.
It can be derived under the requirement that the graph in
\Fig{figure60}
yields the same contribution as the graphs in
\Fig{figure29}.
The S-matrix corresponding to
\Fig{figure60}
appears as
\begin{eqnarray}
S
&=&\nonumber
(-i)^2 \int d^4 x_{u_1} d^4 x_{u_2} d^4
x_{c_1} d^4 x_{c_2} d^4 x_{s_1} d^4 x_{s_2} d^4 x_{d_1} d^4 x_{d_2}
d\omega_{u_1} d \omega_{u_2} d \omega_{d_1} d \omega_{d_2}
\\
&&\nonumber
\times
\bar{\Phi}_\meg(\zhr_{u_1},\zhr_{u_2})
e^{it_{u_1}(\mee_\meg+\omega')} \delta(t_{u_1}-t_{u_2})
\\
&&\nonumber
\times
\frac{i}{2\pi} \sum\limits_{u_1}
\frac{\psi_{u_1}(\zhr_{u_1}) {\bar{\psi}}_{u_1}(\zhr_{c_1})}
     {\omega_{u_1}-\dee_{u_1}(1-i0)}
e^{-i\omega_{u_1}(t_{u_1}-t_{c_1})}
\frac{i}{2\pi} \sum\limits_{u_2} \frac{\psi_{u_2}(\zhr_{u_2})
{\bar{\psi}}_{u_2}(\zhr_{c_2})}
     {\omega_{u_2}-\dee_{u_2}(1-i0)}
e^{-i\omega_{u_2}(t_{u_2}-t_{c_2})}
\\
&&\nonumber
\times
(-i)\Xi(x_{c_1},x_{c_2},x_{s_1},x_{s_2})
\\
&&\nonumber
\times
\frac{i}{2\pi} \sum\limits_{d_1}
\frac{\psi_{d_1}(\zhr_{s_1}) {\bar{\psi}}_{d_1}(\zhr_{d_1})}
     {\omega_{d_1}-\dee_{d_1}(1-i0)}
e^{-i\omega_{d_1}(t_{s_1}-t_{d_1})}
\frac{i}{2\pi} \sum\limits_{d_2} \frac{\psi_{d_2}(\zhr_{s_2})
{\bar{\psi}}_{d_2}(\zhr_{d_2})}
     {\omega_{d_2}-\dee_{d_2}(1-i0)}
e^{-i\omega_{d_2}(t_{s_2}-t_{d_2})}
\\
&&\label{tp2elx050504x1}
\times
e^{-it_{d_1}(\mee_\meg+\omega)}
\delta(t_{d_1}-t_{d_2}) \Phi_\meg(\zhr_{d_1},\zhr_{d_2})
\,.
\end{eqnarray}
In the lowest order of perturbation theory the
equations~\Br{tp2elx050503x2} and \Br{tp2elx050503x3}
follow from
\Eq{tp2elx050504x1}
if we set
\begin{eqnarray}
\Xi(\zhr_{c_1},\zhr_{c_2},\zhr_{s_1},\zhr_{s_2})
&=&\label{eqd061204n6}
2e \gamma^{\mu_1} A^{(k_0,\lambda_0) *}_{\mu_1}(\zhr_{c_1})
\delta(\zhr_{c_1}-\zhr_{s_1}) \delta(\zhr_{c_2}-\zhr_{s_2})
\,.
\end{eqnarray}

\par
Consider now
\Eq{tp2elx050504x1}
with $\Xi$ given by
Eqs.\ \Br{eqd061215n01} and \Br{eqd061204n6}, respectively.
After integration over the time variables we receive
\begin{eqnarray}
S
&=&\nonumber
S^\txt{l} + S^\txt{r}
=
(-i)^3 \left(\frac{i}{2\pi}\right)^4 (2\pi)^3
\int d\omega_{u_1} d\omega_{u_2} d\omega_{d_1} d\omega_{d_2} \Delta_{\ref{eqn080213}}
\\
&&\nonumber
\times
T^{+}_{\meg u_1 u_2}
2e A^{(k_0,\lambda_0) *}_{u_1 d_1}
\delta_{u_2 d_2} T_{d_1 d_2 \meg }
\\
&&\nonumber
\times
[\omega_{u_1}-\dee_{u_1}(1-i0)]^{-1}
[\omega_{u_2}-\dee_{u_2}(1-i0)]^{-1}
\\
&&\label{tp2elx050504x2}
\times
[\omega_{d_1}-\dee_{d_1}(1-i0)]^{-1}
[\omega_{d_2}-\dee_{d_2}(1-i0)]^{-1}
\,,
\end{eqnarray}
where
\begin{eqnarray}
\Delta_{\ref{eqn080213}}
&=&\nonumber
\delta(\mee_\meg+\omega' -\omega_{u_1}-\omega_{u_2})
\delta(\omega_{u_1}+\omega_{u_2}+\omega_0-\omega_{d_1}-\omega_{d_2})
\\
&&\label{eqn080213}
\times
\delta(\omega_{d_1}+\omega_{d_2}-\mee_\meg-\omega)
\,,
\end{eqnarray}
and $\delta_{u_2 d_2}$ is the Kronecker symbol.
The employment of equalities analogous to
\Eq{eqd061204n5}
yields
\begin{eqnarray}
&&\nonumber
\Delta_{\ref{eqn080213}} [\omega_{u_1}-\dee_{u_1}(1-i0)]^{-1}
[\omega_{u_2}-\dee_{u_2}(1-i0)]^{-1}
[\omega_{d_1}-\dee_{d_1}(1-i0)]^{-1}
[\omega_{d_2}-\dee_{d_2}(1-i0)]^{-1}
\\
&&\nonumber
=
\Delta_{\ref{eqn080213}}
[\mee_\meg+\omega'-\dee_{u_1}-\dee_{u_2}]^{-1}
[\mee_\meg+\omega-\dee_{d_1}-\dee_{d_2}]^{-1}
\left(\frac{2\pi}{i}\right)^2
\delta(\omega_{u_2}-\dee_{u_2})\delta(\omega_{d_2}-\dee_{d_2})
\\
&&\label{tp2elx050504x3y1}\label{tp2elx050504x3}
+
\Delta_{\ref{eqn080213}} R
\,.
\end{eqnarray}
The term $\Delta_{\ref{eqn080213}} R$ is again understood as in
\Eq{eq061108n01}
above as shorthand notation for the regular part.

\par
Insertion of
\Eq{tp2elx050504x3}
into
\Eq{tp2elx050504x2}
leads to the expression
\begin{eqnarray}
\hspace{-0.3cm}
S
&=&\nonumber
(-2\pi i) \delta(\omega-\omega_0 -\omega')
\\
&&\label{tp2elx050504x4}
\times
T^{+}_{\meg u_1 u_2}
[\mee_\meg+\omega'-\dee_{u_1}-\dee_{u_2}]^{-1}
\Xi_{u_1 u_2 d_1 d_2}
[\mee_\meg+\omega-\dee_{d_1}-\dee_{d_2}]^{-1}
T_{d_1 d_2 \meg }
\,,
\end{eqnarray}
where
\begin{eqnarray}
\Xi_{u_1 u_2 d_1 d_2}
&=&\label{x050511x3}
2e A^{(k_0,\lambda_0) *}_{u_{1} d_{1}} \delta_{u_{2} d_{2}}
\,.
\end{eqnarray}
We also suppose that $\mee^{(0)}_{\Ini}=\dee_{d_1}  +\dee_{d_2}$,
$\mee^{(0)}_{\Fin}=\dee_{u_1}  +\dee_{u_2}$, otherwise, this term is
absent.
\Eq{tp2elx050504x4}
together with
\Eq{eq080729n06}
gives expression for the amplitude of process
\Eq{tpAIFA}.
With the use of
\Eq{tpuif}
one obtains the expression for the transition amplitude.
In zeroth order the eigenfunctions $\Phi_{\Ini}$,
$\Phi_{\Fin}$ are given by combinations of the Dirac functions in
the \jj coupling scheme.

\par
This was the goal of our derivations in this Section: to express the amplitude in the form equivalent to
\Eq{tpu0}.
This presentation of the amplitude will help us to solve the problem of the transition probabilities for the quasidegenerate states.
For the solution of this problem we need to present all the expressions in the generic matrix form
\Eq{tpu0}.

%
\subsection{Two-electron ion: first order perturbation theory
               (one-photon exchange)}
\label{tpformulas1}
\label{subsubsection03040202}
Now, we go over to the next order corrections to the transition
probabilities and consider the one-photon exchange correction.
This correction is represented by the graph in
\Fig{figure30}(a).
The corresponding S-matrix element can be written as
\begin{eqnarray}
S
&=&\nonumber
(-i)^2 \int d^4 x_1 d^4 x_2
d^4 x_{u_1} d^4 x_{u_2} d^4 x_{c_1} d^4 x_{d_1} d^4 x_{d_2}
d \omega_{u_1} d \omega_{u_2} d \omega_{d_1} d \omega_{d_2}
d \omega_{n}
d\Omega
\\
&&\nonumber
\times
{\bar{\Phi}}_\meg (\zhr_{u_1},\zhr_{u_2})
e^{it_{u_1}(\mee_\meg+\omega')} \delta(t_{u_1}-t_{u_2})
\\
&&\nonumber
\times
\frac{i}{2\pi} \sum\limits_{u_1}
\frac{\psi_{u_1}(\zhr_{u_1}) {\bar{\psi}}_{u_1}(\zhr_{c_1})}
     {\omega_{u_1}-\dee_{u_1}(1-i0)}
e^{-i\omega_{u_1}(t_{u_1}-t_1)}
\frac{i}{2\pi} \sum\limits_{u_2}
\frac{\psi_{u_2}(\zhr_{u_2}) {\bar{\psi}}_{u_2}(\zhr_2)}
     {\omega_{u_2}-\dee_{u_2}(1-i0)}
e^{-i\omega_{u_2}(t_{u_2}-t_2)}
\\
&&\nonumber
\times
(-ie)\gamma^{\mu_{c_1}} A^{(k_0,\lambda_0) *}_{\mu_{c_1}}(\zhr_{c_1})
e^{i\omega_0 t_{c_1}}
\frac{i}{2\pi} \sum\limits_{n}
\frac{\psi_{n}(\zhr_{c_1}) {\bar{\psi}}_{n}(\zhr_1)}
     {\omega_{n}-\dee_{n}(1-i0)}
e^{-i\omega_n (t_1-t_2)}
\\
&&\nonumber
\times
(-ie)^2\frac{i}{2\pi}
\gamma^{\mu_1}\gamma^{\mu_2}
I_{\mu_2 \mu_3}(|\Omega|,r_{12})
e^{-i\Omega(t_1-t_2)}
\\
&&\nonumber
\times
\frac{i}{2\pi} \sum\limits_{d_1}
\frac{\psi_{d_1}(\zhr_{1}) {\bar{\psi}}_{d_1}(\zhr_{d_1})}
     {\omega_{d_1}-\dee_{d_1}(1-i0)}
e^{-i\omega_{d_1}(t_1-t_{d_1})}
\frac{i}{2\pi} \sum\limits_{d_2}
\frac{\psi_{d_2}(\zhr_{2}) {\bar{\psi}}_{d_2}(\zhr_{d_2})}
     {\omega_{d_2}-\dee_{d_2}(1-i0)}
e^{-i\omega_{d_2}(t_2-t_{d_2})}
\\
&&\label{tpSlx1x2el}
\times
e^{-it_{d_1}(\mee_\meg+\omega)}
\delta(t_{d_1}-t_{d_2}) \Phi_\meg(\zhr_{d_1},\zhr_{d_2})
\,,
\end{eqnarray}
where $r_{12}=|\zhr_1-\zhr_2|$ and
the expressions for
$I_{\mu_1 \mu_2}(|\Omega|,r_{12})\equiv
I^\txt{c,t}_{\mu_1 \mu_2}(|\Omega|,r_{12})$
are defined in Coulomb gauge as
\begin{eqnarray}
I_{\mu_1 \mu_2}^{\rm c}(\Omega,r_{12}) \label{ic}
&=&
\frac{\delta_{\mu_1 0} \delta_{\mu_2 0}}{r_{12}}
\,,
\\
I_{\mu_1 \mu_2}^{\rm t}(\Omega,r_{12}) 
&=&\nonumber
-
\left(\frac{\delta_{\mu_1 \mu_2}}{r_{12}}\, e^{i\Omega r_{12}} +
\frac{\partial}{\partial x_1^{\mu_1}}
\frac{\partial}{\partial x_2^{\mu_2}} \frac{1}{r_{12}}\,
\frac{1-e^{i\Omega r_{12}}}{\Omega^2} \right)
\\
&&\label{it} \label{ib}
\times
(1-\delta_{\mu_1 0})(1-\delta_{\mu_2 0})
\end{eqnarray}
or in Feynman gauge as
\begin{eqnarray}
I_{\mu_1 \mu_2}(\Omega,r_{12}) 
&=&
\frac{g_{\mu_1 \mu_2}}{r_{12}}\, e^{i\Omega r_{12}}
\label{itfeynman} \label{ibfeynman}
\,.
\end{eqnarray}
We will also employ the following notation for the matrix element
\begin{eqnarray}
I^{\txt{c,t}}_{a'b'ab}(\Omega)
&=&\label{eq080616n01}
\int d^3 \zhr_1 d^3 \zhr_2 \,
\bpsi_{a'}(\zhr_1) \bpsi_{b'}(\zhr_2)
\gamma^{\mu_1}_{1}\gamma^{\mu_2}_{2}
I^{\txt{c,t}}_{\mu_1 \mu_2}(\Omega,r_{12})
\psi_\oes(\zhr_1) \psi_{b}(\zhr_2)
\,.
\end{eqnarray}
In
\Eq{tpSlx1x2el}
again the additional electron propagator (sum over $n$) is artificially introduced with the same purpose as in the previous Subsection.

\par
Integration over the time variables in
\Eq{tpSlx1x2el}
yields
\begin{eqnarray}
S
&=&\nonumber
(-i)^5 \left(\frac{i}{2\pi}\right)^6
(2\pi)^5
\\
&&\nonumber
\times
\int d^3 \zhr_1 d^3 \zhr_2
d^3 \zhr_{u_1} d^3 \zhr_{u_2}
d^3 \zhr_{c_1}
d^3 \zhr_{d_1} d^3 \zhr_{d_2}
d \omega_{u_1} d \omega_{u_2} d \omega_{d_1} d \omega_{d_2}
d \omega_{n}
d \Omega
\\
&&\nonumber
\times
{\bar{\Phi}}_\meg(\zhr_{u_1},\zhr_{u_2}) \,\,\,
\Delta_{\ref{eq080626n01}}
\\
&&\nonumber
\times
\sum\limits_{u_1} \frac{\psi_{u_1}(\zhr_{u_1})
{\bar{\psi}}_{u_1}(\zhr_{c_1})}
     {\omega_{u_1}-\dee_{u_1}(1-i0)}
\sum\limits_{u_2} \frac{\psi_{u_2}(\zhr_{u_2})
{\bar{\psi}}_{u_2}(\zhr_2)}
     {\omega_{u_2}-\dee_{u_2}(1-i0)}
\\
&&\nonumber
\times
e\gamma^{\mu_{c_1}} A^{(k_0,\lambda_0) *}_{\mu_{c_1}}(\zhr_{c_1})
\sum\limits_{n} \frac{\psi_{n}(\zhr_{c_1})
{\bar{\psi}}_{n}(\zhr_1)}
     {\omega_{n}-\dee_{n}(1-i0)}
\,\,\,
e^2 \gamma^{\mu_1}\gamma^{\mu_2}
I_{\mu_1\mu_2}(|\Omega|,r_{12})
\\
&&\label{tpSlx2x2el}
\times
\sum\limits_{d_1}
\frac{\psi_{d_1}(\zhr_{1}) {\bar{\psi}}_{d_1}(\zhr_{d_1})}
     {\omega_{d_1}-\dee_{d_1}(1-i0)}
\sum\limits_{d_2}
\frac{\psi_{d_2}(\zhr_{2}) {\bar{\psi}}_{d_2}(\zhr_{d_2})}
     {\omega_{d_2}-\dee_{d_2}(1-i0)}
\Phi_\meg(\zhr_{d_1},\zhr_{d_2})
\,,
\end{eqnarray}
where
\begin{eqnarray}
\Delta_{\ref{eq080626n01}}
&=&\nonumber
\delta(\mee_\meg+\omega'-\omega_{u_1}-\omega_{u_2})
\delta(\omega_{u_1}+\omega_{0}-\omega_{n})
\delta(\omega_n-\Omega-\omega_{d_1})
\\
&&\label{eq080626n01}
\times
\delta(\omega_{u_2}+\Omega-\omega_{d_2})
\delta(-\mee_\meg-\omega+\omega_{d_1}+\omega_{d_2})
\,.
\end{eqnarray}
Accordingly, in
\Eq{tpSlx2x2el}
we can set
\begin{eqnarray}
\omega_{u_1}
&=&
\mee_\meg + \omega' - \omega_{u_2}
\,,
\\
\omega_{d_1}
&=&
\mee_\meg + \omega - \omega_{d_2}
\,,
\\
\omega_n
&=&
\mee_\meg + \omega - \omega_{u_2}
\,.
\end{eqnarray}
Investigating the position of resonances near
\begin{eqnarray}
\omega'
&=&\label{x050511x4}
-\mee_\meg+\mee^{(0)}_{\Fin}
\,,
\\
\omega
&=&\label{x050511x5}
-\mee_\meg+\mee^{(0)}_{\Ini}
\,,
\end{eqnarray}
one can separate out those terms in
\Eq{tpSlx2x2el}
that become singular near these resonances with the aid of the following sequence of equations (compare with
\Eq{eqd061204n5})
\begin{eqnarray}
&&\nonumber
\Delta_{\ref{eq080626n01}}
[\omega_{u_1}-\dee_{u_1}(1-i0)]^{-1}
[\omega_{u_2}-\dee_{u_2}(1-i0)]^{-1}
[\omega_{n}-\dee_{n}(1-i0)]^{-1}
\\
&&\nonumber
\times
[\omega_{d_1}-\dee_{d_1}(1-i0)]^{-1}
[\omega_{d_2}-\dee_{d_2}(1-i0)]^{-1}
\\
&&
=
\Delta_{\ref{eq080626n01}}
[\mee_\meg+\omega'-\dee_{u_1}-\dee_{u_2}]^{-1}
\frac{2\pi}{i} \delta(\omega_{u_2}-\dee_{u_2})
[\mee_\meg+\omega-\dee_{n}-\dee_{u_2}]^{-1}
\\
&&\nonumber
\times
\left\{
\frac{2\pi}{i} \delta(\omega_{d_2}-\dee_{d_2})
[\mee_\meg+\omega-\dee_{d_1}-\dee_{d_2}]^{-1}
\right.
\\
&&\nonumber
\left.
-
[\mee_\meg+\omega-\omega_{d_2}-\dee_{d_1}(1-i0)]^{-1}
[-\omega_{d_2}+\dee_{d_2}+i0\dee_{d_2}]^{-1}
\vphantom{\frac{2\pi}{i}}
\right\}
+
\Delta_{\ref{eq080626n01}} R
\,.
\end{eqnarray}
Here $R$ represents the terms which are regular in the vicinity of
the resonances given by
Eqs.\ \Br{x050511x4} and \Br{x050511x5}.
The first term in the curly brackets possesses a singularity (at the resonance
\Eq{x050511x5})
either in the case of $\dee_{d_1}+\dee_{d_2}=\mee^{(0)}_\Ini$
or in the case of $\dee_{n}+\dee_{u_2}=\mee^{(0)}_\Ini$.
The second term becomes singular only in the latter case.

\par
Introducing the variable
\begin{eqnarray}
x
&=&
\omega_{d_2}-\dee_{u_2}
\end{eqnarray}
we rewrite the S-matrix element $S^\txt{ld}$ corresponding to the Feynman graph
\Fig{figure30} (a)
in the form
\begin{eqnarray}
S^\txt{ld}
&=&\nonumber
(-2\pi i)
\delta(\omega'+\omega_0-\omega)
\\
&&\nonumber
\times
T^{+}_{\meg u_1 u_2}
[\mee_\meg+\omega'-\dee_{u_1}-\dee_{u_2}]^{-1}
\sum_{n}
e A^{(k_0,\lambda_0)*}_{u_1 n}
\\
&&\nonumber
\times
\left\{ e^2 I_{n u_2 d_1 d_2}(|\dee_{d_2} - \dee_{u_2}|)
[\mee_\meg+\omega -\dee_n -\dee_{u_2}]^{-1}
[\mee_\meg+\omega - \dee_{d_1} - \dee_{d_2}]^{-1}
\vphantom{\frac{i}{2\pi}} \right.
\\
&&\nonumber
\left.
-
e^2 \frac{i}{2\pi} \int dx \, I_{n u_2 d_1 d_2}(|x|)
[x-\dee_{d_2}+\dee_{u_2}-i0\dee_{d_2}]^{-1}
\right.
\\
&&\nonumber
\times
\left.
[x-\mee_\meg-\omega +\dee_{d_1}+\dee_{u_2}-i0\dee_{d_1}]^{-1}
[\mee_\meg+\omega -\dee_n -\dee_{u_2}]^{-1}
\vphantom{\int}
\right\}
\\
&&\label{x050511x1}
\times
T_{d_1 d_2 \meg}
\,.
\end{eqnarray}
The first term in the curly brackets has usually simple poles at two different points or it has a singularity of the second order if the points coincide.
This term represents the first term of the geometric progression
built for the initial state $\Ini$ (see
\cite{andreev04} 
for details).
Summation of the geometric progression results in a shift of the position of
the resonance corresponding to the initial state
and, accordingly, in
a correction to the eigenvector of the initial state ($\Psi_\Ini$)
(see
\Eq{tpuif}).
As this term is taken into account while we generate the geometric
progression, it does not contribute to the vertex operator
and we can omit it here.

\par
Proceeding in a similar way for the evaluation of the S-matrix element $S^\txt{rd}$, corresponding to the Feynman graph
\Fig{figure30}(b),
we get
\begin{eqnarray}
S^\txt{rd}
&=&\nonumber
(-2\pi i)
\delta(\omega'+\omega_0-\omega)
\\
&&\nonumber
\times
T^{+}_{\meg u_1 u_2}
[\mee_\meg+\omega'-\dee_{u_1}-\dee_{u_2}]^{-1}
\sum_{n}
e A^{(k_0,\lambda_0) *}_{u_2 n}
\\
&&\nonumber
\times
\left\{ e^2 I_{u_1 n d_1 d_2}(|\dee_{u_1} - \dee_{d_1}|)
[\mee_\meg+\omega-\dee_{u_1}-\dee_n]^{-1}
[\mee_\meg+\omega - \dee_{d_1} - \dee_{d_2}]^{-1}
\vphantom{\frac{i}{2\pi}} \right.
\\
&&\nonumber
-
e^2 \frac{i}{2\pi} \int dx \, I_{u_1 n d_1 d_2}(|x|)
[x-\dee_{d_1}+\dee_{u_1}-i0\dee_{d_1}]^{-1}
\\
&&\nonumber
\times
\left.
[x-\mee_\meg-\omega +\dee_{d_2}+\dee_{u_1}-i0\dee_{d_2}]^{-1}
[\mee_\meg+\omega-\dee_{u_1}-\dee_n]^{-1}
\vphantom{\int}
\right\}
\\
&&\label{x050511x2}
\times
T_{d_1 d_2 \meg}
\,.
\end{eqnarray}
Note, that the graphs in
\Fig{figure30}(a)
and
\Fig{figure30}(b)
give equal contributions
(for $\dee_{d_1}+\dee_{d_2}=\mee^{(0)}_{\Ini}$ and
$\dee_{u_1}+\dee_{n}=\mee^{(0)}_{\Ini}$, respectively).
Accordingly, the equality
$
S^\txt{ld}
=
S^\txt{rd}
$
holds.
The vertexes corresponding to the graphs
\Fig{figure30}(a) and (b)
look like
\begin{eqnarray}
\Xi^{(1)\txt{ld}}_{u_1 u_2 d_1 d_2}
&=&\nonumber
\sum\limits_{n}
e A^{(k_0,\lambda_0) *}_{u_1 n}
\delta_{\dee_{n}+\dee_{u_2},\mee^{(0)}_{\Ini}}
\\
&&\nonumber
\left[
-
e^2 \frac{i}{2\pi} \int dx \, I_{n u_2 d_1 d_2}(|x|)
[x-\dee_{d_2}+\dee_{u_2}-i0\dee_{d_2}]^{-1}
\right.
\\
&&\label{x050511x1x2}
\times
\left.
[x-\mee_\meg-\omega +\dee_{d_1}+\dee_{u_2}-i0\dee_{d_1}]^{-1}
\vphantom{\int}
\right]
\\
&=&
(\Xi^{(0)} K^{(1)\txt{ld}})_{u_1 u_2 d_1 d_2}
\,,
\end{eqnarray}
\begin{eqnarray}
\Xi^{(1)\txt{rd}}_{u_1 u_2 d_1 d_2}
&=&\nonumber
\sum\limits_{n}
e A^{(k_0,\lambda_0) *}_{u_2 n}
\delta_{\dee_{n}+\dee_{u_1},\mee^{(0)}_{\Ini}}
\\
&&\nonumber
\times
\left[
-
e^2 \frac{i}{2\pi} \int dx \, I_{u_1 n d_1 d_2}(|x|)
[x-\dee_{d_1}+\dee_{u_1}-i0\dee_{d_1}]^{-1}
\right.
\\
&&\label{x050511x2x2}
\times
\left.
[x-\mee_\meg-\omega +
\dee_{d_2}+\dee_{u_1}-i0\dee_{d_2}]^{-1}
\vphantom{\int}
\right]
\\
&=&
(\Xi^{(0)} K^{(1)\txt{rd}})_{u_1 u_2 d_1 d_2}
\,,
\end{eqnarray}
where $\Xi^{(0)}$ is given by
\Eq{x050511x3}.
Here we have introduced matrices
$K^{(1)\txt{ld}}$ and $K^{(1)\txt{rd}}$ for abbreviation of the matrix elements in the square brackets.
Note, that if the initial state is well isolated the terms which do not match the condition
$\dee_{d_1}+\dee_{d_1}=\mee^{(0)}_{\Ini}$
are smaller by one order of perturbation theory and can be
omitted.
Accordingly, the matrices $K$ become diagonal.

\par
As it was done for the zeroth-order corrections, the expressions
\Br{x050511x1} and \Br{x050511x2}
should be cast into the form
\Eq{tpu0}.
For this purpose consider the graph
Fig.\ 14(a).
The complicated vertex $\Xi$ has been already composed in the zeroth
order
(\Eq{x050511x3}).
Our goal is now to evaluate the interelectron interaction corrections to
the vertex $\Xi$.
For that we will investigate the modification of a generic complicated
vertex $\Xi^\txt{gen}$ after taking into account the interelectron
interaction
(Fig.\ 14(a)).
The S-matrix element corresponding to
Fig.\ 14(a)
in the first order of perturbation theory in the interelectron
interaction looks like
\begin{eqnarray}
S
&=&\nonumber
(-i)^2 \int d^4 x_1 d^4 x_2
d^4 x_{u_1} d^4 x_{u_2} d^4 x_{c_1} d^4 x_{c_2}
d^4 x_{s_1} d^4 x_{s_2}
d^4 x_{d_1} d^4 x_{d_2}
\\
&&\nonumber
\times
d \omega_{u_1} d \omega_{u_2} d \omega_{s_1}
d\omega_{s_2} d \omega_{d_1} d \omega_{d_2} d \Omega\,
{\bar{\Phi}_\meg}(\zhr_{u_1},\zhr_{u_2})
e^{it_{u_1}(\mee_\meg+\omega')} \delta(t_{u_1}-t_{u_2})
\\
&&\nonumber
\times
\frac{i}{2\pi} \sum\limits_{u_1}
\frac{\psi_{u_1}(\zhr_{u_1}) {\bar{\psi}}_{u_1}(\zhr_{c_1})}
     {\omega_{u_1}-\dee_{u_1}(1-i0)}
e^{-i\omega_{u_1}(t_{u_1}-t_{c_1})}
\frac{i}{2\pi} \sum\limits_{u_2}
\frac{\psi_{u_2}(\zhr_{u_2}){\bar{\psi}}_{u_2}(\zhr_{c_2})}
     {\omega_{u_2}-\dee_{u_2}(1-i0)}
e^{-i\omega_{u_2}(t_{u_2}-t_{c_2})}
\\
&&\nonumber
\times
(-i)\Xi^\txt{gen} (\zhr_{c_1},\zhr_{c_2},\zhr_{s_1},\zhr_{s_2})
e^{i\omega_0 t_{c_1}} \delta(t_{c_1}-t_{c_2})
\delta(t_{c_1}-t_{s_1}) \delta(t_{s_1}-t_{s_2})
\\
&&\nonumber
\times
\frac{i}{2\pi} \sum\limits_{s_1}
\frac{\psi_{s_1}(\zhr_{s_1}) {\bar{\psi}}_{s_1}(\zhr_1)}
     {\omega_{s_1}-\dee_{s_1}(1-i0)}
e^{-i\omega_{s_1}(t_{s_1}-t_1)}
\frac{i}{2\pi} \sum\limits_{s_2}
\frac{\psi_{s_2}(\zhr_{s_2}){\bar{\psi}}_{s_2}(\zhr_{2})}
     {\omega_{s_2}-\dee_{s_2}(1-i0)}
e^{-i\omega_{s_2}(t_{s_2}-t_2)}
\\
&&\nonumber
\times
(-ie)^2\frac{i}{2\pi}\gamma^{\mu_1}\gamma^{\mu_2}
I_{\mu_1 \mu_2}(|\Omega|, \zhr_1,\zhr_2)
e^{-i\Omega(t_1-t_2)}
\\
&&\nonumber
\times
\frac{i}{2\pi} \sum\limits_{d_1}
\frac{\psi_{d_1}(\zhr_{1}) {\bar{\psi}}_{d_1}(\zhr_{d_1})}
     {\omega_{d_1}-\dee_{d_1}(1-i0)}
e^{-i\omega_{d_1}(t_1-t_{d_1})}
\frac{i}{2\pi} \sum\limits_{d_2}
\frac{\psi_{d_2}(\zhr_{2}){\bar{\psi}}_{d_2}(\zhr_{d_2})}
     {\omega_{d_2}-\dee_{d_2}(1-i0)}
e^{-i\omega_{d_2}(t_2-t_{d_2})}
\\
&&\label{tpScdx1x2el}
\times
e^{-it_{d_1}(\mee_\meg+\omega)}
\delta(t_{d_1}-t_{d_2}) \Phi_\meg(\zhr_{d_1},\zhr_{d_2})
\,.
\end{eqnarray}
Integration over the time variables yields
\begin{eqnarray}
S
&=&\nonumber
(-i)^5
\left(\frac{i}{2\pi}\right)^{7}
(2\pi)^5
\\\nonumber
&&
\times
\int d^3 \zhr_1 d^3 \zhr_2
d^3 \zhr_{u_1} d^3 \zhr_{u_2} d^3 \zhr_{c_1} d^3 \zhr_{c_2}
d^3 \zhr_{s_1} d^3 \zhr_{s_2} d^3 \zhr_{d_1} d^3 \zhr_{d_2}
d \omega_{u_1} d\omega_{u_2} d\omega_{s_1} d\omega_{s_2}
d\omega_{d_1} d\omega_{d_2} d \Omega
\\
&&\nonumber
\times
{\bar{\Phi}_\meg}(\zhr_{u_1},\zhr_{u_2}) \Delta_{\ref{eq0080626n03}}
\\
&&\nonumber
\times
\sum\limits_{u_1}
\frac{\psi_{u_1}(\zhr_{u_1}) {\bar{\psi}}_{u_1}(\zhr_{c_1})}
     {\omega_{u_1}-\dee_{u_1}(1-i0)}
\sum\limits_{u_2}
\frac{\psi_{u_2}(\zhr_{u_2}) {\bar{\psi}}_{u_2}(\zhr_{c_2})}
     {\omega_{u_2}-\dee_{u_2}(1-i0)}
\Xi^\txt{gen} (\zhr_{c_1},\zhr_{c_2},\zhr_{s_1},\zhr_{s_2})
\\
&&\nonumber
\times
\sum\limits_{s_1}
\frac{\psi_{s_1}(\zhr_{s_1}) {\bar{\psi}}_{s_1}(\zhr_{1})}
     {\omega_{s_1}-\dee_{s_1}(1-i0)}
\sum\limits_{s_2}
\frac{\psi_{s_2}(\zhr_{s_2}) {\bar{\psi}}_{s_2}(\zhr_{2})}
     {\omega_{s_2}-\dee_{s_2}(1-i0)}
e^2 \gamma^{\mu_1}\gamma^{\mu_2}
I_{\mu_1 \mu_2}(|\Omega|, \zhr_1,\zhr_2)
\\
&&\label{tpScdx1x2elx2}
\times
\sum\limits_{d_1}
\frac{\psi_{d_1}(\zhr_{1}) {\bar{\psi}}_{d_1}(\zhr_{d_1})}
     {\omega_{d_1}-\dee_{d_1}(1-i0)}
\sum\limits_{d_2}
\frac{\psi_{d_2}(\zhr_{2}) {\bar{\psi}}_{d_2}(\zhr_{d_2})}
     {\omega_{d_2}-\dee_{d_2}(1-i0)}
\Phi_\meg(\zhr_{d_1},\zhr_{d_2})
\,,
\end{eqnarray}
where
\begin{eqnarray}
\Delta_{\ref{eq0080626n03}}
&=&\nonumber
\delta(\mee_\meg+\omega'-\omega_{u_1}-\omega_{u_2})
\delta(\omega_0+\omega_{u_1}+\omega_{u_2}-\omega_{s_1}-\omega_{s_2})
\delta(-\Omega+\omega_{s_1}-\omega_{d_1})
\\
&&\label{eq0080626n03}
\delta(\Omega+\omega_{s_2}-\omega_{d_2})
\delta(\omega_{d_1}+\omega_{d_2}-\mee_\meg-\omega)
\,.
\end{eqnarray}
Again we use the sequence of equations which separate out the terms
($R$) which are regular near the positions of the resonances
under consideration
(Eqs.\ \Br{x050511x4}, \Br{x050511x5}):
\begin{eqnarray}
&&\nonumber
   \Delta_{\ref{eq0080626n03}}
     [\omega_{u_1}-\dee_{u_1}(1-i0)]^{-1}
     [\omega_{u_2}-\dee_{u_2}(1-i0)]^{-1}
\\
&&\nonumber
\times
     [\omega_{s_1}-\dee_{s_1}(1-i0)]^{-1}
     [\omega_{s_2}-\dee_{s_2}(1-i0)]^{-1}
     [\omega_{d_1}-\dee_{d_1}(1-i0)]^{-1}
     [\omega_{d_2}-\dee_{d_2}(1-i0)]^{-1}
\\
&&\nonumber
=
\Delta_{\ref{eq0080626n03}}
\left(\frac{2\pi}{i}\right)^3
     \delta(\omega_{u_2}-\dee_{u_2})
     \delta(\omega_{s_2}-\dee_{s_2})
     \delta(\omega_{d_2}-\dee_{d_2})
     \delta_{\dee_{u_1}+\dee_{u_2},\mee^{(0)}_\Fin}
     \delta_{\dee_{s_1}+\dee_{s_2},\mee^{(0)}_\Ini}
     \delta_{\dee_{d_1}+\dee_{d_2},\mee^{(0)}_\Ini}
\\\nonumber
&&
\times
     [\mee_\meg+\omega'-\dee_{u_1}-\dee_{u_2}]^{-1}
     [\mee_\meg+\omega-\dee_{s_1}-\dee_{s_2}]^{-1}
     [\mee_\meg+\omega-\dee_{d_1}-\dee_{d_2}]^{-1}
\\\nonumber
&&
+
\Delta_{\ref{eq0080626n03}}
\left(\frac{2\pi}{i}\right)^2
     \delta(\omega_{u_2}-\dee_{u_2})
     \delta(\omega_{s_2}-\dee_{s_2})
     \delta_{\dee_{u_1}+\dee_{u_2},\mee^{(0)}_\Fin}
     \delta_{\dee_{s_1}+\dee_{s_2},\mee^{(0)}_\Ini}
     [\mee_\meg+\omega'-\dee_{u_1}-\dee_{u_2}]^{-1}
\\\nonumber
&&
\times
     [\mee_\meg+\omega-\dee_{s_1}-\dee_{s_2}]^{-1}
     [\mee_\meg+\omega-\omega_{d_2}-\dee_{d_1}(1-i0)]^{-1}
 (-1)[-\omega_{d_2}+\dee_{d_2}+i0\dee_{d_2}]^{-1}
\\\nonumber
&&
+
\Delta_{\ref{eq0080626n03}}
\left(\frac{2\pi}{i}\right)^2
     \delta(\omega_{u_2}-\dee_{u_2})
     \delta(\omega_{d_2}-\dee_{d_2})
     \delta_{\dee_{u_1}+\dee_{u_2},\mee^{(0)}_\Fin}
     \delta_{\dee_{d_1}+\dee_{d_2},\mee^{(0)}_\Ini}
     [\mee_\meg+\omega'-\dee_{u_1}-\dee_{u_2}]^{-1}
\\\nonumber
&&
\times
     [\mee_\meg+\omega-\omega_{s_2}-\dee_{s_1}(1-i0)]^{-1}
 (-1)[-\omega_{s_2}+\dee_{s_2}+i0\dee_{s_2}]^{-1}
     [\mee_\meg+\omega-\dee_{d_1}-\dee_{d_2}]^{-1}
\\
&&
+
\Delta_{\ref{eq0080626n03}}
R
\,.
\end{eqnarray}
Applying this result to
\Eq{tpScdx1x2elx2}
and combining the graphs
Figs.\ \ref{figure60}, 14(a)
(see
\Eq{tp2elx050504x4})
for the S-matrix element $S^\txt{cd}$, corresponding to the Feynman graph
Fig.\ 14(a)
we get
\begin{eqnarray}
S^\txt{cd}
&=&\nonumber
(-2\pi i)
\delta(\omega'+\omega_0-\omega)
\\
&&\nonumber
\times
T^{+}_{\meg u_1 u_2}
[\mee_\meg+\omega'-\dee_{u_1}-\dee_{u_2}]^{-1}
\Xi^\txt{gen}_{u_1 u_2 s_1 s_2}
\\
&&\nonumber
\times
\left\{
\vphantom{\left[\frac{2}{i}\right]^{1/2}}
e^2 I_{s_1 s_2 d_1 d_2}(|\dee_{s_1} -\dee_{d_1}|)
[\mee_\meg+\omega-\dee_{s_1}-\dee_{s_2}]^{-1}
[\mee_\meg+\omega - \dee_{d_1} - \dee_{d_2}]^{-1}
\right.
\\
&&\nonumber
+
\left[
\vphantom{\frac{i}{2\pi}}
(1)_{s_1s_2d_1d_2}
[\mee_\meg+\omega - \dee_{d_1} - \dee_{d_2}]^{-1}
\right.
\\
&&\nonumber
-
e^2 \frac{i}{2\pi} \int dx \, I_{s_1 s_2 d_1 d_2}(|x|)
[x-\dee_{d_1}+\dee_{s_1}-i0\dee_{d_1}]^{-1}
\\
&&\nonumber
\times
[x-\mee_\meg-\omega +
\dee_{d_2}+\dee_{s_1}-i0\dee_{d_2}]^{-1}
[\mee_\meg+\omega-\dee_{s_1}-\dee_{s_2}]^{-1}
\vphantom{\int}
\\
&&\nonumber
-
e^2 \frac{i}{2\pi} \int dx \, I_{s_1 s_2 d_1
d_2}(|x|)
[x-\dee_{d_1}+\dee_{s_1}+i0\dee_{s_1}]^{-1}
\\
&&\label{x050518x1}
\left. \left.
\times
[x-\dee_{d_1}+\mee_\meg+\omega -\dee_{s_2}+i0\dee_{s_2}]^{-1}
[\mee_\meg+\omega-\dee_{d_1}-\dee_{d_2}]^{-1}
\vphantom{\frac{i}{2\pi}}
\right]
\right\}
T_{d_1 d_2 \meg}
\,.
\end{eqnarray}
The first term in the curly brackets has singularities at
\begin{eqnarray}
\omega'
&=&\label{eqx061211n01}
-\mee_\meg+\dee_{u_1}+\dee_{u_2}
\,,
\\
\omega
&=&\label{eqx061211n02}
-\mee_\meg+\dee_{s_1}+\dee_{s_2}
\\
\omega
&=&\label{eqx061211n03}
-\mee_\meg+\dee_{d_1}+\dee_{d_2}
\,.
\end{eqnarray}
It can be considered as the first term of the geometric progression corresponding to the initial state.
This progression can be summed up (see
\cite{andreev04}).
After this the position of the resonance corresponding to
the initial state will include the interelectron interaction
correction (one-photon exchange).

\par
The first term in the square brackets
($(1)_{s_1s_2d_1d_2}=\delta_{s_1 d_1}\delta_{s_2 d_2}$)
represents the contribution
of the graph
\Fig{figure60}.
The terms in the square brackets have singularities given by
Eqs.\ \Br{eqx061211n01}
and by either
\Br{eqx061211n02}
or
\Br{eqx061211n03}.
The last two terms represent the interelectron interaction
correction to the generic vertex $\Xi^\txt{gen}$.
The whole term in the square brackets corresponds both to
the vertex $\Xi^\txt{gen}$ and to the vertex $T_{d_1 d_2 \meg}$, which
represents the process of excitation of the ground state by the photon
$\omega$ with transition to the excited state ($\Ini$).
Accordingly, the contribution of this term to the vertex $\Xi^\txt{gen}$ appears with the power of $1/2$.

\par
Suppose that the initial state is isolated, \ie, the admixture of the other states has a magnitude of the next order
of the perturbation theory.
Then, we can set
\begin{eqnarray}
\dee_{s_1}+\dee_{s_2}
&=&
\dee_{d_1}+\dee_{d_2}=E^{(0)}_{\Ini}
\end{eqnarray}
and omit the summation over $s_1,s_2$ in the square
brackets in
\Eq{x050518x1}.
Accordingly, the vertex $\Xi^\txt{cd}$ with the interelectron
interaction correction given by the graph
Fig.\ 14(a)
will look like
\begin{eqnarray}
\Xi^\txt{cd}_{u_1 u_2 d_1 d_2}
&=&\nonumber
\Xi^\txt{gen}_{u_1 u_2 s_1 s_2}
\left[
\vphantom{\frac{i}{2\pi}}
(1)_{s_1s_2d_1d_2}
\right.
\\
&&\nonumber
-
e^2 \frac{i}{2\pi} \int dx \, I_{s_1 s_2 d_1 d_2}(|x|)
[x-\dee_{d_1}+\dee_{s_1}-i0\dee_{d_1}]^{-1}
\\
&&\nonumber
\times
[x-\mee_\meg-\omega +
\dee_{d_2}+\dee_{s_1}-i0\dee_{d_2}]^{-1}
\vphantom{\int}
\\
&&\nonumber
-
e^2 \frac{i}{2\pi} \int dx \, I_{s_1 s_2 d_1
d_2}(|x|)
[x-\dee_{d_1}+\dee_{s_1}+i0\dee_{s_1}]^{-1}
\\
&&\label{x050518x1x2}
\left.
\times
[x-\dee_{d_1}+\mee_\meg+\omega -\dee_{s_2}+i0\dee_{s_2}]^{-1}
\vphantom{\frac{i}{2\pi}}
\right]^{1/2}
\\
&=&\nonumber
\Xi^\txt{gen}_{u_1 u_2 s_1 s_2}
\left(
\left[
1+K^{(1)\txt{cd}}
\right]^{1/2}
\right)_{s_1 s_2 d_1 d_2}
\,.
\end{eqnarray}
The last equality defines the matrix $K^{(1)\txt{cd}}$.
Note, that the correction factor appears under the square root.
The term in square brackets in
\Eq{x050518x1}
can be equally referred to the vertex $\Xi^\txt{gen}$ and to the vertex $T_{d_1d_2\meg}$.
Accordingly, the part of this term connected with $\Xi^\txt{gen}$, thus contributing to $\Xi^\txt{cd}$, is the square root of this term.

\par
The vertex $\Xi^\txt{cd}$ should be equal to the sum
of the contributions of
Eqs.\ \Br{x050511x3}, \Br{x050511x1x2}, \Br{x050511x2x2}
\begin{eqnarray}
\Xi^\txt{cd}
&=&\nonumber
\Xi^{(0)}
\left(
1
+
K^{(1)\txt{ld}}
+
K^{(1)\txt{rd}}
\right)
\,.
\end{eqnarray}
Accordingly, we derive ($\Xi^\txt{d}=\Xi^\txt{gen}$)
\begin{eqnarray}
\Xi^\txt{d}
&=&\nonumber
\Xi^{(0)}
\left(
1
+
K^{(1)\txt{ld}}
+
K^{(1)\txt{rd}}
\right)
\left(
1+K^{(1)\txt{cd}}
\right)^{-1/2}
\,.
\end{eqnarray}

\par
Employing an expansion
\begin{eqnarray}
(1+x)^{-1/2}
&=&
1- \frac{1}{2}x +\Ordnung(x^2)
\end{eqnarray}
and neglecting the higher order terms we can write
\begin{eqnarray}
\Xi^\txt{d}
&=&
\Xi^{(0)} \left( 1 + K^{(1)\txt{ld}}
+ K^{(1)\txt{rd}} - \frac{1}{2} K^{(1)\txt{cd}}
\right) +e\Ordnung(\alpha^2)
\,.
\end{eqnarray}

\par
Since we are interested only in the corrections of the zeroth and of
the first order we can set
\begin{eqnarray}
\omega
&=&
-\mee_\meg + \mee_{\Ini}
\,.
\end{eqnarray}
The first order correction to the $\Xi$ arise in the case of
the reference state, \ie, when some of the following conditions
are fulfilled
\begin{eqnarray}
\dee_{n}+\dee_{u_2}
&=&\label{eq06117n02}
\dee_{d_1}+\dee_{d_2}
\,,
\\
\dee_{u_1}+\dee_{n}
&=&\label{x050525x04}
\dee_{d_1}+\dee_{d_2}
\,,
\\
\dee_{s_1}+\dee_{s_2}
&=&\label{eq06117n03}
\dee_{d_1}+\dee_{d_2}
\,.
\end{eqnarray}
It is possible to describe the contributions of graphs
\Fig{figure30}(a),(b)
as twice the contribution of the graph
\Fig{figure30}(a).
Accordingly, the reference states are defined by
Eqs.\ \Br{eq06117n02} and \Br{eq06117n03}, respectively.

\par
Thus, we can write
\begin{eqnarray}
K^{(1)\txt{ld}} + K^{(1)\txt{rd}} - \frac{1}{2}
K^{(1)\txt{cd}}
&=&\nonumber
2K^{(1)\txt{ld}} - \frac{1}{2} K^{(1)\txt{cd}}
\\
&=&\nonumber
-2 e^2 \frac{i}{2\pi} \int dx \, I_{n u_2 d_1 d_2}(|x|)
[x-\dee_{d_2}+\dee_{u_2}-i0\dee_{d_2}]^{-2}
\\
&&\nonumber
-
\left\{
-e^2 \frac{i}{2\pi} \int dx \, I_{s_1 s_2 d_1 d_2}(|x|)
[x-\dee_{d_1}+\dee_{s_1}-i0\dee_{d_1}]^{-2} \right.
\\
&&\label{x050519x1}
\left.
-
e^2 \frac{i}{2\pi} \int dx \, I_{s_1 s_2 d_1 d_2}(|x|)
[x-\dee_{d_1}+\dee_{s_1}+i0\dee_{s_1}]^{-2}
\right\}
\,.
\end{eqnarray}
This expression can be simplified with the aid of the following identities
\begin{eqnarray}
&&\nonumber
2 I_{n u_2 d_1 d_2}(|x|) [x-\dee_{d_2}+\dee_{u_2}-i0]^{-2}
\\
&&\nonumber
-
I_{s_1 s_2 d_1 d_2}(|x|) \left\{ [x-\dee_{d_2}+\dee_{u_2}-i0]^{-2}
+ [x-\dee_{d_2}+\dee_{u_2}+i0]^{-2} \right\}
\\
&&\nonumber
=
I_{n u_2 d_1 d_2}(|x|) \left\{ [x-\dee_{d_2}+\dee_{u_2}-i0]^{-2} -
[x-\dee_{d_2}+\dee_{u_2}+i0]^{-2} \right\}
\\
&&
=
\left( \frac{2 \pi}{i} \right)
\frac{\partial}{\partial x} I_{n u_2 d_1 d_2}(|x|)
\delta(x-\dee_{d_2}+\dee_{u_2})
\,.
\end{eqnarray}
Here the formula
\begin{eqnarray}
\frac{1}{(x+i0)^2}-\frac{1}{(x-i0)^2}
&=&
- \frac{2\pi}{i}
\frac{\partial}{\partial x} \delta(x)
\end{eqnarray}
was utilized.

\par
The same procedure should be applied to
\Fig{figure30}(c),(d)
and
Fig.\ 14(b),
where the one-photon exchange is inserted above the emission of
the photon $\omega_{0}$.

\par
Finally, we can write the following expression for the vertex
$\Xi$
\begin{eqnarray}
\Xi
&=&\label{x050520x25}
\Xi^{(0)} + \Xi^{(1)} + e\Ordnung
(\alpha^2)
\,,
\end{eqnarray}
where
\begin{eqnarray}
{ \Xi}^{(0)}_{u_1 u_2 d_1 d_2}
&=&\label{x050520x3}
2e
A^{(k_0,\lambda_0) *}_{u_1 d_1} \delta_{u_2 d_2}
\,,
\\
{ \Xi}^{(1)}_{u_1 u_2 d_1 d_2}
&=&\nonumber
\sum\limits_{n\atop \dee_{n}+\dee_{u_2}=\dee_{d_1}+\dee_{d_2}} e^3
A^{(k_0,\lambda_0)*}_{u_1 n} \left.
\frac{\partial}{\partial x}
 { I}_{n u_2 d_1 d_2}(|x|)\right|_{x=\dee_{u_2}-\dee_{d_2}}
\\
&&\label{x050520x2}
+
\sum\limits_{n\atop \dee_{n}+\dee_{d_2}=\dee_{u_1}+\dee_{u_2}}
e^3
\left. \frac{\partial}{\partial x}
 { I}_{u_1 u_2 n d_2}(|x|)\right|_{x=\dee_{d_2}-\dee_{u_2}}
A^{(k_0,\lambda_0)*}_{n d_1}
\,.
\end{eqnarray}
\Eq{x050520x2}
represents the reducible part of the first-order corrections, \ie, the reference states contribution.

\par
Having constructed a general expression for the vertex $\Xi$
we can apply the formulas derived for the generic graphs
Figs.\ 14 and \ref{figure60}
for evaluating the contributions of the graphs
Figs.\ \ref{figure29} and \ref{figure30}.
Now we can express these contributions via the matrix $\Xi$, what enables us to extend the calculations to the quasidegenerate levels.
To derive the formula for the amplitude
\Eq{tpuif}, we will have to consider separately the case of nondegenerate
levels and the case of quasidegenerate levels, respectively.

%
%
\section{Evaluation of transition probabilities}
\label{tpevaluation}
\label{section-evaluation}
Evaluating the transition probabilities we should distinguish
nondegenerate and quasidegenerate levels.
For the nondegenerate levels standard QED perturbation theory can be applied.
Configurations are called quasidegenerate, if they can not be considered
as being well isolated.
For these configurations the interelectron interaction must be taken into account up to higher orders.
Accordingly, this requires to develop a special technique.

\par
In the next two subsections we derive expression for the amplitude of the scattering process for nondegenerate and quasidegenerate levels, respectively.
In the last subsection we write down the final expressions for transition probabilities suitable for numerical calculations.

\subsection{Nondegenerate levels}
\label{tpevaluation1}
\label{subsection-tp-two-non}
\label{subsection030403}
Here, we will suppose that the initial and the final states are well isolated.
The set of graphs in
Figs.\ \ref{figure29} and \ref{figure30}
should be divided into two subsets: reducible
(containing the reference states) and irreducible.
For the zero-order and the reducible subset of diagrams formulas
Eqs.\ \Br{tpuif}, \Br{x050520x25}
can be applied, where the functions $\Phi_{\Ini}$, $\Phi_{\Fin}$ are given by
a combination of the two electron determinants in
\jj coupling scheme.
For irreducible subsets of diagrams we can apply a procedure described below.

\par
Consider the first terms in the curly brackets in
Eqs.\ \Br{x050511x1} and \Br{x050511x2} 
\begin{eqnarray}
S^\txt{ld}
&=&\label{x050520x20}
(-2\pi i)
\delta(\omega'+\omega_0-\omega) T^{+}_{\meg u_1 u_2}
[\mee_\meg+\omega'-\dee_{u_1}-\dee_{u_2}]^{-1}
\\\nonumber
&&
\times
\left\{ (-1) e^3 {\sum\limits_{n}}' A^{(k_0,\lambda_0) *}_{u_1 n}
[\mee_\meg+\omega -\dee_{u_2}-\dee_n]^{-1}
I_{n u_2 d_1 d_2}(|\dee_{d_2} - \dee_{u_2}|) \right\}
\\\nonumber
&&
\times
[\mee_\meg+\omega - \dee_{d_1} - \dee_{d_2}]^{-1}
T_{d_1 d_2 \meg}
\,,
\end{eqnarray}
\begin{eqnarray}
S^\txt{rd}
&=&\nonumber
(-2\pi i)
\delta(\omega'+\omega_0-\omega) T^{+}_{\meg u_1 u_2}
[\mee_\meg+\omega'-\dee_{u_1}-\dee_{u_2}]^{-1}
\\
&&\nonumber
\times
\left\{ (-1) e^3 {\sum\limits_{n}}' A^{(k_0,\lambda_0) *}_{u_2 n}
[\mee_\meg+\omega -\dee_{u_1}-\dee_n]^{-1} I_{u_1 n d_1
d_2}(|\dee_{u_1} - \dee_{d_1}|) \right\}
\\
&&\label{x050520x21}
\times
[\mee_\meg+\omega - \dee_{d_1} - \dee_{d_2}]^{-1}
T_{d_1 d_2 \meg}
\,.
\end{eqnarray}
The prime at the sum symbol indicates that in
\Eq{x050511x1}
terms for which
$\dee_{d_1}+\dee_{d_2}-\dee_{n}-\dee_{u_2} = 0$ (and in
\Eq{x050511x2}
terms, where $\dee_{d_1}+\dee_{d_2}-\dee_{u_1}-\dee_{n} = 0$) holds
are omitted.
Since the levels $\Ini$, $\Fin$ are well isolated the expressions in the curly brackets in
\Br{x050520x20}, \Br{x050520x21}
can be viewed as corrections to the vertex $\Xi$;
one can also set $\omega=-\mee_\meg+\mee^{(0)}_{\Ini}$
in the vertex.
Thus, we have to take into account the following corrections
\begin{eqnarray}
\Xi^{(1)\mbox{\rm d}}
&=&\nonumber
(-1) e^3
{\sum\limits_{n \atop \dee_{d_1}+\dee_{d_2}-\dee_{n}-\dee_{u_2}
\ne 0}} A^{(k_0,\lambda_0) *}_{u_1 n} [\dee_{d_1} +
\dee_{d_2}-\dee_{u_2}-\dee_n]^{-1}
I_{n u_2 d_1 d_2}(|\dee_{d_2} - \dee_{u_2}|)
\\
&&\label{x050523x01}
\hspace{-0.3cm}
+
(-1) e^3 {\sum\limits_{n \atop
\dee_{d_1}+\dee_{d_2}-\dee_{u_1}-\dee_{n} \ne 0}} A^{(k_0,\lambda_0) *}_{u_2 n}
[\dee_{d_1} + \dee_{d_2}-\dee_{u_1}-\dee_n]^{-1}
I_{u_1 n d_1 d_2}(|\dee_{u_1} - \dee_{d_1}|)
\end{eqnarray}
(this is the contribution of the graphs
\Fig{figure30}(a),(b))
and
\begin{eqnarray}
\Xi^{(1)\mbox{\rm u}}
&=&\nonumber
(-1) e^3
{\sum\limits_{n \atop \dee_{u_1}+\dee_{u_2}-\dee_{n}-\dee_{d_2}
\ne 0}} I_{u_1 u_2 n d_2}(|\dee_{d_2} - \dee_{u_2}|)
[\dee_{u_1}+\dee_{u_2} -\dee_{d_2}-\dee_n]^{-1} A^{(k_0,\lambda_0)*}_{d_1 n}
\\
&&\label{x050523x02}
\hspace{-0.3cm}
+
(-1) e^3 {\sum\limits_{n \atop
\dee_{u_1}+\dee_{u_2}-\dee_{d_1}-\dee_{n} \ne 0}}
I_{u_1 u_2 d_1 n}(|\dee_{u_1} - \dee_{d_1}|)
[\dee_{u_1}+\dee_{u_2}-\dee_{d_1}-\dee_n]^{-1} A^{(k_0,\lambda_0)*}_{d_2 n}
\end{eqnarray}
(this is the contribution of the graphs in
\Fig{figure30}(c),(d)).
Accordingly, for nondegenerate levels the amplitude \Eq{tpuif} is
given by the matrix element of
\begin{eqnarray}
\Xi
&=&\label{x050525x07}
\Xi^{(0)}+\Xi^{(1)}+\Xi^{(1)\mbox{\rm
d}}+\Xi^{(1)\mbox{\rm u}}
\end{eqnarray}
evaluated with the aid of the zeroth-order wave functions corresponding to
the states $\Ini$ and $\Fin$, \ie, by means of two-electron determinants
in the \jj coupling scheme.

%
%
\subsection{Quasidegenerate levels}
\label{tpevaluation2}
\label{subsection-tp-two-quasi}
\label{subsection030404}
In the previous section we introduced the vertex $\Xi$ via the expression
\Eq{tpu0}.
In order to derive the amplitude as defined in
\Eq{tpuif}
the wave functions $\Phi_{\Ini}$, $\Phi_{\Fin}$ have to be constructed.
These functions are eigenvectors of the matrix $V$ which was investigated in
\cite{andreev04}.
Diagonalization of the matrix $V$ is a serious task, because $V$ has infinite dimension.
One possible solution of this problem is the substitution by a large but finite matrix.
Another strategy is the modification of a perturbation theory.
Here we will concentrate on the development of a proper perturbation theory.

\par
The perturbation theory for the case of a nondegenerate level (as well as for the case of the fully degenerate levels) is well known
\cite{landau77b}.
Here we will apply it to the case of the quasidegenerate levels.
Considering $N$ two-electron states $\{\Psi\}$ defined in the \jj coupling scheme we assume that these states are mixing with each other, \ie, they have the same symmetry and the corresponding energy levels are close to each other.
Under such condition the standard perturbation theory may not work and we have to modify it.
These $N$ states compose a set
$\setg = \{\Psi_{i_{\setg}}, i_{\setg}=1,\ldots,N \}$.
The idea is now to build an eigenvector $\Phi_{n_{\setg}}$ corresponding to a state $\Psi_{n_{\setg}} \in \setg$.
We also suppose, that all the other states (beyond the set $\setg$)
are either nonmixing with the state $n_{\setg}$ or their energy
levels are far enough from the level $n_{\setg}$, \ie, that the set $\setg$ is large enough to incorporate all the closely
lying levels.
Then perturbation theory will again work.
Otherwise the set $\setg$ has to be enlarged.
Similar, but not equivalent schemes were considered earlier in the frames of RMBPT
\cite{dzuba96}.
Here, we apply it for the first time in QED.

\par
It is convenient to write the matrix $V$ in a block form
\begin{eqnarray}
V
&=&
\xmatrix{V_{11}}{V_{12}}{V_{21}}{V_{22}}
\,,
\end{eqnarray}
where the block $V_{11}$ is constructed entirely on the states
from the set $\setg$ and the block $V_{22}$ does not contain
states from the set $\setg$.
The matrix $V$ can be decomposed as
\begin{eqnarray}
V
&=&\label{eqd070723n02}
V^{(0)}+\Delta V
\,,
\end{eqnarray}
where $V^{(0)}$ is a diagonal matrix (a sum of the Dirac energies).
The matrix $\Delta V$ contains the small parameter $\alpha$ (the expansion parameter of the QED perturbation theory) and can be
treated as a perturbation.
In what follows we restrict ourselves to the interelectron interaction
corrections.
In the lowest order these corrections reduce to the one photon exchange correction
\begin{eqnarray}
\Delta V
&=&\label{x050525x05}
\sum\limits_{\mbox{\rm g=c,t}} I^{\mbox{\rm g}}(|b-b'|)_{a'b'ab}
\,,
\end{eqnarray}
We can write the matrix $V$ as
\begin{eqnarray}
V
&=&
\xmatrix{V_{11}}{V_{12}}{V_{21}}{V_{22}}
=
\xmatrix{V^{(0)}_{11}+\Delta V_{11}}
        {\Delta V_{12}}{\Delta V_{21}}
        {V^{(0)}_{22}+\Delta V_{22}}
\,,
\end{eqnarray}

\par
The block matrix $V_{11}$ is finite and can be diagonalized numerically according to 
\begin{eqnarray}
V^{\mbox{\rm diag}}_{11}
&=&\label{eqd070723n01}
B^{+}V_{11}B
\,.
\end{eqnarray}
Since in general $V$ is a complex valued symmetrical matrix, \ie, $V_{ij}=V_{ji}$
the matrix $B$ is a complex orthogonal matrix
\begin{eqnarray}
B^{t}B
&=&\label{eqd070206n01}
I
\,.
\end{eqnarray}
Here $I$ is a unit matrix ($I_{ij}=\delta_{ij}$) of the proper
dimension.
The superscript $t$ in
\Eq{eqd070206n01}
means transposition.

\par
Compose a matrix
\begin{eqnarray}
A
&=&
\xmatrix{B}{0}{0}{I}
\,
\end{eqnarray}
which is also an orthogonal matrix
\begin{eqnarray}
A^{t}A
&=&
I
\,.
\end{eqnarray}

\par
Acting by the matrix $A$ on $V$ yields 
\begin{eqnarray}
\tilde{V}
&=&
A^{t}VA
=
\xmatrix{V^{\mbox{\rm diag}}_{11}}
        {B^{t} \Delta V_{12}}{\Delta V_{21} B}
        {V_{22}}
\,.
\end{eqnarray}
Since we have supposed that the required state $n_{\setg}$ is weakly
mixing with the states not included in the set $\setg$, the matrix
$\tilde{V}$ can be diagonalized with the standard procedure
\cite{landau77b}
\begin{eqnarray}
{\tilde{V}}^{\mbox{\rm diag}}
&=&
{\tilde{C}}^{t}{\tilde{V}}{\tilde{C}}
\,,
\end{eqnarray}
where the matrix $\tilde{C}$ can be built order by order.
The zeroth and the first orders of the matrix $\tilde{C}$ look like
\begin{eqnarray}
{\tilde{C}}_{ij}
&=&
{\tilde{C}}^{(0)}_{ij}+{\tilde{C}}^{(1)}_{ij}
=
I_{ij} + \xmatrix{0}
        {\frac{(B^{t}\Delta V_{12})_{ij}}{\mee_j - \mee_i}}
        {\frac{(\Delta V_{21} B)_{ij}}{\mee_j - \mee_i}}
        {\frac{(V_{22})_{ij}}{\mee_j - \mee_i}}
\,.
\end{eqnarray}
The diagonalized matrices $V$ and $\tilde{V}$ coincide, so we can write
\begin{eqnarray}
V^{\mbox{\rm diag}}
&=&
{\tilde{V}}^{\mbox{\rm diag}}
=
(A{\tilde{C}})^{t} V (A{\tilde{C}})
\,.
\end{eqnarray}
Accordingly, an eigenvector $\Phi$ corresponding to a basic function $\Psi$ can be defined as
\begin{eqnarray}
\Phi
&=&
A{\tilde{C}}\Psi
\,.
\end{eqnarray}
Now we represent the state $n_{\setg}\in\setg$ in terms of a perturbation expansion
\begin{eqnarray}
\Phi_{n_{\setg}}
&=&\label{x050525x06}
A{\tilde{C}}\Psi_{n_{\setg}}
=
\sum\limits_{k_{\setg}\in\setg}
B_{k_{\setg} n_{\setg}} \Psi^{(0)}_{k_{\setg}}
+ \sum\limits_{{k\notin\setg}\atop {l_{\setg}\in\setg} }
(\Delta V_{21})_{k l_{\setg}}
\frac{B_{l_{\setg}n_{\setg}}}
{E^{(0)}_{n_{\setg}}-E^{(0)}_{k}}
\Psi^{(0)}_{k}
\,.
\end{eqnarray}
An expression for $\Delta V_{21}$ is given by
\Eq{x050525x05}.
Summation over index $k$ means the summation over all two-electron
configurations (\jj coupling scheme) including
the negative part of the Dirac spectrum (not included in
the set $\setg$).
The employment of the \jj coupling scheme is not
obligatory here.

\par
In case when the investigated state $n_{\setg}$ is well isolated nondegenerate level and the set $\setg$ consists only of this single state, \ie, $\setg=\{\Psi_{n_\setg}\}$ the matrix $B$ is just a one-dimensional unit matrix.
It is easy to ensure that formula
(\ref{x050525x06})
together with
\Eq{x050520x25}
gives the same result as
\Eq{x050525x07}
(taking into account only the zeroth- and first-order corrections).

\par
We again would like to point out the QED effects that are now taken into account in the framework of the LPA, and which are missing in the Relativistic Many Body Perturbation Theory (RMBPT)
\cite{johnson95}:
The first is the inclusion of the retardation (see
\Eq{x050525x05}),
the second is the account of the negative part of the Dirac spectrum
(summation over $k$ in
\Eq{x050525x06}
and over $n$ in
Eqs.\ \Br{x050523x01}, \Br{x050523x02})
and the third is the incorporation of a nonzero contribution of ${\Xi}^{(1)}$
in
\Eq{x050520x3}
(reference state contribution).

\par
The amplitude ($U$) of the scattering process for quasidegenerate levels is given by
\Eq{tpuif}
where the eigenvectors are defined by
\Eq{x050525x06}
and the vertex operator is given by
Eqs.\ \Br{x050520x3}, \Br{x050520x2}.

%
%
\subsection{Transition probability}
\label{tpevaluation3}
Based on the scattering amplitude $U$ of the process $\Ini\to\Fin$ with emission of the photon $\omega$ the transition probability between $\Ini$ and $\Fin$ states is given by formula
\begin{eqnarray}
W
&=&\label{eq080615n01}
\sum\limits_{\lambda}
\int\frac{d^3\zhk}{(2\pi)^3}\,
(2\pi)|U|^2
\delta(\mee_\Fin+\omega-\mee_\Ini)
=
\frac{\omega^2}{(2\pi)^2}
\sum\limits_{\lambda}
\int d{\zhnu}\,
|U|^2
\,,
\end{eqnarray}
where $\zhnu=\zhk/|\zhk|$.
$\mee_\Ini$, $\mee_\Fin$ are the energies of the initial and final states, respectively.
These energies comprise the Dirac energies and the one-photon exchange corrections.
For quasidegenerate levels they are given by the corresponding eigenvalues of the matrix $V$
(\Eq{eqd070723n02}).
The photon frequency $\omega=|\zhk|$ should be set equal to $\omega=\mee_\Ini-\mee_\Fin$.
\Eq{eq080615n01}
defines the full transition probability, \ie, integration over all momenta of the photon ($\zhk$) and summation over all polarizations of the photon ($\lambda$) is performed.

\par
The integration over $\zhk$ and summation over $\lambda$ are taken analytically.
In
\Eq{eq080615n01}
only the photon wave functions depend on $\zhk$ and $\lambda$.
Accordingly, in a very general way we can consider the case when
\begin{eqnarray}
U^{(0)}
&=&
A^{(k,\lambda)*}_{n_1n_2}
\\
&=&
\int d^3\zhr\, \bar{\psi}_{n_1}(\zhr)\gamma^{\mu} A^{(k,\lambda)*}_{\mu}(\zhr)\psi_{n_2}(\zhr)
\,,
\end{eqnarray}
where $A^{(k,\lambda)}_{\mu}(\zhr)$ is given by
Eqs.\ \Br{photonwavefunction0}, \Br{photonwavefunction}.
The corresponding expression for the transition probability can be written as
\cite{akhiezer65b}
\begin{eqnarray}
W^{(0)}
&=&\label{eq080615n10}
\frac{\omega^2}{(2\pi)^2}
\sum\limits_{jm}
\left\{
\left|(A^{(E)*}_{jm}(\zhr,\omega))_{n_1 n_2}\right|^2
+
\left|(A^{(M)*}_{jm}(\zhr,\omega))_{n_1 n_2}\right|^2
\right\}
\,.
\end{eqnarray}
Here, the notation introduced in 
\Eq{defmea}
is employed.
The summations run over the angular momenta of the photon ($j$) and projections ($m$).
The 4-vector $A^{(M,E)\mu}=(V,{\zh A})$ corresponds to magnetic ($M$) and electric ($E$) photons, respectively.
In the case of the magnetic photons
\begin{eqnarray}
{V}^{(M)}_{jm}(\zhr,\omega)
&=&\label{eq080615n02}
0
\,,
\\
{\zh A}^{(M)}_{jm}(\zhr,\omega)
&=&\label{eq080615n03}
\sqrt{\frac{2\pi}{\omega}}
g_{j}(\omega r)
{\zh Y}_{jjm}(\zhn)
\,.
\end{eqnarray}
With the appropriate choice of the gauge for the electric photons we can write
\begin{eqnarray}
{V}^{(E)}_{jm}(\zhr,\omega)
&=&\label{eq080615n04}
0
\\
\hspace{-0.3cm}
{\zh A}^{(E)}_{jm}(\zhr,\omega)
&=&\label{eq080615n05}
\sqrt{\frac{2\pi}{\omega}}
\left\{
\sqrt{\frac{j}{2j+1}}
g_{j+1}(\omega r)
{\zh Y}_{jj+1m}(\zhn)
-
\sqrt{\frac{j+1}{2j+1}}
g_{j-1}(\omega r)
{\zh Y}_{jj-1m}(\zhn)
\right\}
\,.
\end{eqnarray}
In
Eqs.\ \Br{eq080615n03}, \Br{eq080615n05}
the radial functions
\begin{eqnarray}
g_l(x)
&=&
4\pi
\sqrt{\frac{\pi}{2x}} J_{l+1/2}(x)
\end{eqnarray}
involve Bessel functions $J_{l+1/2}(x)$ is the  of the first kind
\cite{abramowitz},
${\zh Y}_{jlm}$ ($l=j-1,j,j+1$) denotes the vector spherical harmonics
\cite{varshalovich,akhiezer65b}
depending on angles $\zhn=\zhr/|\zhr|$.
Formulas \Br{eq080615n04}, \Br{eq080615n05}
correspond to the photon wave function given by
Eqs.\ \Br{photonwavefunction0}, \Br{photonwavefunction},
specified within the ``transverse'' gauge
\cite{johnson95}.

\par
For the nonrelativistic limit the more convenient gauge is represented by the transformation ${\zh A}\to {\zh A} + {\zhnu} \chi(\zhk,t)$, $V\to V + \chi(\zhk,t)$ with
\begin{eqnarray}
\chi(\zhk,t)
&=&
\delta(\omega-|\zhk|)
\sqrt{\frac{j+1}{j}}
Y_{jm}(\zhnu)
e^{-i\omega t}
\,,
\end{eqnarray}
where $Y_{jm}(\zhnu)$ is the spherical harmonics
\cite{varshalovich}.
This transformation affects only electric photons.
Accordingly, in the nontransverse gauge the 4-vector $A^{(E)}$ appears as
\begin{eqnarray}
{V}^{(E)}_{jm}(\zhr,\omega)
&=&
i
\sqrt{\frac{2\pi}{\omega}}
\sqrt{\frac{j+1}{j}}
g_{j}(\omega r)
{Y}_{jm}(\zhn)
\,,
\\
{\zh A}^{(E)}_{jm}(\zhr,\omega)
&=&
\sqrt{\frac{2\pi}{\omega}}
\sqrt{\frac{2j+1}{j}}
g_{j+1}(\omega r)
{\zh Y}_{jj+1m}(\zhn)
\,.
\end{eqnarray}
In the work
\cite{johnson95}
this gauge is referred as ``length'' gauge.

\par
Comparing
\Eq{eq080615n01}
and
\Eq{eq080615n10}
we can express the transition probability in terms of the corresponding scattering amplitudes $U^{(E,M)}_{jm}$ as
\begin{eqnarray}
W
&=&\label{eq080615n11}
\frac{\omega^2}{(2\pi)^2}
\sum\limits_{jm}
\left\{
\left|U^{(E)}_{jm}(\zhr,\omega)\right|^2
+
\left|U^{(M)}_{jm}(\zhr,\omega)\right|^2
\right\}
\,,
\end{eqnarray}
where $A^{(k,\lambda)}$ are substituted by $A^{(M,E)}_{jm}$, respectively.
This expression was applied for the numerical calculations of the transition probabilities.

\par
The modified amplitudes $U^{(E,M)}_{jm}$ are derived within perturbation theory.
As we take into account only the corrections up to zeroth and first order, \ie,
\begin{eqnarray}
U^{(E,M)}_{jm}
&=&\label{eq080615n30}
U^{(E,M)(0)}_{jm}
+
U^{(E,M)(1)}_{jm}
+\ldots
\,,
\end{eqnarray}
then, the squared absolute values of $U^{(E,M)}_{jm}$ read
\begin{eqnarray}
\left|U^{(E,M)}_{jm}\right|^2
&=&\label{eq080615n31}
\left|U^{(E,M)(0)}_{jm}\right|^2
+
2\real\left\{U^{(E,M)(0)}_{jm}U^{(E,M)(1)}_{jm}\right\}
+
\left|U^{(E,M)(1)}_{jm}\right|^2
+\ldots
\,.
\end{eqnarray}
The last term in
\Eq{eq080615n31}
already corresponds to a corrections of second order and can be, accordingly, disregarded in the calculations.
However, this term may serve as an estimate for the magnitude of the higher-orders corrections (\ie, for the error magnitude).
The contributions of this term are given in the tables as $\Delta W^{(2+)}$.

%
%
\section{Numerical methods}
\label{methods}
In the numerical calculations an ion is considered to be enclosed into a spherical box with the radius $R=60/(\alpha Z)$ (in the relativistic units), where $\alpha$ is the fine-structure constant, $Z$ is the nuclear charge.
The size of the box reflects the size of the volume, where the physical processes of the interest (photon emission, interelectron interaction) mainly occur for the two-electron ions with high and intermediate $Z$.
Hence, the electron spectrum becomes discrete.
The Dirac spectrum in the external field of the nucleus is constructed in terms of B-splines
\cite{johnson88p307,shabaev04}.
We used B-splines of order 8 and a grid with 50 knots.

\par
Expression
\Eq{x050525x06}
for the eigenvectors $\Phi_{n_{\setg}}$ involves the zeroth- and first-order terms of the perturbation expansion.
The matrix $B$ required for the calculation of the eigenvectors $\Phi_{n_{\setg}}$ was generated perturbatively.
For a given set $\setg$, these two perturbation series are independent.

\par
The matrix $V$ employed in the construction of the matrix $B$ was borrowed from our work
\cite{andreev04},
where it was evaluated up to the second order with respect to the interelectron interaction corrections.
The matrix $\Delta V_{21}$ involved in
\Eq{x050525x06}
is given by
\Eq{x050525x05}
and includes only the first order of the interelectron interaction corrections.
It was calculated in the present work.

\par
The spatial integration in the matrix elements of the type
Eqs.\ \Br{defmea} and \Br{eq080616n01}
is performed in spherical coordinates.
The integration over the angular variables can be performed analytically, while the integration over the radial variables is performed numerically.
For the numerical integration we employed Gauss-Legendre quadratures, which yield a numerical accuracy of our calculations about $0.03\,\%$.

%
%
\section{Numerical results and discussion}
\label{results}
\label{section0305}
\label{section-tp-numerical}
In
Tables~\ref{tab0201}-\ref{tab0204}
we present numerical results
for $M1$, $M2$ and $E1$-transition probabilities
for low-lying two-electron configurations in HCI.
The values are given in units $s^{-1}$ and the digits in square brackets refer to the power of $10$.

\par
In
Tables~\ref{tab0201} and \ref{tab0202}
we consider transition probabilities between the
{$(1s2s)\, \confcsa$} configuration
and the ground {$(1s1s)\, \confaso$} configuration
and between
{$(1s2p_{3/2})\, \confcpb$} configuration
and {$(1s1s)\, \confaso$}
configuration with emission of magnetic $M1$ and $M2$ photons,
respectively.
By $W$ we denote the transition probability evaluated in this work.
The frequency of the emitted photon is set equal to $\omega=\mee_\Ini-\mee_\Fin$, where $\mee_\Ini$, $\mee_\Fin$ are the energies of the initial and final states, respectively.
In the case of the nondegenerate levels they comprise the Dirac energies together with one-photon corrections.
Accordingly, we do not include the radiative corrections.
The contributions of the negative energy states to the amplitude are included according to
Eqs.\ \Br{x050523x01}, \Br{x050523x02}
when performing summation ($n$) over the entire Dirac spectrum.
Investigation of the contribution due to the negative energy part of the continuum was performed in
\cite{indelicato96,derevianko98,indelicato04p062506}.
For $Z\ge 18$ the set $\setg$ contains configurations in the \jj coupling scheme built on $1s$, $2s$-electrons for {$(1s2s)\, \confcsa$} configuration and on $1s$, $2p_{3/2}$-electrons for {$(1s2p_{3/2})\, \confcpb$} configuration, respectively.
For $Z\le 10$, due to the poore convergence of the perturbation theory, the set $\setg$ contains $4000$ configurations.
Whenever available we compare our results with data obtained in other works.
The work
\cite{drake71}
presents the first relativistic calculation of transition probabilities for the $(1s2s)\,\confcsa\to(1s1s)\,\confaso$ transition.
The paper by Johnson \etal
\cite{johnson95}
provides a comprehensive review, where the transition probabilities are tabulated for all $Z$ values.
However, this work is performed neglecting QED effects such as retardation and the contribution arising from the derivative in the vertex
operator.
The work
\cite{indelicato04p062506}
is performed within the framework of the two-time Green's function method,
which is a full QED approach, too.
However, in this work only the nondegenerate two-electron configurations are
considered.
The present status of measurements of the transition probabilities can be found in
\cite{trabert08}.
In
Table~V
we give our numerical results for transition probabilities between {$(1s2s)\, \confcsa$} configuration
and the ground state for nuclear charge numbers $10\le Z\le100$.
Digits in square brackets indicate the accuracy of the measured values.

\par
In
Tables~\ref{tab0203},\ref{tab0204}
we present numerical results for $E1$-transition probabilities
between
$(1s2p)\,\confcpa,\,\confapa$
and ground
$(1s1s) \,\confaso$
two-electron configurations.
This provides the first exact QED calculation of the transition
probabilities for the quasidegenerate configurations.
The calculation is performed within the ``transverse'' and ``nontransverse'' gauges for the emitted photons:
$W_{\Vg}$ and $W_{\Lg}$, respectively.
The photon propagator was specified in the Coulomb gauge.
In the case of quasidegenerate levels the energies $\mee_\Ini$, $\mee_\Fin$ include also interelectron interaction corrections of the second order.
We considered one- and two-photon exchange corrections to the energy levels,
taken from
\cite{andreev04},
and one photon exchange corrections to the eigenvector
$\Phi_{n_{\setg}}$ in
\Eq{x050525x06}.
The contributions of the negative energy states to the amplitude are included in
\Eq{x050525x06}
in the summation over $k$ (over the complete Dirac spectrum but the set $\setg$)
and in the matrix $V$ (see
\Eq{eqd070723n02})
in the two-photon exchange corrections.
In the columns $W^\txt{1ph}_{\Lg,\Vg}$
we give the transition probabilities calculated with only one photon exchange correction taken into account, \ie, $W_{\Lg,\Vg}$ is recalculated where set $\Delta V^{(2)}= 0$ in
\Eq{x050525x05}.
The columns $W^{\txt{RMBPT}}_{\Lg,\Vg}$ display results of our recalculation of the $W_{\Lg,\Vg}$ values within RMBPT.
In the columns $\Delta W^{(2+)}_{\Lg,\Vg}$ an estimate for the interelectron interaction corrections of higher orders (see the end of
Subsection~\ref{tpevaluation3})
is given.
The blank fields in the columns $\Delta W^{(2+)}_{\Lg,\Vg}$ express that the corresponding values are smaller than the level 
of accuracy of the calculation.
In the last two columns we give the results of Drake
\cite{drake79}
(the application of the unified method)
and the RMBPT calculations by Johnson \etal
\cite{johnson95}.
While in
\cite{drake79}
the ``transverse'' gauge was used, the calculations performed in
\cite{johnson95}
for $(1s2p)\,\confcpa,\,\confapa$ configurations utilized a ``nontransverse'' gauge.
Digits in square brackets again denote powers of $10$.

\par
The diagonalization of the matrix $V_{11}$
(\Eq{eqd070723n01})
implies that we partly take into account the photon exchange corrections 
to all orders.
This violates the gauge invariance and,
accordingly, explains the deviation between $W_{\Lg}$ and $W_{\Vg}$.
This deviation also helps us to estimate the contribution of
the higher order terms in the expansions
Eqs.\ \Br{x050525x06}, \Br{x050525x05}.
This contribution is larger for small $Z$ values, where the convergence
of the perturbation theory in the interelectron interaction is poorer.

\par
The difference between the data in the columns $W_{\Lg,\Vg}$
and $W^\txt{RMBPT}_{\Lg,\Vg}$
determines the nonradiative QED corrections.
For small $Z$ values the considered configurations are strongly mixed.
The transition probabilities for $\confcpa$ levels
are very sensitive to
the mixing matrix $B$ what explains
the large value of the QED corrections for the small $Z$ values.
Note, that the transition probability for the $\confcpa$ level
is by several orders of magnitude smaller, than for the $\confapa$ level.
This means that the relative correction to the decay of the $\confcpa$ level
due to the change of the matrix $B$ is essentially larger than the correction to the decay of the level $\confapa$.
For a large numbers of $Z$ values the mixing of the configurations is small and QED corrections appear mainly as QED corrections to 
$(\Delta V_{21})_{kl_{\setg}}$ in the function $\Phi_{n_{\setg}}$ 
(see \Eq{x050525x06}.

\par
The perturbation expansion employed for the construction of the matrix $V$
(\Eq{eqd070723n02})
and the one applied in the diagonalization of $V$
(\Eq{x050525x06})
are different.
The comparison of the columns $W_{\Lg,\Vg}$ and $W^\txt{1ph}_{\Lg,\Vg}$ reveals the importance of two-photon corrections to the matrix $V$ for the convergence of the series of the perturbation theory.
If we enlarge the set $\setg$ by excited configurations, the values for $W^\txt{1ph}_{\Lg,\Vg}$ would approach the ones for $W_{\Lg,\Vg}$.
However, in order to achieve good agreement for small $Z$, we would have to include about $4000$ significant configurations in the set $\setg$.
The matrix $V$ is well investigated for the purpose of evaluating the energies of the configurations
\cite{andreev04,artemyev05},
so the evaluation of the mixing matrix $B$ in higher orders is a more efficient
technique, rather than any enlargement of the set $\setg$.
The technique presented for the calculation of the transition probabilities is a rigorous QED procedure, which allows for systematic improvements of the accuracy of the calculation by taking into account corrections of higher orders.

\par
The accuracy of the presented calculations is determined by the accuracy of the numerical methods, by the contribution of the omitted orders of the perturbation theories and by the radiative corrections which are not considered here.
The relative accuracy of the numerical calculation is set to $0.03\%$.
Contribution of the omitted orders of the perturbation theories can be estimated as difference between the values calculated within the different gauges, \ie,  difference between the $W_{\Lg}$ and $W_{\Vg}$ columns.
However, this is a very rough estimation, because one of the gauges may present better convergence than the other.
Partly, the contribution of the omitted orders of the perturbation theories can be estimated by the values in columns $\Delta W^{(2+)}_{\Lg,\Vg}$; they show that the ``nontransverse'' gauge gives considerably better convergence.
To estimate the order of magnitude of the radiative corrections we suppose that they have the same order as the other QED effects, \ie, as the difference between the values in $W_{\Lg,\Vg}$ and $W^\txt{RMBPT}_{\Lg,\Vg}$ columns, respectively.
Accordingly, the values of the transition probabilities calculated within the ``nontransverse'' gauge ($W_{\Lg}$) present the most accurate data for the transition probabilities.
The estimate of inaccuracy of the data is indicated by digits in round brackets.

\par
Concluding, we can state that at present this paper provides the most extensive and the most accurate calculations of the transition probabilities in HCI with intermediate nuclear charge numbers $Z$.
The inclusion of radiative corrections into the LPA (which is underway) would yield the most rigorous and powerful approach to the calculation of the transition probabilities for HCI with an utmost precision.

\begin{acknowledgments}
The authors acknowledge financial support from DFG, GSI and the RFBR grant 08-02-00026-a.
\end{acknowledgments}

%
\appendix
\section{Adiabatic S-matrix}
\label{appendixa}
Adiabatic S-matrix is a modified common S-matrix where adiabatic exponent $e^{-\addi |t|}$ is inserted in every vertex.
The adiabatic parameter $\addi$ is an infinitesimal quantity ($\addi\to+0$).
The presence of the adiabatic exponent switches off interaction with the electromagnetic field at $t=\pm\infty$.
In this appendix we show that the singularities present in the common S-matrix (see
Section~\ref{section02})
vanish completely in the adiabatic S-matrix.
The singularities arise when one makes insertions into the outer electron lines of the Feynman graphs.

\par
\begin{figure}
\centerline{\mbox{\epsfysize=90pt \epsffile{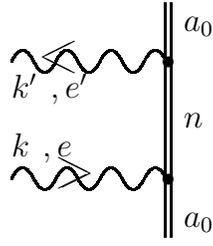}}}
\caption{\label{figure161}
Feynman graph, describing the photon scattering on an atomic electron.
The wavy lines with the arrows describe the absorption and emission of photons with momenta $k$, $k'$ and polarizations $e$, $e'$, respectively.
The double solid line denotes the electron in the field of the nucleus, $\oeg$ corresponds to the ground electron state.
}
\end{figure}
\begin{figure}
\centerline{\mbox{\epsfysize=120pt \epsffile{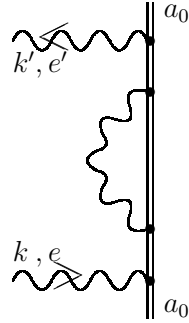}}}
\caption{\label{figure164}
Feynman graph corresponding to the electron self-energy insertion into the electron propagator in
\Fig{figure161}.
The wavy line denotes the virtual photon.
The other notations are the same as in
\Fig{figure161}.
}
\end{figure}
\begin{figure}
\centerline{\mbox{\epsfysize=150pt \epsffile{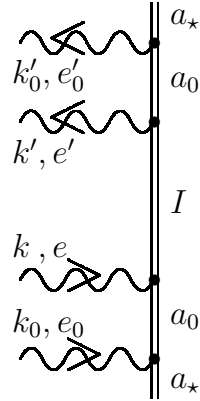}}}
\caption{\label{figure85}
The process of double-photon scattering on the artificial ``lower than ground'' state $\oegg$.
Notations are the same as in
\Fig{figure161}.
}
\end{figure}
\begin{figure}
\centerline{\mbox{\epsfysize=100pt \epsffile{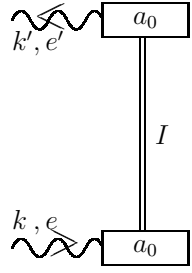}}}
\caption{\label{figure82}
The Feynman graph representing the process of elastic photon
scattering on one-electron ion.
The double solid line denotes the electron in the field of the nucleus (Furry picture of QED).
The boxes with the wavy lines describe the absorption and emission of the photons by electron the ground state.
The letter $\Ini$ at the internal electron line implies the resonance approximation, where only the resonant state $\Ini$ remains in the sum over the intermediate states in the electron propagator.
}
\end{figure}
\begin{figure}
\centerline{\mbox{\epsfysize=120pt \epsffile{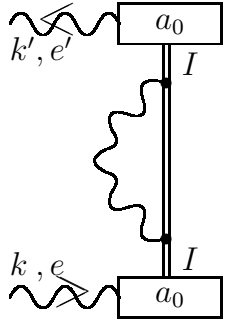}}}
\caption{\label{figure83}
The insertion of the one-loop electron self-energy in the internal electron line in
\Fig{figure82}.
The wavy line describes the virtual photon.
The other notations are the same as in
\Fig{figure82}.
}
\end{figure}
\begin{figure}
\centerline{\mbox{\epsfysize=120pt \epsffile{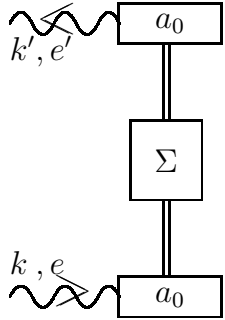}}}
\caption{\label{figure84}
The Feynman graph, illustrating
\Eq{ln06}.
The box with the letter $\Sigma$ inside corresponds to the infinite number of successive insertions of the type
\Fig{figure83}.
}
\end{figure}
\begin{figure}
\centerline{\mbox{\epsfysize=100pt \epsffile{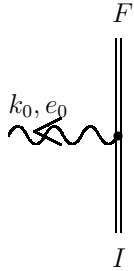}}}
\caption{\label{figure77}
The Feynman graph representing the process of the photon emission.
The labels $\Ini$ and $\Fin$ correspond to the initial and final states.
}
\end{figure}
\begin{figure}
\centerline{\mbox{\epsfysize=150pt \epsffile{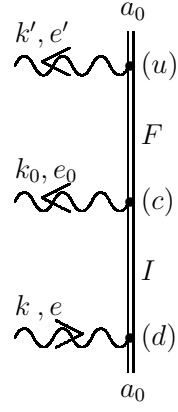}}}
\caption{\label{figure512}
The Feynman graphs representing the process of photon emission in the LPA.
This graph incorporates the graph in
\Fig{figure77}.
The upper, central and down vertices are specified by corresponding subscripts ($u$), ($c$) and ($d$), respectively.
}
\end{figure}
\begin{figure}
\centerline{\mbox{\epsfysize=130pt \epsffile{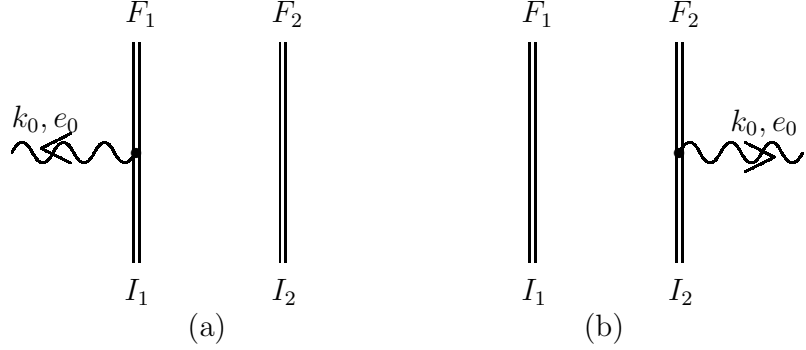}}}
\caption{\label{figure80}
The Feynman graphs representing the single photon transition in a two-electron ion to lowest order in $\alpha$.
The notations are the same as in
Figs.\ \ref{figure82}-\ref{figure512}.
$\Ini_i$ $(i=1,2)$ and $\Fin_i$ $(i=1,2)$ denote the initial and final states for the two electrons.
}
\end{figure}
\begin{figure}
\centerline{\mbox{\epsfysize=150pt \epsffile{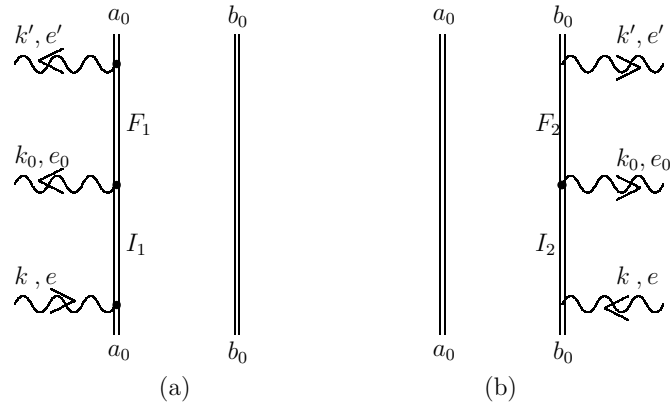}}}
\caption{\label{figure81}
The Feynman graphs representing the process of elastic photon
scattering on the two-electron ion within the LPA.
These graphs incorporate the graphs in
\Fig{figure80}.
See in the text the notations $a_0, b_0$.
}
\end{figure}
\begin{figure}
\centerline{\mbox{\epsfysize=160pt \epsffile{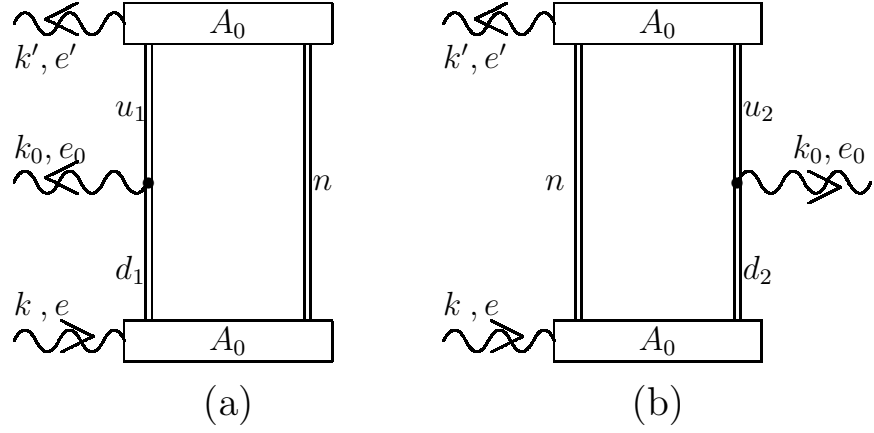}}}
\caption{\label{figure29}
The Feynman graphs representing the process of elastic photon
scattering on the two-electron ion.
The boxes with the letter $\meg$ inside and wavy lines depicts complicated vertices describing the absorption and emission of a photon by the two-electron ion.
The photon line in the center denotes the emission of a photon with frequency $\omega_0=|\zhk_0|$ corresponding to the transition energy from
the initial to the final two-electron state.
}
\end{figure}
\begin{figure}
\centerline{\mbox{\epsfysize=160pt \epsffile{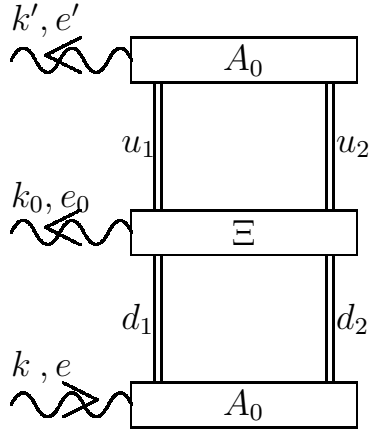}}}
\caption{\label{figure60}
The Feynman graph representing the process of elastic photon
scattering on the two-electron ion.
In addition to
\Fig{figure29}
the box with letter $\Xi$ inside and with an external photon indicates the complicated vertex for the emission of a photon $\omega_0$ corresponding to the transition from
the initial to the final state ($\Ini\to \Fin$).
}
\end{figure}
\begin{figure}
\centerline{\mbox{\epsfysize=340pt \epsffile{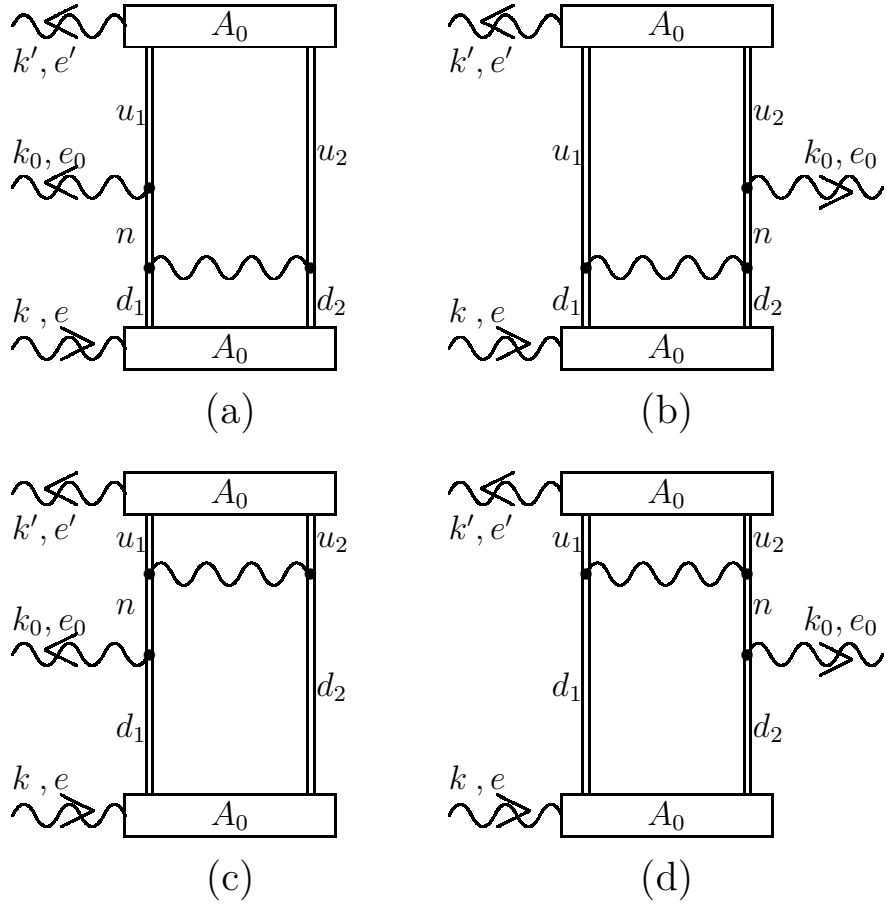}}}
\caption{\label{figure30}
The Feynman graph representing the process of elastic photon
scattering on the two-electron ion.
This graph includes the one photon exchange correction and,
accordingly, it represents the next order of the perturbation theory (photon exchange correction) compared to the graph
\Fig{figure29}.
}
\end{figure}

In the present paper we will consider the one-electron ions and the insertions of the self-energy operator.
In the lowest order of QED perturbation theory S-matrix element corresponding to the process of elastic photon scattering on the one-electron ions is given by the Feynman graph in
\Fig{figure161}.
We consider the case when electron in the ground state $\oeg$ absorbs photon $(k,\lambda)$, then emits photon $(k',\lambda')$ and decays back to the ground state.
According to energy conservation law $\omega=\omega'$.
This graph gives no singularities, accordingly, the adiabatic S-matrix element coinsides with the common S-matrix element
\begin{eqnarray}
S^{(0,0)}
&=&
S^{(0,0)}_{\addi}
=
(-2\pi i)\delta(\omega'-\omega)
e^2
\sum\limits_{n}
\frac{A^{(k,\lambda) *}_{\oeg n }
A^{(k,\lambda)}_{n \oeg}}
{\omega'+\dee_\oeg-\dee_n}
\,.
\end{eqnarray}
The superscripts at the S-matrices indicate the number of insertions of the self-energy operator into the upper and lower external electron lines, respectively.
Here there are no insertions.

\par
In the next orders of perturbation theory we have to make insertions of the self-energy operators into the electron lines.
The insertions into the internal lines yield no singularities.
They result in the energy shift of the excited atomic states and were investigated in
\cite{andreev01}.
For simplicity of the derivation we omit them.
Accordingly, we consider the insertions into the outer electron lines.

\par
The $N$ insertions of electron self-energy operators into the lower electron line are depicted in
Fig.\ 15.
After integration over time variables with employment of equality
\begin{eqnarray}
\int^{+\infty}_{-\infty} dt\,\, e^{-\addi |t|+iat}
&=&\label{tp080716n04}
i\left[
\frac{1}{a+i\addi} + \frac{1}{-a+i\addi}
\right]
\end{eqnarray}
the corresponding adiabatic S-matrix element is given by
\begin{eqnarray}
S^{(0,N)}_{\addi}
&=&\nonumber
\int d^3 \zhr_{u}
d^3 \zhr_{d_{1}}\ldots d^3 \zhr_{d_{2N}}
d\omega_{n}
d\omega_{d_{1}}\ldots d\omega_{d_{N}}
d\omega_{s_{1}}\ldots d\omega_{s_{N}}
d\Omega_{1}\ldots d\Omega_{N}
\,
{\bar{\psi}_\oeg}(\zhr_{u})
\\
&&\nonumber
\times
(-ie)\gamma^{\mu_{u}}
A^{(k',\lambda') *}_{\mu_{u}}(\zhr_{u})
\frac{i}{2\pi}
\sum\limits_{n}
\frac{\psi_{n}(\zhr_{u}){\bar{\psi}}_{n}(\zhr_{d_1})}
     {\omega_{n}-\dee_{n}(1-i0)}
\\
&&\nonumber
\times
(-ie)\gamma^{\mu_{d_1}}A^{(k,\lambda)}_{\mu_{d_1}}(\zhr_{d_1})
(i)^2
\\
&&\nonumber
\times
\left[
\frac{1}{\dee_\oeg+\omega'-\omega_n+i\addi}
+
\frac{1}{-\dee_\oeg-\omega'+\omega_n+i\addi}
\right]
\\
&&\nonumber
\times
\left[
\frac{1}{\omega_n-\omega-\omega_{d_1}+i\addi}
+
\frac{1}{-\omega_n+\omega+\omega_{d_1}+i\addi}
\right]
\\
&&\nonumber
\times
\frac{i}{2\pi}
\sum\limits_{d_1}
\frac{\psi_{d_1}(\zhr_{d_1}){\bar{\psi}}_{d_1}(\zhr_{d_2})}
     {\omega_{d_1}-\dee_{d_1}(1-i0)}
(-ie)\gamma^{\mu_{d_2}}
\frac{i}{2\pi}
\sum\limits_{s_1}
\frac{\psi_{s_1}(\zhr_{d_2}){\bar{\psi}}_{s_1}(\zhr_{d_3})}
     {\omega_{s_1}-\dee_{s_1}(1-i0)}
\\
&&\nonumber
\times
I_{\mu_{d_2}\mu_{d_3}}(|\Omega_1|,r_{d_{23}})
(-ie)\gamma^{\mu_{d_3}}
\psi_\oeg(\zhr_{d_3})
(i)^2
\\
&&\nonumber
\times
\left[
\frac{1}{\omega_{d_1}-\omega_{s_1}-\Omega_1+i\addi}
+
\frac{1}{-\omega_{d_1}+\omega_{s_1}+\Omega_1+i\addi}
\right]
\\
&&\nonumber
\times
\left[
\frac{1}{\omega_{s_1}+\Omega_1-\omega_{d_2}+i\addi}
+
\frac{1}{-\omega_{s_1}-\Omega_1+\omega_{d_2}+i\addi}
\right]
\\
&&\nonumber
\cdots
\\
&&\nonumber
\times
\frac{i}{2\pi}
\sum\limits_{d_k}
\frac{\psi_{d_k}(\zhr_{d_{2k-1}}){\bar{\psi}}_{d_k}(\zhr_{d_{2k}})}
     {\omega_{d_k}-\dee_{d_k}(1-i0)}
(-ie)\gamma^{\mu_{d_{2k}}}
\frac{i}{2\pi}
\sum\limits_{s_k}
\frac{\psi_{s_k}(\zhr_{d_{2k}}){\bar{\psi}}_{s_k}(\zhr_{d_{2k+1}})}
     {\omega_{s_k}-\dee_{s_k}(1-i0)}
\\
&&\nonumber
\times
I_{\mu_{d_{2k}}\mu_{d_{2k+1}}}(|\Omega_k|,r_{d_{2k},d_{2k+1}})
(-ie)\gamma^{\mu_{d_{2k+1}}}
(i)^2
\\
&&\nonumber
\times
\left[
\frac{1}{\omega_{d_k}-\omega_{s_k}-\Omega_k+i\addi}
+
\frac{1}{-\omega_{d_k}+\omega_{s_k}+\Omega_k+i\addi}
\right]
\\
&&\nonumber
\times
\left[
\frac{1}{\omega_{s_k}+\Omega_k-\omega_{d_{k+1}}+i\addi}
+
\frac{1}{-\omega_{s_k}-\Omega_k+\omega_{d_{k+1}}+i\addi}
\right]
\\
&&\nonumber
\cdots
\\
&&\nonumber
\times
\frac{i}{2\pi}
\sum\limits_{d_N}
\frac{\psi_{d_N}(\zhr_{d_{2N-1}}){\bar{\psi}}_{d_N}(\zhr_{d_{2N}})}
     {\omega_{d_N}-\dee_{d_N}(1-i0)}
(-ie)\gamma^{\mu_{d_{2N}}}
\frac{i}{2\pi}
\sum\limits_{s_N}
\frac{\psi_{s_N}(\zhr_{d_{2N}}){\bar{\psi}}_{s_N}(\zhr_{d_{2N+1}})}
     {\omega_{s_N}-\dee_{s_N}(1-i0)}
\\
&&\nonumber
\times
I_{\mu_{d_{2N}}\mu_{d_{2N+1}}}(|\Omega_N|,r_{d_{2N},d_{2N+1}})
(-ie)\gamma^{\mu_{d_{2N+1}}}
\psi_\oeg(\zhr_{d_{2N+1}})
(i)^2
\\
&&\nonumber
\times
\left[
\frac{1}{\omega_{d_N}-\omega_{s_N}-\Omega_N+i\addi}
+
\frac{1}{-\omega_{d_N}+\omega_{s_N}+\Omega_N+i\addi}
\right]
\\
&&\label{tp080721n03}
\times
\left[
\frac{1}{\omega_{s_N}+\Omega_N-\dee_\oeg+i\addi}
+
\frac{1}{-\omega_{s_N}-\Omega_N+\dee_\oeg+i\addi}
\right]
\,.
\end{eqnarray}
In order to make the derivations shorter we will neglect the negative energy part of the Dirac spectrum (\ie, we suppose that $\dee_n>0$, $\dee_d>0$), since negative energy terms do not generate singularities.

\par
Consider separately the integral over $\omega$-variables and designate it as $F$.
The integrand of $F$ includes all the terms of
\Eq{tp080721n03}
depending on the $\omega$-variables: the fractions in the square brackets and the denominators originating from the electron propagators.
Integration over $\omega_n$ and $\omega_{s_{1\ldots N}}$ yields
\begin{eqnarray}
F
&=&\nonumber
\int
d\omega_{d_{1}}\ldots d\omega_{d_{N}}\,
\frac{2\pi}{i}
\frac{1}
     {\omega+\omega_{d_1}-\dee_{n}+i\addi}
\\
&&\nonumber
\times
\left[
\frac{1}{\dee_\oeg+\omega'-\omega-\omega_{d_1}+2i\addi}
+
\frac{1}{-\dee_\oeg+\omega'+\omega+\omega_{d_1}+2i\addi}
\right]
\\
&&\nonumber
\times
\frac{2\pi}{i}
\frac{1}
     {(\omega_{d_1}-\dee_{d_1}(1-i0))
	(\omega_{d_2}-\dee_{s_1}-\Omega_1+i\addi)}
\\
&&\nonumber
\times
\left[
\frac{1}{\omega_{d_1}-\omega_{d_2}+2i\addi}
+
\frac{1}{-\omega_{d_1}+\omega_{d_2}+2i\addi}
\right]
\\
&&\nonumber
\cdots
\\
&&\nonumber
\times
\frac{2\pi}{i}
\frac{1}
	{(\omega_{d_{k}}-\dee_{k}(1-i0))
	(\omega_{d_{k+1}}-\dee_{s_{k}}-\Omega_{k}+i\addi)}
\\
&&\nonumber
\times
\left[
\frac{1}{\omega_{d_{k}}-\omega_{d_{k+1}}+2i\addi}
+
\frac{1}{-\omega_{d_{k}}+\omega_{d_{k+1}}+2i\addi}
\right]
\\
&&\nonumber
\cdots
\\
&&\nonumber
\times
\frac{2\pi}{i}
\frac{1}
	{(\omega_{d_N}-\dee_{d_N}(1-i0))
	(\dee_\oeg-\dee_{s_{N}}-\Omega_{N}+i\addi)}
\\
&&\label{eq080721n05}
\times
\left[
\frac{1}{\omega_{d_N}-\dee_\oeg+2i\addi}
+
\frac{1}{-\omega_{d_N}+\dee_\oeg+2i\addi}
\right]
\,.
\end{eqnarray}
Integrations in
\Eq{eq080721n05}
can be performed recursively with the use of equality
\begin{eqnarray}
&&\nonumber
\int d\omega_{d_N}\,
\frac{2\pi}{i}
\frac{1}
	{(\omega_{d_{N-1}}-\dee_{d_{N-1}}(1-i0))
	(\omega_{d_{N}}-\dee_{s_{N-1}}-\Omega_{N-1}+i\addi)}
\\
&&\nonumber
\times
\left[
\frac{1}{\omega_{d_{N-1}}-\omega_{d_{N}}+2i\addi}
+
\frac{1}{-\omega_{d_{N-1}}+\omega_{d_{N}}+2i\addi}
\right]
\\
&&\nonumber
\times
\frac{2\pi}{i}
\frac{1}
	{(\omega_{d_N}-\dee_{d_N}(1-i0))
	(\dee_\oeg-\dee_{s_{N}}-\Omega_{N}+i\addih)}
\\
&&\nonumber
\times
\left[
\frac{1}{\omega_{d_N}-\dee_\oeg+2i\addih}
+
\frac{1}{-\omega_{d_N}+\dee_\oeg+2i\addih}
\right]
\\
&=&\nonumber
\left(\frac{2\pi}{i}\right)^3
\frac{1}{(\omega_{d_{N-1}}-\dee_{d_{N-1}}(1-i0))
	(\dee_\oeg-\dee_{d_{N-1}}-\Omega_{N-1}+2i\addih+i\addi)}
\\
&&\nonumber
\times
\left[
\frac{1}{\omega_{d_{N-1}}-\dee_\oeg+2i\addih+2i\addi}
+
\frac{1}{-\omega_{d_{N-1}}+\dee_\oeg+2i\addih+2i\addi}
\right]
\\
&&\label{eq080721n06}
\times
\frac{1}{(\dee_\oeg-\dee_{d_{N}}+2i\addih)(\dee_\oeg-\dee_{s_{N}}-\Omega_{N}+i\addih)}
+R_{\addi,\addih}
\,,
\end{eqnarray}
where $\lim\limits_{\addi,\addih\to0}R_{\addi,\addih}=0$.
After integration we get the following expression for
\Eq{eq080721n05}
\begin{eqnarray}
F
&=&
\left(
\frac{2\pi}{i}
\right)^{2N+1}
\\\nonumber
&&
\times
\left\{
\frac{1}{(\dee_\oeg-\dee_{s_{N}}-\Omega_{N}+2i\addi)}
\cdots
\frac{1}{(\dee_\oeg-\dee_{s_{N-k}}-\Omega_{N-k}+(2k+1)i\addi)}
\cdots
\right.
\\\nonumber
&&
\times
\left.
\cdots
\frac{1}{(\dee_\oeg-\dee_{s_{1}}-\Omega_{1}+(2N-1)i\addi)}
\right\}
\\\nonumber
&&
\times
\frac{1}{(\dee_\oeg-\dee_{d_{N}}+2i\addi)
\cdots
(\dee_\oeg-\dee_{d_{N-k+1}}+2ki\addi)
\cdots
(\dee_\oeg-\dee_{d_{1}}+2Ni\addi)
}
\\\nonumber
&&
\times
\frac{1}{\omega+\dee_\oeg-\dee_{n}+(2N+1)i\addi}
\\\label{eq080728n01}
&&
\times
\left[
\frac{1}{\omega-\omega'+2(N+1)i\addi}
+
\frac{1}{-\omega+\omega'+2(N+1)i\addi}
\right]
+R_{\addi}
\,,
\end{eqnarray}
where $\lim\limits_{\addi\to0}R_{\addi}=0$.
Function $F$ is singular at $\addi\to 0$ when
$\dee_{d_k}=\dee_\oeg$, $k=1,\ldots,N$.
The term in square brackets in
\Eq{eq080728n01}
can be written as
\begin{eqnarray}
&&
\left[
\frac{1}{\omega-\omega'+2(N+1)i\addi}
+
\frac{1}{-\omega+\omega'+2(N+1)i\addi}
\right]
\\\nonumber
&&
=
\frac{1}{2(N+1)}
\left[
\frac{1}{\frac{\omega-\omega'}{2(N+1)}+i\addi}
+
\frac{1}{-\frac{\omega-\omega'}{2(N+1)}+i\addi}
\right]
\\
&&
=
\frac{1}{2(N+1)}
\,
\left(
\frac{2\pi}{i}
\right)
\delta\left(\frac{\omega-\omega'}{2(N+1)}\right)
=
\left(
\frac{2\pi}{i}
\right)
\delta(\omega-\omega')
\,.
\end{eqnarray}

\par
Let us restrict ourselves to the case when $\dee_{d_k}=\dee_\oeg$ for every $k=1,\ldots,N$
(the derivations for the cases when some of $\dee_{d_k}\ne\dee_\oeg$ can be performed by analogy).
Accordingly, we write
\Eq{eq080728n01}
as
\begin{eqnarray}
F
&=&\nonumber
\left(
\frac{2\pi}{i}
\right)^{2N+1}
\\\nonumber
&&
\times
\left\{
\frac{1}{(\dee_\oeg-\dee_{s_{N}}-\Omega_{N}+i0)
\cdots
(\dee_\oeg-\dee_{s_{N-k}}-\Omega_{N-k}+i0)
\cdots
(\dee_\oeg-\dee_{s_{1}}-\Omega_{1}+i0)
}
\right\}
\\\nonumber
&&
\times
\left(\frac{1}{2i\addi}\right)^{N}
\frac{1}{N!}
\\\nonumber
&&
\times
\frac{1}{\omega+\dee_\oeg-\dee_{n}+(2N+1)i\addi}
\\\label{eq080728n02}
&&
\times
\left(
\frac{2\pi}{i}
\right)
\delta(\omega-\omega')
+R_{\addi}
\,.
\end{eqnarray}

\par
With employment of
\Eq{eq080728n02}
we can write
\Eq{tp080721n03}
as
\begin{eqnarray}
S^{(0,N)}_{\addi}
&=&\nonumber
(-2\pi i)\delta(\omega-\omega') e^2
\\
&&\nonumber
\times
\sum\limits_{n}
\frac{
A^{(k',\lambda') *}_{\oeg n}
A^{(k,\lambda)}_{n\oeg}
}
{(\omega+\dee_\oeg-\dee_n+(2N+1)i\addi)}
\frac{1}{N!}
\left(\frac{\Sigmareg_{\oeg\oeg}(\dee_{\oeg})}{2i\addi}\right)^N
\\
&&\label{tp080717n06}
+
R_\addi
\,,
\end{eqnarray}

\par
Employing the asymptotic ($\addi\to +0$) equality
\begin{eqnarray}
\sum\limits_{N=0}^{\infty}
\frac{1}{(x+Ni\addi)N!}
\left(
\frac{\Delta}{i\addi}
\right)^N
&=&\label{eq080728n03}
\frac{1}{x+\Delta}
\exp\left(\frac{\Delta}{i\addi}\right)
\,,
\end{eqnarray}
where $|x|>|\Delta|$ we can write
\begin{eqnarray}
\sum\limits_{N=0}^{\infty}
S^{(0,N)}_{\addi,N}
&=&
(-2\pi i)\delta(\omega-\omega') e^2
\\
&&\nonumber
\times
\left[
\sum\limits_{n}
\frac{
A^{(k',\lambda') *}_{\oeg n}
A^{(k,\lambda)}_{n\oeg}
}
{\omega+\dee_\oeg+\Sigmareg_{\oeg\oeg}(\dee_{\oeg})-\dee_n}
+
R_\addi
\right]
\\
&&\label{tp080725n01}
\times
\exp\left(\frac{\Sigmareg_{\oeg\oeg}(\dee_{\oeg})}{2i\addi}\right)
\,.
\end{eqnarray}
Although the condition $|x|>|\Delta|$ is necessary for
\Eq{eq080728n03},
we employ
\Eq{eq080728n03}
for
\Eq{tp080717n06}
for any $\omega$.
This is considered as an analytical continuation of
\Eq{tp080717n06}
to the area close to the resonance and, accordingly, the entire complex  plane ($\omega$).
This analytical continuation was discussed in
\cite{andreev08pr}.

\par
If we insert the self-energy operator into the upper ($N_u$ times) and lower ($N_d$ times) outer electron lines, the similar derivations yield
\begin{eqnarray}
S^{(N_u, N_d)}_{\addi}
&=&\nonumber
(-2\pi i)\delta(\omega-\omega') e^2
\\
&&\nonumber
\times
\frac{1}{N_u!}
\left(\frac{\Sigmareg_{\oeg\oeg}(\dee_{\oeg})}{2i\addi}\right)^{N_u}
\sum\limits_{n}
\frac{
A^{(k',\lambda') *}_{\oeg n}
A^{(k,\lambda)}_{n\oeg}
}
{(\omega+\dee_\oeg-\dee_n+(2N+1)i\addi)}
\frac{1}{N_d!}
\left(\frac{\Sigmareg_{\oeg\oeg}(\dee_{\oeg})}{2i\addi}\right)^{N_d}
\\
&&\label{tp080725n02}
+
R_\addi
\,,
\end{eqnarray}
The value of $N$ in the denominator can be set equal to $N_u$ or $N_d$ without changing the final result since it influences only the terms $R_\addi$ which disappear in the asymptotics ($\addi\to+0$).
Finally, we get
\begin{eqnarray}
S_\addi
&=&
\sum\limits_{N_u,N_d=0}^{\infty}
S^{(N_u,N_d)}_{\addi}
\\
&=&
(-2\pi i)\delta(\omega-\omega') e^2
\\
&&\nonumber
\times
\sum\limits_{n}
\frac{
A^{(k',\lambda') *}_{\oeg n}
A^{(k,\lambda)}_{n\oeg}
}
{\omega+\dee_\oeg+\Sigmareg_{\oeg\oeg}(\dee_{\oeg})-\dee_n}
\\
&&\label{tp080725n03}
\times
\exp\left(\frac{\Sigmareg_{\oeg\oeg}(\dee_{\oeg})}{i\addi}\right)
\,.
\end{eqnarray}
As the regularized self-energy matrix element for the ground state ($\oeg$) has no imaginary part, the absolute value of the exponent in
\Eq{tp080725n03}
reads
\begin{eqnarray}
\left|
\exp\left(\frac{\Sigmareg_{\oeg\oeg}(\dee_{\oeg})}{i\addi}\right)
\right|
&=&
1
\,.
\end{eqnarray}
Accordingly, the absolute value of the amplitude is given by
\begin{eqnarray}
|U|
&=&\label{eq080728n05}
e^2
\left|
\sum\limits_{n}
\frac{
A^{(k',\lambda') *}_{\oeg n}
A^{(k,\lambda)}_{n\oeg}
}
{\omega+\oee_\oeg-\dee_n}
\right|
\,.
\end{eqnarray}
The regularized self-energy matrix element in the denominator is a correction to the energy of the ground state: $\oee_\oeg=\dee_\oeg+\Sigmareg_{\oeg\oeg}(\dee_{\oeg})$.

\par
In
\Eq{eq080728n02}
we considered only the case when
$\dee_{d_k}=\dee_\oeg$ for every $k=1,\ldots,N$.
The cases when some of $\dee_{d_k}\ne\dee_\oeg$ correspond to the insertions of the second and higher orders self-energy corrections of the ``loop-after-loop'' type.
The case when all $\dee_{d_k}\ne\dee_\oeg$ gives the correction to the wave function of the ground state electron $\oeg$.

\par
The presented derivations show that the employment of the adiabatic theory allows for the insertions into the outer electron lines within the LPA.
They also justify the introduction of the vertex functions $\Phi_\oeg$, $\bar\Phi_\oeg$ ($\Phi_\meg$, $\bar\Phi_\meg$ for two-electron ions) in Section~\ref{tpformulas}.
The energies of the ground state can be considered to be the full energies, \ie, with all the corrections included.

\par
Finally, we would like to note that the goal of employment of the adiabatic approach was the justification of the LPA backgrounds.
Formally, the ground state can be investigated within the same matrix formulation of the LPA as the excited states which allows to employ the general technique developed in
Section~\ref{tpevaluation}.

\newpage

%

\newpage

%
\begin{table}
\caption{
$M1$-transition probabilities ($\mbox{\rm s}^{-1}$) between
{ $(1s2s)\,\confcsa$}
and { $(1s1s)\,\confaso$}
configurations.
The digits in square brackets denote the power of $10$.
}
\begin{tabular}{rrrrrr}
\hline
\multicolumn{1}{c}{$Z$}&
\multicolumn{1}{c}{$W$}&
Ref.~\cite{drake71}
&
Ref.~\cite{johnson95}
&
Ref.~\cite{indelicato04p062506}
&
\multicolumn{1}{c}{Experiment}
\\
\hline
6&4.867(11)[\phantom{0}1]&4.856[\phantom{0}1]&4.860[\phantom{0}1]&&4.857(11)[\phantom{0}1]${}^a$\\
10&1.097(7)[\phantom{0}4]&1.087[\phantom{0}4]&1.092[\phantom{0}4]&&1.105(18)[\phantom{0}4]${}^b$\\
&&&&&1.0905(48)[\phantom{0}4]${}^c$\\
12&7.324(30)[\phantom{0}4]&7.243[\phantom{0}4]&7.293[\phantom{0}4]&&7.35(26)[\phantom{0}4]${}^d$\\
16&1.430(5)[\phantom{0}6]&1.408[\phantom{0}6]&1.426[\phantom{0}6]&&1.422(8)[\phantom{0}6]${}^e$\\
18&4.798(18)[\phantom{0}6]&4.709[\phantom{0}6]&4.787[\phantom{0}6]&&\\
26&2.078(6)[\phantom{0}8]&2.002[\phantom{0}8]&2.075[\phantom{0}8]&&\\
30&8.987(21)[\phantom{0}8]&&8.981[\phantom{0}8]&8.993[\phantom{0}8]&\\
50&1.727(2)[11]&&1.726[11]&1.729[11]&\\
54&3.852(5)[11]&&3.846[11]&3.856[11]&3.92(12)[11]${}^f$\\
70&5.980(10)[12]&&5.968[12]&5.983[12]&\\
90&9.468(23)[13]&&9.439[13]&9.469[13]&\\
100&3.193(1)[14]&&3.181[14]&&\\
\hline
\multicolumn{3}{l}{${}^a$ Schmidt \etal \cite{schmidt94}.}\\
\multicolumn{3}{l}{${}^b$ Wargelin \etal \cite{wargelin93}.}\\
\multicolumn{3}{l}{${}^c$ Tr\"abert \etal \cite{trabert99}.}\\
\multicolumn{3}{l}{${}^d$ Stefanelli \etal \cite{stefanelli95}.}\\
\multicolumn{3}{l}{${}^e$ Crespo L\'opez-Urrutia \etal \cite{crespo06}.}\\
\multicolumn{3}{l}{${}^f$ Marrus \etal \cite{marrus89}.}\\
\end{tabular}
\\
\label{tab0201}
\end{table}

%
%
\begin{table}
\caption{
$M2$-transition probabilities ($\mbox{\rm s}^{-1}$) between
{ $(1s2p_{3/2}) \,\confcpb$}
and { $(1s1s)\,\confaso$}
configurations.
}
\begin{tabular}{rrrr}
\hline
\multicolumn{1}{c}{$Z$}&
\multicolumn{1}{c}{$W$}&
Ref.~\cite{johnson95}&
Ref.~\cite{indelicato04p062506}\\
\hline
 5&5.016(4)[\phantom{0}3]&5.014[\phantom{0}3]& \\
10&2.258(3)[\phantom{0}6]&2.257[\phantom{0}6]& \\
18&3.145(1)[\phantom{0}8]&3.141[\phantom{0}8]&\\
26&6.515(2)[\phantom{0}9]&6.510[\phantom{0}9]&\\
30&2.104(1)[10]&2.104[10]&2.105[10]\\
50&1.365(1)[12]&1.365[12]&1.366[12]\\
54&2.560(3)[12]&2.560[12]&\\
70&2.148(1)[13]&2.146[13]&2.148[13]\\
90&1.720(1)[14]&1.718[14]&1.721[14]\\
100&4.165(3)[14]&4.156[14]&\\
\hline
\end{tabular}
\label{tab0202}
\end{table}

%
%
\begin{table}
\caption{
$E1$-transition probabilities ($\mbox{\rm s}^{-1}$) between
{$(1s2p) \,\confcpa$}
and {$(1s1s) \,\confaso$}
configurations.
}
\begin{tabular}{rrrrrrrrrrr}
\hline
\multicolumn{1}{c}{$Z$}&
\multicolumn{1}{c}{$W^{\txt{1ph}}_{\Lg}$}&
\multicolumn{1}{c}{$W^{\txt{1ph}}_{\Vg}$}&
\multicolumn{1}{c}{$W^{\txt{RMBPT}}_{\Lg}$}&
\multicolumn{1}{c}{$W^{\txt{RMBPT}}_{\Vg}$}&
\multicolumn{1}{c}{$W_{\Lg}$}&\multicolumn{1}{c}{$\Delta W^{(2+)}_{\Lg}$}&
\multicolumn{1}{c}{$W_{\Vg}$}&\multicolumn{1}{c}{$\Delta W^{(2+)}_{\Vg}$}&
Ref.\ \cite{drake79}&
Ref.\ \cite{johnson95}
\\
\hline
10&3.096[ 9]&2.917[ 9] &5.211[ 9]&4.963[ 9] &5.351(140)[ 9]&0.003[ 9]&5.095[ 9]&0.15[ 9] &5.356[ 9]&5.356[ 9]\\
18&1.391[12]&1.370[12] &1.793[12]&1.772[12] &1.799(6)[12]&         &1.777[12]&0.011[12]&1.800[12]&1.799[12]\\
26&3.925[13]&3.898[13] &4.482[13]&4.419[13] &4.421(61)[13]&         &4.396[13]&0.011[13]&4.425[13]&4.421[13]\\
30&1.160[14]&1.154[14] &1.258[14]&1.254[14] &1.251(7)[14]&         &1.246[14]&0.002[14]&1.252[14]&1.251[14]\\
40&6.846[14]&6.826[14] &7.047[14]&7.041[14] &7.013(34)[14]&         &6.997[14]&0.007[14]&7.017[14]&7.011[14]\\
50&2.104[15]&2.101[15] &2.132[15]&2.133[15] &2.123(9)[15]&         &2.120[15]&0.001[15]&2.123[15]&2.120[15]\\
60&4.838[15]&4.832[15] &4.874[15]&4.879[15] &4.855(19)[15]&         &4.850[15]&0.002[15]&4.853[15]&4.845[15]\\
70&9.472[15]&9.463[15] &9.523[15]&9.538[15] &9.497(26)[15]&         &9.489[15]&0.003[15]&9.480[15]&9.460[15]\\
80&1.672[16]&1.671[16] &1.680[16]&1.683[16] &1.674(6)[16]&         &1.673[16]&         &1.672[16]&1.668[16]\\
92&3.007[16]&3.005[16] &3.020[16]&3.027[16] &3.008(12)[16]&         &3.006[16]&         &         &2.994[16]\\
\hline
\end{tabular}
\label{tab0203}
\end{table}

%
%
\begin{table}
\caption{
$E1$-transition probabilities ($\mbox{\rm s}^{-1}$) between
{$(1s2p) \,\confapa$}
and {$(1s1s) \,\confaso$}
configurations.
}
\begin{tabular}{rrrrrrrrrrr}
\hline
\multicolumn{1}{c}{$Z$}&
\multicolumn{1}{c}{$W^{\txt{1ph}}_{\Lg}$}&
\multicolumn{1}{c}{$W^{\txt{1ph}}_{\Vg}$}&
\multicolumn{1}{c}{$W^{\txt{RMBPT}}_{\Lg}$}&
\multicolumn{1}{c}{$W^{\txt{RMBPT}}_{\Vg}$}&
\multicolumn{1}{c}{$W_{\Lg}$}&\multicolumn{1}{c}{$\Delta W^{(2+)}_{\Lg}$}&
\multicolumn{1}{c}{$W_{\Vg}$}&\multicolumn{1}{c}{$\Delta W^{(2+)}_{\Vg}$}&
Ref.\ \cite{drake79}&
Ref.\ \cite{johnson95}
\\
\hline
10&8.607[12]&8.071[12] &8.538[12]&8.107[12] &8.538(14)[12]&0.007[12]&8.103[12]&0.27[12] &8.851[12]&8.851[12]\\
18&1.069[14]&1.052[14] &1.061[14]&1.048[14] &1.061(1)[14]&         &1.047[14]&0.007[14]&1.071[14]&1.070[14]\\
26&4.611[14]&4.578[14] &4.551[14]&4.529[14] &4.553(2)[14]&         &4.526[14]&0.01[14] &4.570[14]&4.566[14]\\
30&7.857[14]&7.815[14] &7.747[14]&7.723[14] &7.754(7)[14]&         &7.720[14]&0.016[14]&7.773[14]&7.763[14]\\
40&2.234[15]&2.227[15] &2.211[15]&2.209[15] &2.215(4)[15]&         &2.210[15]&0.002[15]&2.216[15]&2.212[15]\\
50&5.096[15]&5.085[15] &5.064[15]&5.066[15] &5.074(10)[15]&         &5.065[15]&0.004[15]&5.071[15]&5.057[15]\\
60&1.013[16]&1.012[16] &1.009[16]&1.010[16] &1.011(2)[16]&         &1.010[16]&         &1.010[16]&1.006[16]\\
70&1.819[16]&1.817[16] &1.813[16]&1.816[16] &1.816(3)[16]&         &1.814[16]&         &1.813[16]&1.805[16]\\
80&3.010[16]&3.007[16] &3.002[16]&3.009[16] &3.008(6)[16]&         &3.006[16]&0.001[16]&3.000[16]&2.986[16]\\
92&5.046[16]&5.043[16] &5.034[16]&5.049[16] &5.045(11)[16]&         &5.041[16]&0.001[16]&         &5.001[16]\\
\hline
\end{tabular}
\label{tab0204}
\end{table}

%


%
%
\begin{table}
\caption{
$M1$-transition probabilities ($\mbox{\rm s}^{-1}$) between
{ $(1s2s)\,\confcsa$}
and { $(1s1s)\,\confaso$}
configurations.
The digits in square brackets denote the power of $10$.
}
\begin{tabular}{rr|rr|rr|rr|rr}
\hline
\multicolumn{1}{c}{$Z$}&
\multicolumn{1}{c|}{$W$}&
\multicolumn{1}{c}{$Z$}&
\multicolumn{1}{c|}{$W$}&
\multicolumn{1}{c}{$Z$}&
\multicolumn{1}{c|}{$W$}&
\multicolumn{1}{c}{$Z$}&
\multicolumn{1}{c|}{$W$}&
\multicolumn{1}{c}{$Z$}&
\multicolumn{1}{c}{$W$}
\\
\hline
&&21&2.332(6)[7]&41&2.221(2)[10]&61&1.386(2)[12]&81&2.922(7)[13]\\
&&22&3.757(9)[7]&42&2.847(2)[10]&62&1.644(3)[12]&82&3.357(8)[13]\\
&&23&5.923(13)[7]&43&3.629(2)[10]&63&1.949(4)[12]&83&3.825(10)[13]\\
&&24&9.158(11)[7]&44&4.600(3)[10]&64&2.301(4)[12]&84&4.388(12)[13]\\
&&25&1.391(3)[8]&45&5.804(3)[10]&65&2.716(5)[12]&85&5.007(12)[13]\\
&&26&2.078(6)[8]&46&7.283(4)[10]&66&3.191(6)[12]&86&5.679(14)[13]\\
&&27&3.058(6)[8]&47&9.102(5)[10]&67&3.745(7)[12]&87&6.499(16)[13]\\
&&28&4.438(7)[8]&48&1.132(1)[11]&68&4.387(8)[12]&88&7.350(19)[13]\\
&&29&6.357(9)[8]&49&1.402(2)[11]&69&5.136(10)[12]&89&8.400(22)[13]\\
10&1.097(7)[4]&30&8.987(21)[8]&50&1.727(2)[11]&70&5.980(10)[12]&90&9.468(23)[13]\\
11&2.966(13)[4]&31&1.259(2)[9]&51&2.125(2)[11]&71&6.980(12)[12]&91&1.081(3)[14]\\
12&7.324(30)[4]&32&1.742(2)[9]&52&2.600(3)[11]&72&8.101(15)[12]&92&1.216(3)[14]\\
13&1.678(7)[5]&33&2.388(2)[9]&53&3.172(4)[11]&73&9.419(18)[12]&93&1.387(4)[14]\\
14&3.609(13)[5]&34&3.244(3)[9]&54&3.852(5)[11]&74&1.089(2)[13]&94&1.555(4)[14]\\
15&7.354(23)[5]&35&4.367(4)[9]&55&4.671(7)[11]&75&1.260(2)[13]&95&1.774(5)[14]\\
16&1.430(5)[6]&36&5.831(5)[9]&56&5.637(8)[11]&76&1.455(3)[13]&96&2.004(6)[14]\\
17&2.667(9)[6]&37&7.726(6)[9]&57&6.789(11)[11]&77&1.677(4)[13]&97&2.262(7)[14]\\
18&4.798(18)[6]&38&1.016(1)[10]&58&8.144(13)[11]&78&1.931(4)[13]&98&2.550(8)[14]\\
19&8.358(15)[6]&39&1.327(1)[10]&59&9.753(17)[11]&79&2.220(5)[13]&99&2.875(10)[14]\\
20&1.414(4)[7]&40&1.722(1)[10]&60&1.163(2)[12]&80&2.548(6)[13]&100&3.193(10)[14]\\
\hline
\end{tabular}
\\
\label{xtab0202}
\end{table}

%
%


\newpage
%
%

%
%

%
%

%
%

%
%

%
%

%
%

%
%

%
%

%
%

%
%

%
%

%
%

%
%
\begin{figure}
\centerline{\mbox{\epsfysize=170pt \epsffile{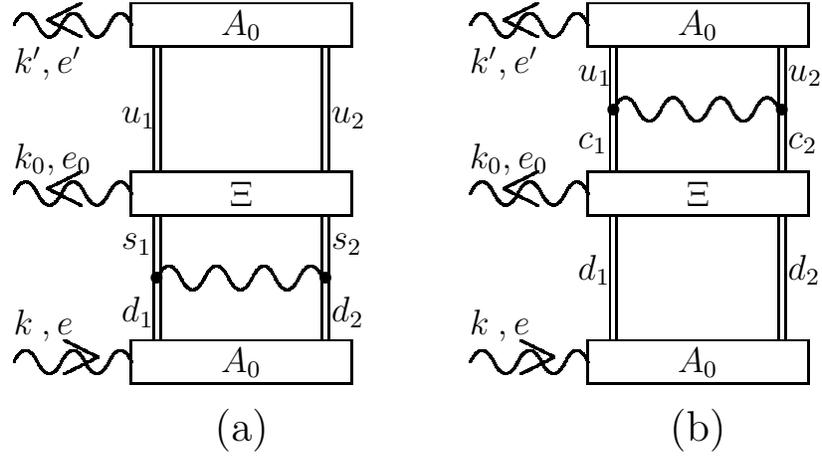}}}
\caption{\label{figure31}
The Feynman graphs representing the process of elastic photon
scattering on the two-electron ion.
In the lowest order of the perturbation theory these graphs reduce to the graphs
\Fig{figure30}.
}
\end{figure}

%
%
\begin{figure}
\centerline{\mbox{\epsfysize=220pt \epsffile{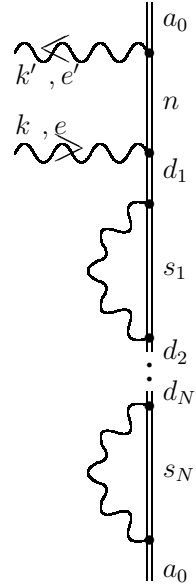}}}
\caption{\label{figure163}
The Feynman graph representing the process of elastic photon
scattering on the one-electron ion.
Multiple insertions of the self-energy operator into the lower outer electron line are made in the framework of the adiabatic approximation.
The break in the electron lines denotes the possible multiple insertions.
}
\end{figure}

%
%

\newpage

%

%

%

\end{document}